\documentstyle[preprint,aps,rotating,epsfig]{revtex}

\def\ETslash{\not{\hbox{\kern-4pt $E_T$}}}

\def\ra{\rightarrow}

\def\MWW {M_{WW} }
\def\WWWW {W_L W_L \ra W_L W_L}

\def\WPWM{ W^+(\ra \ell^+\nu) W^-(\ra q_1 \bar q_2) }
\def\WPZ{ W^+(\ra \ell^+ \nu) Z^0(\ra q \bar q) }
\def\WPWP{ W^+(\ra \ell^+ \nu) W^+(\ra \ell^+ \nu) }
\def\etc{ {\it etc.}}
\def\dis{\displaystyle}

\def\be{\begin{equation}}
\def\ee{\end{equation}}
\def\bea{\begin{eqnarray}}
\def\eea{\end{eqnarray}}

\def\D0{D\O~}

\def\ifb{ {\rm fb}^{-1} }

\def\ra{\rightarrow}

\def\dbrackl{\left[\mkern-5mu\left[}
\def\dbrackr{\right]\mkern-5mu\right]}
\def\Dbrack#1{\dbrackl#1\dbrackr}

\newfont{\titlefont}{cmbx10 scaled 1270}      
\newfont{\titlefonta}{cmbx10 scaled 1100}      
\catcode`\@=11
\long\def\@makefntext#1{
\protect\noindent \hbox to 3.2pt {\hskip-.9pt
$^{{\ninerm\@thefnmark}}$\hfil}#1\hfill}                
\def\@makefnmark{\hbox to 0pt{$^{\@thefnmark}$\hss}}  
\def\ps@myheadings{\let\@mkboth\@gobbletwo
\def\@oddhead{\hbox{}
\rightmark\hfil\ninerm\thepage}
\def\@oddfoot{}\def\@evenhead{\ninerm\thepage\hfil
\leftmark\hbox{}}\def\@evenfoot{}
\def\sectionmark##1{}\def\subsectionmark##1{}}
\renewcommand{\thefootnote}{\fnsymbol{footnote}}

\renewenvironment{thebibliography}[1]
    {\begin{list}{\arabic{enumi}.}
    {\usecounter{enumi}\setlength{\parsep}{0pt}
\setlength{\leftmargin 1.25cm}{\rightmargin 0pt}
     \setlength{\itemsep}{0pt} \settowidth
    {\labelwidth}{#1.}\sloppy}}{\end{list}}
\newcounter{itemlistc}
\newcounter{romanlistc}
\newcounter{alphlistc}
\newcounter{arabiclistc}

\newcommand{\tcaption}[1]{
        \refstepcounter{table}
        \setbox\@tempboxa = \hbox{\footnotesize Table~\thetable. #1}
        \ifdim \wd\@tempboxa > 6in
           {\begin{center}
        \parbox{6in}{\footnotesize\baselineskip=15pt Table~\thetable. #1}
            \end{center}}
        \else
             {\begin{center}
             {\footnotesize Table~\thetable. #1}
              \end{center}}
        \fi}
 1 
 1  
 1  
 1  
 1  
 1  
 1  

\font\ninerm=cmr9

\evensidemargin=\oddsidemargin  
\renewcommand{\baselinestretch}{1}
\def\today{\number\day
           \space\ifcase\month\or
             January\or February\or March\or April\or May\or June\or
             July\or August\or September\or October\or November\or December\fi
           \space\number\year}

\evensidemargin=\oddsidemargin  
\def\doublespaced{\baselineskip=\normalbaselineskip\multiply
    \baselineskip by 150\divide\baselineskip by 100}
\doublespaced

\begin{document}

\addtolength{\textheight}{2.4cm}
\addtolength{\topmargin}{0.1cm}

\begin{titlepage}
\thispagestyle{empty}
\renewcommand{\thefootnote}{\fnsymbol{footnote}}
\setcounter{footnote}{0}
\begin{flushright}
March, 1997      \hfill      {\small DESY-97-056}\\ 
{hep-ph/9704276} \hfill      {\small TUIMP-TH-97/84}\\              
{\small MSUHEP-70105}
\end{flushright}
\vspace{0.5cm}

\centerline{\large\sc
                 Global Analysis for Probing Electroweak Symmetry}
\baselineskip=20pt
\centerline{\large\sc 
                   Breaking Mechanism at High Energy Colliders
\footnote{
~~Lectures given by C.--P. Yuan at the CCAST Workshop on 
`` {\it Physics at TeV Energy Scale} '', July 15-26, 1996, Beijing, China.}}

\vspace*{1.2cm}
\centerline{
{\sc Hong-Jian He }~$^{(a)}$,~~~~ 
{\sc Yu-Ping Kuang}~$^{(b)}$,~~~~
{\sc C.--P. Yuan  }~$^{(c)}$ }
  
\vspace*{0.2cm}

\baselineskip=17pt
\centerline{\it
$^{(a)}$ Theory Division, Deutsches Elektronen-Synchrotron DESY} 
\baselineskip=14pt
\centerline{\it D-22603 Hamburg , Germany }
\vspace*{0.2cm}
\centerline{\it
$^{(b)}$ 
CCAST ( World Laboratory ), P.O.Box 8730, Beijing 100080, China }
\baselineskip=14pt
\centerline{\it
Institute of Modern Physics, Tsinghua University, Beijing 100084, China}
\vspace*{0.2cm}
\centerline{\it
$^{(c)}$
{ Department of Physics and Astronomy, Michigan State University } }
\baselineskip=14pt
\centerline{\it East Lansing, Michigan 48824, USA }

\renewcommand{\baselinestretch}{1}
                     \begin{abstract} 
\noindent
We review our recent global analysis for probing the electroweak symmetry
breaking (EWSB) mechanism by the universal effective Lagrangian
description. After summarizing and commenting upon the current bounds for 
the EWSB parameters, we focus on testing them at the forthcoming high 
energy colliders such as the CERN LHC and the future linear colliders.  
We develop a precise electroweak power counting method
(\`{a} la Weinberg) and formulate 
the longitudinal-Goldstone boson equivalence 
as a necessary criterion for sensitively probing the EWSB sector.
Armed with these, we systematically estimate and classify the contributions 
of {\it all} next-to-leading order (NLO) bosonic operators to various 
scattering processes at the high energy colliders. 
The experimental signatures at these colliders are also analyzed. 
Furthermore, the {\it complementarity} of different scattering processes 
via different colliders for a complete probe of all these NLO effective 
operators is demonstrated.\\
PACS number(s): 11.30.Qc, 11.15.Ex, 12.15.Ji, 14.70.--e     
                     \end{abstract}

\end{titlepage}

\addtolength{\textheight}{-2.4cm}
\addtolength{\topmargin}{0.0cm}
\addtolength{\textheight}{0.5cm}

\normalsize
\baselineskip=19.5pt
\setcounter{footnote}{00}
\renewcommand{\thefootnote}{\arabic{footnote}}
\renewcommand{\baselinestretch}{1}

\newpage

\noindent
{\bf Contents:} \hfill Pages  \\[0.4cm]
{\bf 1.} Introduction \dotfill 3 \\[0.2cm]
{\bf 2.} Effective Lagrangians for Strongly Interacting 
EWSB Sector \dotfill 4 \\[0.2cm]
{\bf 3.} Current Bounds on the EWSB Parameters \dotfill 7 \\[0.2cm]
{\bf 4.} Precise Electroweak Power Counting Method \dotfill 13 \\[0.2cm]
 {\it 4.1. Generalizing Weinberg's 
 power counting rule to SSB gauge theories} \dotfill 13\\
 {\it 4.2. Constructing the precise electroweak 
 power counting method}  \dotfill 15\\
 {\small\it 4.2.1. Non-decoupling scenario (without Higgs boson)}
 \dotfill 15\\
 {\small\it 4.2.2. Decoupling scenario (with Higgs boson)} 
 \dotfill 18\\[0.2cm]
{\bf 5.} Longitudinal-Goldstone Boson Equivalence Theorem: 
   from mathematical formulation to
   its physical content for probing EWSB \dotfill 22 \\[0.2cm]
 {\it 5.1. ET and the Higgs mechanism} \dotfill 22 \\
 {\it 5.2. ET and its multiplicative modifications} \dotfill 27\\
 {\it 5.3. ET and its additive modifications} \dotfill 34 \\
 {\small\it 5.3.1. The Longitudinal-transverse ambiguity and the
 precise formulation of the ET} \dotfill 34 \\
 {\small\it 5.3.2. On the kinematic singularity} \dotfill  37\\
 {\it 5.4. Formulating the ET as a Criterion for probing the EWSB} 
  \dotfill 42 \\[0.2cm]
{\bf 6.} Global Classification for the S-matrix Elements \dotfill 45\\[0.2cm]
  {\it 6.1. Power Counting Hierarchy} \dotfill 45\\
  {\it 6.2. Classifications for $VV$-fusions and 
  $f\bar{f}'$-annihilations} \dotfill 48\\
  {\it 6.3. A comparative analysis for the LHC and LCs} \dotfill 50\\[0.2cm]
{\bf 7.} Global Analysis at the Event Rate Level \dotfill  54\\[0.2cm]
  {\it 7.1. Effective-$W$ approximation} \dotfill 54 \\
  {\it 7.2. At the LHC} \dotfill 57\\
  {\small\it 7.2.1. Preliminaries   }  \dotfill 57\\
  {\small\it 7.2.2. Analyzing the model-independent contributions to the 
   event rates  }  \dotfill  59\\
  {\small\it 7.2.3. Estimating the sensitivities for probing the 
   model-dependent operators }    \dotfill  60\\
  {\it 7.3. At the LCs } \dotfill 63\\\\[0.2cm]

\noindent
{\small\it (Continued)} \hfill  Pages\\[0.7cm]
{\bf 8.} How to Study TeV $\WWWW$ 
Interactions Experimentally \dotfill 64\\[0.2cm]
  {\it 8.1. Signal}  \dotfill 64\\
  {\it 8.2. Backgrounds} \dotfill 65\\
  {\it 8.3. How to distinguish signal from backgrounds} \dotfill 68\\
  {\it 8.4. Various models} \dotfill 72\\
  {\small\it 8.4.1. A TeV scalar resonance} \dotfill 72\\
  {\small\it 8.4.2. A TeV vector resonance} \dotfill 74\\
  {\small\it 8.4.3. No resonance }  \dotfill 74 \\[0.2cm]
{\bf 9.} Conclusion  \dotfill 76  \\[0.2cm]
Acknowledgments \dotfill 79  \\[0.2cm]
References \dotfill  80 \\[0.4cm]
Table Captions \dotfill 87\\
Figure Captions \dotfill 89\\
Table I-VII  \dotfill 91-99\\
Figure 1-12  \dotfill 100-115

\newpage
\addtolength{\topmargin}{0.2cm}
\addtolength{\textheight}{-0.4cm}
\baselineskip=19.5pt
\renewcommand{\baselinestretch}{1}

\noindent
{\bf 1. Introduction}
\vspace{0.3cm}

The Standard Model (SM) has successfully passed all available 
experimental tests with high precisions, but still leaves its electroweak 
symmetry breaking (EWSB) sector undetermined. 
Due to Veltman's screening theorem~\cite{screening}, the current data, 
allowing the SM Higgs boson mass to range 
from $65.2$\,GeV to about $O(1)$\,TeV~\cite{LEP2},
tell us little about the EWSB mechanism.
The light resonance(s) originating from the EWSB sector with mass(es) 
well below the TeV scale can exist possibly in the SM and necessarily in 
its supersymmetric extensions.
In such cases,  these particles should be 
detected~\cite{discovery,wwlhc,wwnlc}
at the high energy colliders such as the CERN Large Hadron Collider (LHC),
a proton-proton collider,  
and the future electron (and photon) Linear Colliders (LC)~\cite{lcws93},
even though the current direct experimental searches so far are all negative.
If the EWSB is, however, driven by strong interactions with no new 
resonance well below the TeV scale,  then it will be a {\it greater
challenge} to future colliders to decisively probe the EWSB mechanism. 
This latter case is what our present global analysis will focus on.

At the scale below new heavy resonances, 
the EWSB sector can be parametrized by means of the
electroweak chiral Lagrangian (EWCL) in which the 
$SU(2)_L \otimes U(1)_Y$ gauge symmetry is nonlinearly realized.
Without experimental observation of any
new light resonance in the EWSB sector, this effective
field theory approach provides the most economic description of
the possible new physics effects 
and is thus {\it complementary} to those specific model 
buildings~\cite{peskin-snow}.
In the present analysis, taking this 
general EWCL approach, we shall concentrate on 
studying the effective bosonic operators among which 
the leading order operators are universal 
and the next-to-leading-order (NLO) operators describe 
the model-dependent new effects.

In performing such a global analysis, in contrast
to just studying a few operators in the literature, we 
need to estimate the contributions of all the 
NLO operators to various high energy scattering processes. For this
purpose, we construct   
a precise electroweak power counting rule
for the EWCL formalism through a natural generalization of Weinberg's
counting method for non-linear sigma model~\cite{wei}. 
This simple power counting rule proves
to be extremely convenient and useful for our global 
analysis~\cite{global1,global2}.

Furthermore, we analyze how the longitudinal-Goldstone boson equivalence
theorem (ET) is deeply rooted in the underlying (elementary or dynamical) 
Higgs mechanism and demonstrate under what conditions this {\it equivalence}
can be manifest in the physical processes. The intrinsic
connection of the longitudinal-Goldstone boson {\it equivalence} to probing
the EWSB sector is thus revealed~\cite{et3}. 
We theoretically formulate this {\it equivalence} as
a {\it necessary criterion} for sensitively probing the EWSB sector.
Armed with these, we globally analyze and classify the sensitivities of 
all the effective operators for probing the EWSB mechanism
at the high energy colliders~\cite{et3,global1,global2}. 
This global classification will be 
performed at both the $S$-matrix element and event-rate levels. 
We note that at the event-rate level the sensitivity of a collider will
also depend on the detection efficiency for suppressing 
the backgrounds to observe the specific decay mode of the final state
for each given process (cf. Ref.~\cite{wwlhc}). 
Some general features of the event structure in the 
experimental signals and backgrounds
will be discussed and summarized (cf. Sec.~8).

\vspace{0.7cm}
\noindent
{\bf 2.  Effective Lagrangians for Strongly Interacting EWSB Sector}
\vspace{0.3cm}
       
The electroweak chiral Lagrangian (EWCL)  
gives the most economical description of the EWSB sector below the
scale of new heavy resonance and can be 
constructed as follows~\cite{app,peccei}:
$$
\begin{array}{ll}
{\cal L}_{\rm eff}& 
= \displaystyle\sum_n 
\ell_n\displaystyle\frac{f_\pi~^{r_n}}{\Lambda^{a_n}}
{\cal O}_n({\bf W_{\mu\nu},B_{\mu\nu}},D_\mu U,U,f,\bar{f})
= {\cal L}_{\rm G} + {\cal L}_{\rm S} + {\cal L}_{\rm F}
\end{array}
\eqno(2.1)                          
$$
where 
$$
\begin{array}{l}
D_{\mu}U =\partial_{\mu}U + ig{\bf W}_{\mu}U -ig^{\prime}U{\bf B}_{\mu}~,
\\[0.25cm]
U=\exp [i\tau^a\pi^a/f_\pi ]~,~~~
{\bf W}_{\mu}\equiv W^a_{\mu}\displaystyle\frac{\tau^a}{2}~,~~~
{\bf B}_{\mu}\equiv B_{\mu}\displaystyle\frac{\tau^3}{2}~,\\[0.20cm]
{\bf W}_{\mu\nu} =\partial_{\mu}{\bf W}_{\nu}-\partial_{\nu}{\bf W}_{\mu}
                      +ig [{\bf W}_\mu , {\bf W}_\nu ] ~,\\[0.20cm]
B_{\mu\nu}= \partial_\mu B_\nu - \partial_\nu B_\mu~,\\[0.20cm]           
\end{array}
$$
$\pi^a$ is the would-be Goldstone boson (GB) field and  
$f$($\bar{f}$) is the SM fermion with mass $~m_{f}\leq 
O(m_t)\simeq O(M_W)~$.
$~{\cal L}_{\rm G}~$, $~{\cal L}_{\rm S}~$ and $~{\cal L}_{\rm F}~$ denote
 gauge boson kinetic terms, scalar boson interaction terms
(containing GB self-interactions and gauge-boson-GB interactions), 
and fermion interaction terms, respectively.
Here we concentrate on probing 
new physics from all possible bosonic operators so that we shall 
not include the next-to-leading order fermionic operators 
in $~{\cal L}_F~$. 
For clearness, we have factorized out the dimensionful parameters 
$~f_\pi~$ and $~\Lambda~$ in the coefficients so that the dimensionless 
factor $~\ell_n~$ is of  $~O(1)~$. This makes our definitions of the
$~\ell_n$'s different from the $~\alpha_i$'s in Ref.~\cite{app}
by a factor of $~(f_{\pi}/\Lambda )^2~$.
We note that  $~f_\pi~$  and $~\Lambda~$ 
are the two essential scales in any effective Lagrangian
that describes the spontaneously broken symmetry.  The former determines the 
symmetry breaking scale while the latter determines the scale at which
new resonance(s) besides the light fields 
(such as the SM
weak bosons, would-be Goldstone bosons and fermions) may appear. 
In the non-decoupling scenario, 
the effective cutoff scale $\Lambda$ cannot be arbitrarily large:
$~~\Lambda = \min (M_{SB},\Lambda_0 )\leq\Lambda_0~~$ \cite{georgi}, 
where $~~\Lambda_0\equiv 4\pi f_\pi\simeq 3.1$~TeV and
$~M_{SB}~$ is the mass of the 
lightest new resonance in the EWSB sector.
In (2.1), $~~ r_n=4+a_n-D_{{\cal O}_n} ~$, where $~D_{{\cal O}_n}
={\rm dim}({\cal O}_n)~$.  
The power factor $~\Lambda^{a_n}~$ associated with each operator
$~{\cal O}_n~$ can be counted by the naive dimensional analysis 
(NDA)~\cite{georgi}\footnote{ ~In this paper, 
the NDA is only used to count the 
$~\Lambda$-powers  ($~\Lambda^{a_n}~$) associated with the 
operators $~{\cal O}_n~$'s in the chiral Lagrangian (2.1). 
This is irrelevant to the derivation of the
power counting rule for $~D_E~$ in the following (3.2.4).}.
~For the bosonic part of EWCL, we have~\cite{app}:
$$
\begin{array}{ll}
{\cal L}_{\rm G} &
 =~ -\frac{1}{2}{\rm Tr}({\bf W}_{\mu\nu}{\bf W}^{\mu\nu})
              -\frac{1}{4}B_{\mu\nu}B^{\mu\nu}  ~~,\\[0.25cm]
{\cal L}_{\rm S} & 
 = {\cal L}^{(2)}+{\cal L}^{(2)\prime}+
             \displaystyle\sum_{n=1}^{14} {\cal L}_n ~~,\\[0.25cm]
{\cal L}^{(2)} & =
   \frac{f_\pi^2}{4}{\rm Tr}[(D_{\mu}U)^\dagger(D^{\mu}U)]   ~~,\\[0.20cm]
{\cal L}^{(2)\prime} & =\ell_0 (\frac{f_\pi}{\Lambda})^2~\frac{f_\pi^2}{4}
               [ {\rm Tr}({\cal T}{\cal V}_{\mu})]^2 ~~,\\[0.2cm]
{\cal L}_1 & = \ell_1 (\frac{f_\pi}{\Lambda})^2~ \frac{gg^\prime}{2}
B_{\mu\nu} {\rm Tr}({\cal T}{\bf W^{\mu\nu}}) ~~,\\[0.2cm]
{\cal L}_2 & = \ell_2 (\frac{f_\pi}{\Lambda})^2 ~\frac{ig^{\prime}}{2}
B_{\mu\nu} {\rm Tr}({\cal T}[{\cal V}^\mu,{\cal V}^\nu ]) ~~,\\[0.2cm]
{\cal L}_3 & = \ell_3 (\frac{f_\pi}{\Lambda})^2 ~ig
{\rm Tr}({\bf W}_{\mu\nu}[{\cal V}^\mu,{\cal V}^{\nu} ]) ~~,\\[0.2cm]
{\cal L}_4 & = \ell_4 (\frac{f_\pi}{\Lambda})^2 
                     [{\rm Tr}({\cal V}_{\mu}{\cal V}_\nu )]^2 ~~,\\ [0.2cm] 
{\cal L}_5 & = \ell_5 (\frac{f_\pi}{\Lambda})^2 
                     [{\rm Tr}({\cal V}_{\mu}{\cal V}^\mu )]^2 ~~,\\  [0.2cm]
{\cal L}_6 & = \ell_6 (\frac{f_\pi}{\Lambda})^2 
[{\rm Tr}({\cal V}_{\mu}{\cal V}_\nu )]
{\rm Tr}({\cal T}{\cal V}^\mu){\rm Tr}({\cal T}{\cal V}^\nu) ~~,\\[0.2cm]
{\cal L}_7 & = \ell_7 (\frac{f_\pi}{\Lambda})^2 
[{\rm Tr}({\cal V}_\mu{\cal V}^\mu )]
{\rm Tr}({\cal T}{\cal V}_\nu){\rm Tr}({\cal T}{\cal V}^\nu) ~~,\\[0.2cm]
{\cal L}_8 & = \ell_8 (\frac{f_\pi}{\Lambda})^2~\frac{g^2}{4} 
[{\rm Tr}({\cal T}{\bf W}_{\mu\nu} )]^2  ~~,\\[0.2cm]
\end{array}
$$

$$
\begin{array}{ll}
{\cal L}_9 & = \ell_9 (\frac{f_\pi}{\Lambda})^2 ~\frac{ig}{2}
{\rm Tr}({\cal T}{\bf W}_{\mu\nu}){\rm Tr}
        ({\cal T}[{\cal V}^\mu,{\cal V}^\nu ]) ~~,\\[0.2cm]
{\cal L}_{10} & = \ell_{10} (\frac{f_\pi}{\Lambda})^2\frac{1}{2}
[{\rm Tr}({\cal T}{\cal V}^\mu){\rm Tr}({\cal T}{\cal V}^{\nu})]^2 ~~,\\[0.2cm]
{\cal L}_{11} & = \ell_{11} (\frac{f_\pi}{\Lambda})^2 
~g\epsilon^{\mu\nu\rho\lambda}
{\rm Tr}({\cal T}{\cal V}_{\mu}){\rm Tr}
({\cal V}_\nu {\bf W}_{\rho\lambda}) ~~, \\[0.2cm]
{\cal L}_{12} & = \ell_{12}(\frac{f_\pi}{\Lambda})^2 ~2g
                    {\rm Tr}({\cal T}{\cal V}_{\mu}){\rm Tr}
                  ({\cal V}_\nu {\bf W}^{\mu\nu}) ~~,\\[0.2cm]
{\cal L}_{13} & = \ell_{13}(\frac{f_\pi}{\Lambda})^2~ 
      \frac{gg^\prime}{4}\epsilon^{\mu\nu\rho\lambda}
      B_{\mu\nu} {\rm Tr}({\cal T}{\bf W}_{\rho\lambda}) ~~,\\[0.2cm]     
{\cal L}_{14} & = \ell_{14} (\frac{f_\pi}{\Lambda})^2~\frac{g^2}{8} 
\epsilon^{\mu\nu\rho\lambda}{\rm Tr}({\cal T}{\bf W}_{\mu\nu})
{\rm Tr}({\cal T}{\bf W}_{\rho\lambda})   ~~,\\[0.2cm]
\end{array}
\eqno(2.2)                                         
$$

\noindent
where
$~{\cal V}_{\mu}\equiv (D_{\mu}U)U^\dagger~$,  
and $~{\cal T}\equiv U\tau_3 U^{\dagger} ~$.
There is certain arbitrariness in choosing the complete set of operators 
which can be related to another set after applying the equation of motion,
but this will not affect the physical results~\cite{eom}.
Eq.~(2.2) contains fifteen bosonic 
NLO effective operators among which
there are twelve $CP$-conserving operators 
($~{\cal L}^{(2)\prime},{\cal L}_{1\sim 11}~$) and three $CP$-violating
operators (~${\cal L}_{12\sim 14}~$).
Furthermore, the operators $~{\cal L}_{6,7,10}~$ 
violate custodial $SU(2)_C$ symmetry 
(~even after $g^{\prime}$ being turned off~) in contrary to
 $~{\cal L}_{4,5}~$ which contain $SU(2)_C$-invariant 
 pure GB interactions.
The coefficients ( $\ell_n$'s ) of all the above operators are 
model-dependent and carry 
information about possible new physics beyond the SM.
The dimension-$2$ custodial $SU(2)_C$-violating operator 
$~{\cal L}^{(2)\prime} ~$  has a coefficient at most of 
$~O(f_\pi^2/\Lambda^2 )~$ since it is  proportional to
$~\delta\rho \approx O(m_t^2/(16\pi^2 f_\pi^2))
\approx O(f_\pi^2 /\Lambda^2 )~$ for the top Yukawa coupling being 
of $~O(1)~$. 

In the non-decoupling scenario~\cite{georgi,review},
the coefficients for all 
NLO dimension-$4$ operators are suppressed 
by a factor $~(f_\pi/\Lambda )^2\approx 1/(16\pi^2)~$ 
relative to that of the universal dimension-$2$ operator ${\cal L}^{(2)}$,
because of the derivative expansion in terms of $~(D_\mu/\Lambda )^2~$.
If we ignore the small $CP$-violating effects from the 
 Cabibbo-Kobayashi-Maskawa mixings in the lowest order 
fermionic Lagrangian $~{\cal L}_{\rm F}~$,
all the one-loop level new divergences generated from 
$~{\cal L}_{\rm G}+{\cal L}_{\rm F}+{\cal L}^{(2)}~$ are thus $CP$-invariant. 
Therefore, the $CP$-violating operators $~{\cal L}_{12\sim 14}~$ are 
actually {\it decoupled} at this level, and their
coefficients can have values significantly larger or smaller than that from
the naive dimensional analysis~\cite{georgi}.
Since the true mechanism for $CP$-violation remains un-revealed,  we
shall consider in this paper the coefficients $~\ell_{12\sim 14}~$
to be around of $~O(1)~$.\\\\

\vspace{0.5cm}
\noindent
{\bf 3. Current Bounds on the EWSB Parameters}
\vspace{0.3cm}

Before proceeding further, we analyze the current constraints on these EWCL 
parameters in this section which is useful for our study on future
high energy colliders.
 First, the coefficients $~\ell_{0,1,8}~$ are 
related to the low energy $S,T,U$ parameters~\cite{STU} through the
oblique corrections:
$$
\begin{array}{l}
\ell_0~ =\displaystyle\left(\frac{\Lambda}
          {\Lambda_0}\right)^2 8\pi^2\alpha T 
        =\left(\frac{\Lambda}{\Lambda_0}\right)^2 8\pi^2 \delta\rho 
        =\displaystyle\left(\frac{\Lambda}{\Lambda_0}\right)^2 
         \frac{32\pi^3\alpha}{c_{\rm w}^2s_{\rm w}^2M_Z^2}
         [\Pi_{11}^{\rm new}(0)-\Pi_{33}^{\rm new}(0)] ~~,   \\[0.35cm]
\ell_1~ = -\displaystyle\left(\frac{\Lambda}{\Lambda_0}\right)^2 \pi S 
        = \displaystyle\left(\frac{\Lambda}{\Lambda_0}\right)^2 
          8\pi^2 \Pi_{3Y}^{{\rm new}~\prime}(0) ~~,\\[0.3cm]
\ell_8~ = -\displaystyle\left(\frac{\Lambda}{\Lambda_0}\right)^2 \pi U 
        = \displaystyle\left(\frac{\Lambda}{\Lambda_0}\right)^2 
          16\pi^2[\Pi_{33}^{{\rm new}~\prime}(0)
          -\Pi_{11}^{{\rm new}~\prime}(0)] ~~,
\end{array}
\eqno(3.1)                        
$$
where $~\Lambda_0\equiv 4\pi f_\pi\simeq 3.1$~TeV, and 
$~s_{\rm w}=\sin\theta_{\rm W},~c_{\rm w}=\cos\theta_{\rm W},
~\alpha =\frac{e^2}{4\pi}~$ are measured at $Z$-pole 
in the $\overline{\rm MS}$
scheme. The factor $~\left(\frac{\Lambda}{\Lambda_0}\right)^2~$ in (3.1)
reduces to one for the case  $~\Lambda =\Lambda_0~$.
The newest updated global fit to the low energy data gives \cite{PDG}:
$$
\begin{array}{lll}
S & = & -0.26 \pm 0.16  ~~,\\
T & = & ~~0.13\pm  0.20   ~~,\\
U & = & ~~0.07\pm 0.42    ~~,
\end{array}                         
\eqno(3.2)
$$ 
for $~s^2_{\rm w}=0.2314\pm 0.0002$ and $m_t=173\pm 6~$GeV. 
For the present analysis, we have specified the reference value of 
the SM Higgs mass as $~m_H=1$~TeV. 
Results for other values of $m_H$ can be found in Ref.~\cite{PDG}.
The inclusion or exclusion of ~$\delta R_b^{\rm new}$~ in the global
fit affects (3.2) very little~\cite{PDG}.
Since the experimental errors in (3.4) are still quite large, the
parameters ~$T$~ and ~$U$~ ($~\ell_0$ and $~\ell_8~$) can 
be either positive or negative within $1\sigma$ error. 
The parameter $~S~$ ($~\ell_1$~) 
is only about $~-1.63\sigma~$ below the SM value. 
This is even smaller than the present $R_b$ anomaly 
which is $~+1.86\sigma~$ above the SM value~\cite{PDG}. (Another recent
global fit~\cite{warsaw} gives $~+1.75\sigma~$ for the $R_b$ anomaly.) 

From (3.1) and (3.2), for $\Lambda\simeq 3.1$~TeV,
we get the following $2\sigma$ ($95.5\%$~C.L.) constraints on $~\ell_{0,1,8}~$
at the scale $~\mu =M_Z$:
$$
\begin{array}{rllll}
-0.17 & \leq & \ell_0  & \leq & 0.32  ~~,\\
-0.19 & \leq & \ell_1  & \leq & 1.82  ~~, \\
-2.86 & \leq & \ell_8  & \leq & 2.42  ~~,
\end{array}
\eqno(3.3)                        
$$
which allow ~$\ell_{0,1,8}$~ to be either positive or negative around of
$~O(1)~$.  We can further
deduce the bounds at the TeV scale ({\it e.g.}, $\mu =1$~TeV)
by incorporating the running effects from the 
renormalization $\log$-terms. 
By the one-loop renormalization calculation~\cite{long}, 
we find, for instance, the coefficients
$~\ell_{0,1,8}~$ are running as:
$$
\begin{array}{ll}  \left\{ \begin{array}{l} 
\ell_0 (\mu )=\ell_0^{b} - 
              \displaystyle\left(\frac{\Lambda}{\Lambda_0}\right)^2 
              \frac{3}{4}(g^{\prime})^2
              \left(\frac{1}{\hat\epsilon} + c_0 \right) ~,\\[0.3cm] 
\ell_1 (\mu )=\ell_1^{b} - 
              \displaystyle\left(\frac{\Lambda}{\Lambda_0}\right)^2 
              \frac{1}{6}
              \left(\frac{1}{\hat\epsilon} + c_1\right) ~,\\[0.3cm] 
\ell_8 (\mu )=\ell_8^{b} - 
             \displaystyle\left(\frac{\Lambda}{\Lambda_0}\right)^2 c_8 ~;
\end{array}    \right.
&
\Longrightarrow ~~\left\{ \begin{array}{l} 
\ell_0 (\mu^{\prime})=\ell_0(\mu ) + 
              \displaystyle\left(\frac{\Lambda}{\Lambda_0}\right)^2 
              \frac{3}{4}(g^{\prime})^2\ln\frac{\mu^{\prime}}{\mu}
              ~,\\[0.3cm] 
\ell_1 (\mu^{\prime})=\ell_1(\mu ) + 
              \displaystyle\left(\frac{\Lambda}{\Lambda_0}\right)^2 
              \frac{1}{6}\ln\frac{\mu^{\prime}}{\mu} 
               ~,\\[0.3cm] 
\ell_8 (\mu^{\prime})=\ell_8(\mu ) ~; 
\end{array}   \right. 
\end{array}
\eqno(3.4)                            
$$
where 
$~~\displaystyle
\frac{1}{\hat\epsilon}\equiv\frac{1}{4-n}- \ln\mu
~~$  and the superscript ``$~^{b}~$''  denotes the bare quantity. 
In (3.4), the $c_i~$'s are finite constants
which depend on the subtraction scheme and are irrelevant to the running
of $\ell_n(\mu )~$'s. From (3.3) and (3.4) we can deduce the
$2\sigma$ bounds at other scales, {\it e.g.}, at $\mu =1$~TeV 
for $\Lambda=\Lambda_0=3.1~$TeV,
$$
\begin{array}{rllll}
-0.06 & \leq & \ell_0(1~{\rm TeV}) & \leq & 0.55  ~~,\\
 0.21 & \leq & \ell_1(1~{\rm TeV}) & \leq & 2.22  ~~, \\
-2.86 & \leq & \ell_8(1~{\rm TeV}) & \leq & 2.42  ~~,
\end{array}
\eqno(3.5)                        
$$
where the ranges for ~$\ell_{0,1}$~ are slightly moved toward positive
direction due to the running effects. 
(3.5) shows that $~\ell_{0,1,8}~$  are 
allowed to be around of $O(1)$ except that the
parameter space for $~\ell_0$~ is
about a factor of ~$4\sim 5$~ smaller than the others.
All those NLO coefficients $\ell_n$'s in (2.2) 
varies for different underlying theories and thus must be
{\it independently tested} since the real underlying theory is unknown
and these operators are inequivalent by the equation of motion. 
For example, the $SU(2)_C$-violating operator
${\cal L}^{(2)\prime}$ (containing two ${\cal T}=U\tau_3U^\dagger$~'s)
is constrained to be slightly below $O(1)$ within $2\sigma$ bound
[cf. (3.3)(3.5)], while the updated data still allow 
the coefficients of other $SU(2)_C$-violating
operators such as ${\cal L}_1$ (containing one ${\cal T}$ operator) and  
${\cal L}_8$ (containing two ${\cal T}$ operators) 
to be around of $O(1)$ [cf. (3.3) and (3.5)].

Besides the bounds from the oblique corrections,
the tests on triple gauge boson couplings (TGCs) 
at LEP and Tevatron~\cite{TGC} impose further constraints on more operators 
at the tree level.  
In the conventional notation~\cite{TGC,TGC1}, the TGCs are parameterized as
$$
\begin{array}{ll}
\displaystyle\frac{{\cal L}_{WWV}}{g_{WWV}}= &
ig^V_1 [W^+_{\mu\nu}W^{-\mu}V^\nu - W^{-}_{\mu\nu}W^+_{\mu}V^\nu ]
+i\kappa_V W^+_\mu W^-_\nu V^{\mu\nu} + \frac{i\lambda_V}{\Lambda^2}
W^+_{\mu\nu}W^{-\nu}_{~~\rho}V^{\rho\mu}\\
& -g_4^V W^+_\mu W^-_\nu
  [\partial^\mu V^\nu + \partial^\nu V^\mu ]
 + g_5^V\epsilon^{\mu\nu\rho\lambda}[W^+_\mu\partial_\rho W^-_\nu
                              -W^-_\nu\partial_\rho W^+_\mu ]V_\lambda\\
& + i\tilde{\kappa}_V W^+_\mu W^-_\nu \tilde{V}^{\mu\nu}
+ i\frac{\tilde{\lambda}_V}{\Lambda^2}W^+_{\mu\nu}W^{-\nu}_{~~\rho}
  \tilde{V}^{\rho\mu} ~~,\\[0.4cm]
& V= Z, ~{\rm or }~ \gamma~,~~
  W^{\pm}_{\mu\nu}=\partial_{\mu}W^\pm_\nu-\partial_\nu W^\pm_\mu~,~~
 \tilde{V}_{\mu\nu} =\frac{1}{2}\epsilon_{\mu\nu\rho\lambda}V^{\rho\lambda}~.
\end{array}
\eqno(3.6)                  
$$
We summarize the tree level contributions from all $15$ NLO
order operators listed in (2.2) to the TGCs defined in (3.6) as follows:
$$
\begin{array}{l}
g_1^Z-1 \equiv \Delta g_1^Z = \displaystyle\frac{f^2_\pi}{\Lambda^2}
        \left[\frac{1}{c^2-s^2}\ell_0 +\frac{e^2}{c^2(c^2-s^2)}\ell_1
                                 +\frac{e^2}{s^2c^2}\ell_3 \right] ~~,~~~
g_1^\gamma -1 =0 ~~,\\[0.4cm]
g_4^Z=-\displaystyle\frac{f^2_\pi}{\Lambda^2}
\frac{e^2}{{\rm s}^2_{\rm w}{\rm c}^2_{\rm w}}\ell_{12}~~,~~~
g_4^\gamma =0 ~~,\\[0.4cm]
g_5^Z =\displaystyle\frac{f^2_\pi}{\Lambda^2}\frac{e^2}{s^2c^2}\ell_{11} ~~,
~~~g_5^\gamma =0 ~~,\\[0.4cm]
\kappa_Z -1\equiv\Delta\kappa_Z = \displaystyle\frac{f^2_\pi}{\Lambda^2}
\left[\frac{1}{c^2-s^2}\ell_0 +
 \frac{2e^2}{c^2-s^2}\ell_1 -\frac{e^2}{c^2}\ell_2
 +\frac{e^2}{s^2}(\ell_3 -\ell_8 +\ell_9 ) \right]~~,\\[0.5cm]
\kappa_\gamma -1 \equiv \Delta\kappa_\gamma
 =\displaystyle\frac{f^2_\pi}{\Lambda^2}\frac{e^2}{s^2}
    \left[ -\ell_1 +\ell_2 +\ell_3 -\ell_8 +\ell_9 \right]~~,\\[0.5cm]
\tilde{\kappa}_Z = \displaystyle\frac{f^2_\pi}{\Lambda^2}
\left[\frac{e^2}{{\rm c}_{\rm w}^2}\ell_{13}-
      \frac{e^2}{{\rm s}_{\rm w}^2}\ell_{14}\right]~~,~~~
\tilde{\kappa}_\gamma = -\displaystyle\frac{f^2_\pi}{\Lambda^2}
    \frac{e^2}{{\rm s}_{\rm w}^2}[\ell_{13}+\ell_{14}] ~~,
\end{array}
\eqno(3.7)                     
$$
which coincide with Ref.~\cite{app2} after taking into account the
difference in defining the coefficients.
(3.7) shows that, at the first non-trivial order 
[i.e., $O\left(\frac{f_\pi^2}{\Lambda^2}\right)$ ],  
only operators $~{\cal L}^{(2)\prime}~$ and
$~{\cal L}_{1,2,3,8,9,11\sim 14}~$ 
can contribute to anomalous triple gauge
couplings while $~{\cal L}^{(4)}_{4,5,6,7,10}~$ do not. 
Among $~{\cal L}^{(2)\prime}~$ and
$~{\cal L}_{1,2,3,8,9,11\sim 14}~$, $~{\cal L}^{(2)\prime}~$ and
$~{\cal L}_{1,8}~$ can be constrained by the oblique corrections
[cf. (2.5) and (2.7)] so that we are left with seven operators
$~{\cal L}^{(4)}_{2,3,9,11,12,13,14}~$. There are just seven independent
relations in (3.7) by which the coefficients
$~\ell_{2,3,9,11,12,13,14}~$ can be independently determined in principle
if the seven TGC parameters $~g^Z_{1,4,5}~$ and $~\kappa^{Z,\gamma},~
\tilde{\kappa}^{Z,\gamma}~$ can all be measured. 
{}From (3.7), we derive
$$
\begin{array}{l}
\ell_2 = \displaystyle\frac{1}{e^2}
 \frac{s^2_{\rm w}c^2_{\rm w}}{c^2_{\rm w}-s^2_{\rm w}}\ell_0
 +\frac{c^2_{\rm w}}{c^2_{\rm w}-s^2_{\rm w}}\ell_1 +
 \displaystyle\frac{\Lambda^2}{f_\pi^2}
 \frac{s^2_{\rm w}c^2_{\rm w}}{e^2}
 (\Delta\kappa_\gamma -\Delta\kappa_Z) ~~,\\[0.4cm]
\ell_3 = -\displaystyle\frac{1}{e^2}
 \frac{s^2_{\rm w}c^2_{\rm w}}{c^2_{\rm w}-s^2_{\rm w}}\ell_0
 -\frac{c^2_{\rm w}}{c^2_{\rm w}-s^2_{\rm w}}\ell_1 +
 \displaystyle\frac{\Lambda^2}{f_\pi^2}
 \frac{s^2_{\rm w}c^2_{\rm w}}{e^2}\Delta g^Z_1 ~~,\\[0.4cm]
\ell_9 =\displaystyle\ell_8 + \frac{\Lambda^2}{f_\pi^2}
  \frac{s^2_{\rm w}c^2_{\rm w}}{e^2}\left(
 \Delta\kappa_Z+\frac{s^2_{\rm w}}{c^2_{\rm w}}\Delta\kappa_\gamma
 -\Delta g_1^Z\right) ~~,\\[0.4cm]
\ell_{11}=\displaystyle\frac{\Lambda^2}{f_\pi^2} 
 \frac{s^2_{\rm w}c^2_{\rm w}}{e^2}g^Z_5 ~~,\\[0.4cm]
\ell_{12}=-\displaystyle\frac{\Lambda^2}{f_\pi^2} 
 \frac{s^2_{\rm w}c^2_{\rm w}}{e^2}g^Z_4 ~~,\\[0.4cm]
 \ell_{13}=\displaystyle\frac{\Lambda^2}{f_\pi^2}
 \frac{s^2_{\rm w}c^2_{\rm w}}{e^2}
(\tilde{\kappa}_Z-\tilde{\kappa}_\gamma ) ~~,\\[0.4cm]
 \ell_{14}=-\displaystyle\frac{\Lambda^2}{f_\pi^2}
 \frac{s^2_{\rm w}c^2_{\rm w}}{e^2}
\left(\tilde{\kappa}_Z+\frac{s^2_{\rm w}}{c^2_{\rm w}}
      \tilde{\kappa}_\gamma \right)     ~~.
\end{array}
\eqno(3.8)                
$$
Inputting the experimentally measured $~S,~T,~U~$ and 
$~g^{Z,\gamma}_{1,4,5}~$, $\kappa_{Z,\gamma}$~ and
$~\tilde{\kappa}_{Z,\gamma}~$, we can derive constraints 
on all $\ell_n$~'s from (3.1) and (3.8).
For example, a recent global fit at LEP~\cite{TGC} gives the following 
$1\sigma$ (i.e. $68.27\%$ confidence level) bounds, for allowing 
only one TGC to be nonzero each time:
$$
\begin{array}{c}
-0.064 ~\leq ~ \Delta g^Z_1 ~\leq   ~-0.002 ~~,~~~
-0.070 ~\leq ~ \lambda_\gamma ~\leq ~-0.002 ~~,~~~
0.004 ~\leq ~\lambda_Z ~\leq ~ 0.094 ~~,\\
-0.046 ~\leq ~\Delta\kappa_Z ~\leq ~0.042 ~~,~~~
0 ~\leq ~ \Delta\kappa_\gamma ~\leq ~0.112  ~~.
\end{array}
\eqno(3.9)                    
$$
{}From the above result
 we can estimate the constraints on $~\ell_{2,3,9}~$ as
$$
\begin{array}{lllll}
-12.1\left(\frac{\Lambda}{\Lambda_0}\right)^2 & \leq & \ell_2 &
 \leq & 32.3\left(\frac{\Lambda}{\Lambda_0}\right)^2   ~~,\\[0.30cm]
-18.5\left(\frac{\Lambda}{\Lambda_0}\right)^2 & \leq & \ell_3 &
 \leq & 0.61\left(\frac{\Lambda}{\Lambda_0}\right)^2   ~~,\\[0.30cm]
-13.3\left(\frac{\Lambda}{\Lambda_0}\right)^2 & \leq & \ell_9 &
 \leq & 18.5\left(\frac{\Lambda}{\Lambda_0}\right)^2   ~~,
\end{array}
\eqno(3.10)                    
$$
where $~\Lambda \leq \Lambda_0 \equiv 4\pi f_\pi \simeq 3.1$~TeV.

At the FermiLab Tevatron, the TGCs can be directly measured at the tree
level instead of at the loop level. 
For instance, the CDF group gives,
at the $95\%$ confidence level (C.L.)~\cite{TGC}, 
$$
\begin{array}{cl}
-1.1~<~ \Delta\kappa_V ~<~1.3 ~~, &
~~~(~{\rm for}~\lambda_V=\Delta g^V_1 =0~)~,\\
-0.8~<~ ~~\lambda_V~      ~<~0.8 ~~, &
~~~(~{\rm for}~\Delta\kappa_V=\Delta g^V_1 =0~)~,\\
-1.2~<~ \Delta g^Z_1   ~<~1.2 ~~, &
~~~(~{\rm for}~\lambda_V=\Delta\kappa_V =0~)~,\\
\end{array}
\eqno(3.11)        
$$
where $~\Delta\kappa_\gamma =\Delta\kappa_Z~
~{\rm and} ~ \lambda_\gamma =\lambda_Z~$
are assumed. Thus we can estimate the $95\%$ C.L.
constraints on $~\ell_{3,9}~$  as
$$
\begin{array}{cc}
-346\left(\frac{\Lambda}{\Lambda_0}\right)^2 ~<~ \ell_3 ~<~
 346\left(\frac{\Lambda}{\Lambda_0}\right)^2 ~~,~~~~ &
-412\left(\frac{\Lambda}{\Lambda_0}\right)^2 ~<~ \ell_9 ~<~
 488\left(\frac{\Lambda}{\Lambda_0}\right)^2 ~~,
\end{array}
\eqno(3.12)         
$$
which gives, for $~\Lambda =2$~TeV~,
$$
\begin{array}{cc}
-145 ~<~ \ell_3 ~<~ 145 ~~,~~~ &
-173 ~<~ \ell_9 ~<~ 204 ~~,~~~(~\Lambda =2~{\rm TeV}~)~~.
\end{array}
\eqno(3.12a)         
$$
As shown above, the indirect $1\sigma$ bounds from LEP/SLC allow  
$~\ell_{2,3,9}$  to be around of $O(10)$, 
and the direct $95\%$ C.L. bounds from 
Tevatron on $~\ell_{3,9}~$ are also too weak 
to be useful in discriminating different dynamical models
whose effects to these coefficients $\ell_n$~'s are theoretically 
expected to be of ~$O(1)$~\cite{georgi}. 

Since the operators $~{\cal L}^{(4)}_{4,5,6,7,10}~$
contain only quartic vertices, they cannot be constrained at tree level
by any low energy data. The current experiments can
only constrain these operators at one-loop level 
[i.e., of $~O(1/\Lambda^4)$].
By calculating the one-loop logarithmic contributions 
(with all the constant terms ignored) to the low energy data
from these operators, one can roughly estimate the indirect
experimental bounds on their coefficients~\cite{dv,ebo}.
Since the ignored constant terms are of the same order of magnitude
as the logarithmic contributions, we should keep in mind that 
some uncertainties (like a factor of $2$ to $3$ or so) may naturally exist
in these estimated bounds.
It was found in Ref.~\cite{dv} that, at the $90\%$ C.L., 
the LEP data constraints
(allowing only one non-zero coefficient at a time) are
$$
\begin{array}{ll}
-11 ~<~ \ell_4 ~<~ 11 ~~,~~ &  -28 ~<~ \ell_5 ~<~ 26 ~~,
\end{array}
\eqno(3.13)               
$$
for the cut-off scale $~\Lambda = 2$~TeV. In another recent 
study~\cite{ebo}, for $~\Lambda = 2$~TeV and $m_t=170$~GeV, 
the following LEP constraints are derived at the $90\%$ C.L.:
$$
\begin{array}{c}
-3.97 ~\leq ~ \ell_4 ~\leq ~ 19.83 ~~,~~~
-9.91 ~\leq ~ \ell_5 ~\leq ~ 50.23 ~~,\\
-0.66 ~\leq ~ \ell_6 ~\leq ~ 3.50 ~~,~~~ 
-5.09 ~\leq ~ \ell_7 ~\leq ~ 25.78 ~~,~~~
-0.67 ~\leq ~ \ell_{10} ~\leq ~ 3.44 ~~.
\end{array}
\eqno(3.14)               
$$
The above results show that the low energy bounds on
the $SU(2)_C$-violating operators ${\cal L}_{6,10}$
are close to their theoretical expectation for $~\ell_n\sim O(1)~$,
which are stronger than that for ${\cal L}_{7}$ and
the $SU(2)_C$-conserving operators ${\cal L}_{4,5}$ 
(when turning off the $U(1)_Y$ gauge coupling).
Thus, ${\cal L}_{6,10}$ are more sensitive to the low energy data.
But these numerical values should not be taken too seriously
(except as a useful guideline) since 
all non-logarithmic contributions are ignored in the above estimates
and the correlations among different operators are not considered
for simplicity. We also note that the sensitivities of the 
$SU(2)_C$-violating operators to the low energy data
do {\it not} have a naive power-like dependence
on the number of $~{\cal T}$-operators. (3.14) shows that
the operators $~{\cal L}_{6,10}~$ (containing two and four 
$~{\cal T}$-operators, respectively) have quite similar sensitivities
to the low energy data and their bounds are much stronger than that for
$~{\cal L}_{7}~$ (containing two $~{\cal T}$~'s). 
When further looking at the LEP $68.27\%$~C.L. bounds (3.10)
for the triple gauge boson couplings (TGCs) from the $SU(2)_C$-violating 
operators $~{\cal L}_{2,9}~$, 
we find that they are similar and are both much weaker than the
$90\%$~C.L. bounds for quartic couplings from
$~{\cal L}_{6,10}~$ in (3.14) despite $~{\cal L}_{2,9;6,10}~$ containing
one, two, two and four $~{\cal T}$-operators, respectively.
Intuitively, it would be natural to expect that
the $SU(2)_C$-violating operators may get stronger bounds than the 
$SU(2)_C$-conserving ones [as implied in (3.14) for the quartic couplings  
of $~{\cal L}_{6,10}~$] when we consider the $SU(2)_C$
as a good approximate symmetry. But the real situation
is more involved. From the LEP bounds (3.10) and the Tevatron bounds
(3.12,12a), we see that, for TGCs, the $SU(2)_C$-violating operators 
$~{\cal L}_{2,9}~$ have weaker bounds than that of 
the $SU(2)_C$-conserving operator $~{\cal L}_3~$. 
Ref.~\cite{dv} also estimated the $90\%$~C.L. LEP bounds 
for $~{\cal L}_{2,3}~$ as $~~-47 < \ell_2 < 39~~$ 
and $~~-8 < \ell_3 < 11 ~~$ which impose stronger constraint on 
$~\ell_3~$.
[The relations $~L_{9R}=-2\ell_2~,~~L_{9L}=-2\ell_3~,~~L_5=\ell_4~,~~
L_4=\ell_5~$ have been used to translate the eq.~(39) of Ref.~\cite{dv} 
into our notations.]

In summary, the results in (3.10), (3.12,12a), (3.13) and (3.14) 
indicate that the current bounds on
~$\ell_{2,3,9}$~ and ~$\ell_{4,5,7}$~ are still too weak.
Concerning the $SU(2)_C$-violating operators, the bound on $~\ell_0~$
($T$) is most stringent while that on $~\ell_{1,8,6,10}~$ are all
around $~O(1)~$ (or larger) as shown in (3.5) and (3.14).
However, the updated constraints on other $SU(2)_C$-violating operators
$~{\cal L}_{2,9,7}~$ can be of $~O(10)~$ or larger, implying that
the current low energy tests do not well probe the 
$SU(2)_C$-violation effects from these operators.
Although LEPII and the upgraded Tevatron are expected to 
improve the current bounds somewhat, to further improve 
the precision on $\ell_n$~'s and to fully probe the EWSB sector require
finding the most sensitive
high energy scattering processes to {\it independently test}
{\it all} those coefficients in (2.2) at the LHC and the future 
linear colliders (LC).

\vspace{0.7cm}
\noindent
{\bf 4.  The Precise Electroweak Power Counting Method}
\vspace{0.3cm}

In this section, we develop the electroweak power counting
method to precisely and separately count the power dependences on 
the energy $E$ and all relevant mass scales. We have included both
non-decoupling scenario (without Higgs boson) and the decoupling scenario
(with Higgs boson).
This method can correctly estimate any high energy scattering amplitude
and is proven to be extremely convenient for our global analysis on
classifying the complete set of NLO operators (instead of just a few
of them). 
Its phenomenological applications will be given in Sec.~6 and 7.

\vspace{0.5cm}
\noindent
{\bf 4.1.} 
Generalizing Weinberg's power counting rule to SSB gauge theories
\vspace{0.2cm}

 Weinberg's power counting method  
was derived for only counting the energy dependence
in the un-gauged nonlinear $\sigma$-model as a description of low energy
QCD interaction~\cite{wei}. 
But some of its essential features are very general:~
{\bf (i).} The total dimension $~D_T~$ of 
an $S$-matrix element $~T~$ is determined 
by the number of external lines and the space-time dimension; 
{\bf (ii).}   Assume that all mass poles 
in the internal propagators of $~T~$ are much smaller
than the typical energy scale $E$ of $~T~$, then the total
dimension $~D_m~$ of the $E$-independent coupling constants included in 
$~T~$ can be directly counted 
according to the type of vertices contained. 
Hence, the total $E$-power $~D_E~$ for 
$~T~$ is given by $~~D_E= D_T -D_m~~$. 

Here, we shall make a natural generalization of 
Weinberg's power counting method 
for the EWCL in which, except the light SM
gauge bosons, fermions and would-be GB's, all possible heavy fields have
been integrated out.
It is clear that in this case the above conditions 
{\bf (i)} and {\bf (ii)} are satisfied. 
The total dimension of an $L$-loop $S$-matrix element $T$ is
$$
D_T = 4-e~~, 
\eqno(4.1)                                          
$$
where $~~e=e_B+e_F~~$, and $e_B$ ($e_F$) is
 the number of external bosonic  
(fermionic) lines. Here the dimensions of the external spinor wave 
functions are already included in $D_T$. 
For external fermionic lines, we only 
count the SM fermions with masses 
$~~m_f\leq m_t \sim O(M_W)\ll E ~$. So the 
spinor wave function of each external fermion 
will contribute an energy factor
$~E^{1/2}~$ for $~~E\gg m_f~$, where the spinor wave functions are 
normalized as 
$~~ \bar{u}(p,s)u(p,s^\prime )=2m_f\delta_{ss^\prime}~$, etc. 

Let us label the different types of vertices by an index $n$. If
the vertex of type $n$ contains $b_n$ bosonic lines, 
$f_n$ fermionic lines and $d_n$ derivatives, 
then the dimension of the $~E$-independent 
effective coupling constant in $T$ is 
$$
D_m= \displaystyle\sum_n {\cal V}_n\left(4-d_n-b_n-\frac{3}{2}f_n\right) ~~,
\eqno(4.2)                              
$$
where ${\cal V}_n$ is the number of vertices of type $n$. 
Let $i_B$ and $i_F$ be the numbers of 
internal bosonic and fermionic lines, 
respectively. ( $i_B$ also includes possible internal ghost lines.)
Define  $~~i=i_B+i_F~~$,
we have, in addition, the following general relations
$$
\begin{array}{l}
\displaystyle\sum_n b_n{\cal V}_n =2i_B+e_B~~,~~~~
\displaystyle\sum_n f_n{\cal V}_n =2i_F+e_F ~~,~~~~
L=1+i-\displaystyle\sum_n {\cal V}_n ~~.
\end{array}
\eqno(4.3)
$$                                          
These can further simplify the terms in (4.2).

Note that external vector-boson lines may cause extra contributions to the 
power of $~E~$ in $~D_E~$ due to the $E$-dependence of their polarization 
vectors since each longitudinal polarization vector $\epsilon_L^\mu$ 
is of $O(E/M_W)$ for $~E\gg M_W~$.
Thus, if we simply count all external $V_L$-lines directly,
the relation between $D_E,D_T$ and $D_m$ will become 
$~~~D_E=D_T-D_m+e_L-e_v~~~$, where $e_L$ 
and  $e_v$ denote the numbers of external 
$V_L$ and $v^a$ lines, respectively. [As shown in (5.17), each external
$v^a$-line is a gauge-line $~V_\mu^a~$ suppressed by the factor
$~v^\mu = O(M_W/E)~$.]
However, when this relation is applied to the $V_L$-amplitudes with $D_m$
given in (4.2), it does not lead to the correct results. To see 
this, let us take the $V_LV_L \rightarrow V_LV_L$ scattering amplitude as an 
example, in which $e_L=4$ and $e_v=e_F=0$. 
To lowest order of the EWCL, the leading
powers of $E$ in the amplitudes $T[V^{a_1}_L, \cdots ,V^{a_4}_L]$,
$T[\pi^{a_1}, \cdots ,\pi^{a_4}]$ and the $B$-term [cf. (5.17)]
are $E^4$, $E^2$ and $E^0$, respectively. 
This is not consistent with the prediction of the equivalence theorem (ET) [cf.
eq.(5.17)]. 
The reason for this inconsistency is that this naive power counting for the
$V_L$-amplitude only gives the leading $E$-power for individual Feynman
diagrams. It does not reflect the fact that gauge invariance
causes the cancellations of the $E^4$-terms between different diagrams, and 
leads to the final $E^2$-dependence of the whole $V_L$-amplitude.
Thus directly counting the external $V_L$-lines in the $V_L$-amplitudes for
$~D_E~$ does not give the correct answer. This problem can be elegantly
solved by implementing the ET identity (5.17). We see that the power counting 
of the GB-amplitude plus the $B$-term 
does give the correct $E$-dependence because, unlike in the $V_L$-amplitude, 
there is generally no large $E$-power cancellations in 
the GB-amplitudes and the $B$-term. 
Therefore based upon the ET identity (5.17),
the correct counting of the powers of $E$ for the $V_L$-amplitude
can be given by counting the corresponding GB-amplitude plus 
the $B$-term. Thus, in the following generalized power 
counting rule, we do not directly count 
the the external $V_L$-lines in a given diagram. Instead, they will be counted 
through counting the RHS of the ET identity (5.17). 
We shall therefore drop the $e_L$ term in the above relation between $D_E,D_T$
and $D_m$, and {\it make the convention that the number of external
vector-boson lines $e_V$ counts only the number of external $V_T$-lines and 
photon lines}. Then from (4.1), (4.2) and (4.3), the feasible formula 
for the leading energy power in $T$ is
$$
D_E  ~= D_T-D_m-e_v
     ~= 2L +2 + \sum_n {\cal V}_n
      \left(d_n +\frac{1}{2}f_n-2\right)- e_v ~~.
\eqno(4.4)                                                     
$$      
This is just the Weinberg's counting rule~\cite{wei} 
in its generalized form
with the gauge boson, 
ghost and fermion fields and possible $v_\mu$-factors 
included. (4.4) is clearly valid for
any gauge theory satisfying the above conditions 
{\bf (i)} and {\bf (ii)}.

\vspace{0.5cm}
\noindent
 {\bf 4.2.} Constructing the precise electroweak power counting method
\vspace{0.25cm}

\noindent
{\it 4.2.1. Non-decoupling scenario (without Higgs boson)}
\vspace{0.2cm}

We want to {\it separately} count the power dependences of the amplitudes
on the energy $E$, the cutoff scale 
$\Lambda$ of the EWCL and the Fermi scale (vacuum expectation value)
$f_\pi = 246$\,GeV ($\sim M_W, m_t$).\footnote{ ~This is 
essentially different from the previous counting result in the literature
for the heavy Higgs SM ~\cite{hveltman} where only the {\it sum} 
of the powers of $E$ and $m_H$ has been counted.}~
This is {\it crucial} for correctly 
estimating the order of magnitude of an amplitude
at any given order of perturbative calculation.
For instance, an amplitude of order 
$~\frac{E^2}{f_\pi^2}~$ differs by two orders
of magnitude from an amplitude of order $~\frac{E^2}{\Lambda^2}~$ 
in spite that they have the same
$E$-dependence. Also,  the amplitudes $~\frac{E^2}{f_\pi^2}$~ and 
~$\frac{E^2}{f_\pi^2}\frac{E^2}{\Lambda^2}$~ have the {\it same} sum
for the $E$ and $\Lambda$ powers,
but are clearly of different orders in magnitude. E.g.,
in the typical case $E=1$~TeV and $\Lambda\simeq 4\pi f_\pi\simeq 3.1~$TeV, 
they differ by a large factor $\sim 10$~.
Since the weak-boson mass ~$M_W=g f_\pi/2$~ and the fermion mass 
$~m_f=y_f f_\pi/\sqrt{2}~$, we can 
count them through powers of the coupling constants $g$ and $y_f$ 
and the vacuum expectation value $f_\pi$.
The $SU(2)$ weak gauge
coupling $g$ and the top quark Yukawa coupling $y_t$ are around of $O(1)$
and thus will not significantly affect the order of magnitude estimates.
The electromagnetic $U(1)_{\rm em}$ coupling 
$~e=g\sin\theta_W~$ is smaller than $g$ by
about a factor of $2$. The Yukawa couplings of all
light SM fermions other than the top quark are negligibly small.
In our following precise counting rule, the dependences on coupling
constants $~g,~g^{\prime} ({\rm or}~ e)~$ and $~y_t~$ are included, 
while all the light
fermion Yukawa couplings [ $y_f~(\neq y_t) \ll 1$ ] are ignored.

To correctly estimate the magnitude of each given amplitude $T$,
besides counting the power of $E$, it is crucial to also {\it
separately} count the power dependences on the two typical mass scales
of the EWCL: the vacuum expectation value $f_\pi$ and the effective  
cutoff scale $\Lambda$. If the powers of $f_\pi$ and $\Lambda$ are not
separately counted, $\Lambda/f_\pi \simeq 4\pi >12$ will be mistakenly
counted as 1. This can make the estimated results off by orders of 
magnitudes.

Consider the $S$-matrix element $T$ at the $L$-loop order.
Since we are dealing with a spontaneously broken gauge theory which has
 a nonvanishing vacuum expectation value $f_\pi$,  
 $T$ can always be written as $~f_{\pi}^{D_T}~$ 
times some dimensionless function
of $~E,~\Lambda$, and $f_\pi~$, etc.  
The $E$-power dependence has been given by our generalized Weinberg
formula (4.4). 
We now  count the power of $\Lambda$.
The $~\Lambda$-dependence in $T$ can only come from two sources:
\begin{description}
\item{(i).} From tree vertices:
$T$ contains $~{\cal V}=\displaystyle\sum_n {\cal V}_n~$ vertices, 
each of which contributes a factor
$~1/\Lambda^{a_n}~$ so that the total factor from ${\cal V}$-vertices is
$~1/\left(\Lambda^{\sum_n a_n}\right)~$; 
\item{(ii).}  From loop-level: Since each loop brings
a factor $~(1/4\pi)^2 = (f_{\pi}/\Lambda_0 )^2~$, the total
$~\Lambda$-dependence from loop contribution is $~~1/\Lambda_0^{2L}~~$,
where $~\Lambda_0\equiv 4\pi f_\pi \geq \Lambda ~$.

\end{description}
Hence the total $~\Lambda$-dependence given by the above two sources
is $~~ 1/\left(\Lambda^{\sum_na_n}\Lambda_0^{2L}\right)~~$, which
reduces to $~~ 1/\left(\Lambda^{\sum_na_n +2L}\right)~~$ in the case
$~\Lambda \simeq \Lambda_0 =4\pi f_\pi ~$.
For generality, we shall explicitly keep the loop factor 
$~(1/4\pi )^{2L}= (f_{\pi}/\Lambda_0 )^{2L} ~$ in eq.~(4.5)
because $~\Lambda =\min (M_{SB}, \Lambda_0 )~$ can be 
somehow lower than $~\Lambda_0=4\pi f_\pi \approx 3.1$~TeV for strongly
coupled EWSB sector, as indicated by model buildings.
{}From the above discussion, we conclude the following
precise counting rule for $~T~$:
$$
\begin{array}{l}
T= c_T f_\pi^{D_T}\displaystyle 
\left(\frac{f_\pi}{\Lambda}\right)^{N_{\cal O}}
\left(\frac{E}{f_\pi}\right)^{D_{E0}}
\left(\frac{E}{\Lambda_0}\right)^{D_{EL}}
\left(\frac{M_W}{E}\right)^{e_v} H(\ln E/\mu)~~,\\[0.5cm]
N_{\cal O}=\displaystyle\sum_n a_n~,~~
D_{E0}=2+\displaystyle\sum_n {\cal V}_n\left(d_n+\frac{1}{2}f_n-2\right)~, 
~~ D_{EL}=2L~,~~ \Lambda_0 =4\pi f_\pi ~, \\
\end{array}
\eqno(4.5)                                                  
$$ 
where the dimensionless coefficient $~c_T~$ contains possible powers of
gauge couplings ($~g,g^\prime~$) 
and Yukawa couplings ($~y_f~$) from the vertices in $~T~$. 
$~H$ is a function of $~\ln (E/\mu )~$ which arises from 
loop integrations in the standard 
dimensional regularization~\cite{app,georgi1}  and is insensitive to $E$. 
Here, $\mu$ denotes the relevant renormalization scale for loop corrections. 
In Ref.~\cite{georgi1}, it has been specially emphasized 
that the dimensional regularization supplemented 
by the minimal subtraction scheme is most
convenient for loop calculations in the effective Lagrangian formalism.

It is useful to 
give the explicit and compact form of $~D_{E0}~$ in (4.5) 
for the lowest order EWCL 
$~{\cal L}_{\rm G}+{\cal L}^{(2)}+{\cal L}_{\rm F}~$.\footnote{ 
 ~It is straightforward to include the
higher order operators in the EWCL for
counting $D_{E0}$ although the possible vertices in this case are
more complicated.}~~ Expanding the interaction terms 
in $~{\cal L}_{\rm G}+{\cal L}^{(2)}+{\cal L}_{\rm F}~$, we find
$$
\begin{array}{l}
\sum_n {\cal V}_n  \equiv {\cal V}   =
 {\cal V}_F +{\cal V}_d^{(V)} + {\cal V}_\pi 
 + {\cal V}_4^{VVVV} + {\cal V}^{VV-\pi} 
             + {\cal V}^{c\bar{c}-\pi} ~~,\\[0.25cm]
\sum_n d_n{\cal V}_n = {\cal V}_d^{(V)} +2{\cal V}_{\pi} ~~,
\end{array}
\eqno(4.6)                                             
$$
with
$$
\begin{array}{l}
{\cal V}_F = {\cal V}_3^{F\bar{F}V} + {\cal V}^{F\bar{F}-\pi} ~~,\\[0.25cm]
{\cal V}_d^{(V)}\equiv 
{\cal V}_3^{\pi\pi V}+ 
\sum_{n=1}^{\infty}{\cal V}_{2n+2}^{V\pi^{2n+1}}
+{\cal V}_3^{VVV} 
            +{\cal V}^{c\bar{c}V}~~,\\[0.25cm]
{\cal V}_\pi     \equiv \sum_{n=2}^{\infty}{\cal V}_{2n}^{\pi} ~~.
\end{array}
\eqno(4.7)                                             
$$
In the above equations,
$~{\cal V}_\pi~$ denotes the number
of vertices with pure GB self-interactions;
$~{\cal V}^{F\bar{F}-\pi}~$ and $~{\cal V}^{c\bar{c}-\pi}~$
denote the numbers of fermion-GB vertices and ghost-GB vertices,
 respectively;  $~{\cal V}^{VV-\pi}~$ denotes the $~V$-$V$-$\pi^n~$
( $n\geq 1$ ) vertices; and $~{\cal V}_3^{F\bar{F}V}~$ 
denotes the three-point vertex $F$-$\bar{F}$-$V$, etc. 
(Note that $~{\cal V}^{c\bar{c}-\pi}~$  vanishes in the Landau gauge 
because of the decoupling of GB fields from ghost fields \cite{app}.)  
Hence, for $~{\cal L}_{\rm G}+{\cal L}^{(2)}+{\cal L}_{\rm F}~$, 
the $~D_{E0}~$ factor in (4.5) is
$$
D_{E0}=2- \left( {\cal V}_d^{(V)} +{\cal V}_F +2{\cal V}_4^{VVVV}+
           2{\cal V}^{VV-\pi} + 2{\cal V}^{c\bar{c}-\pi}\right) ~~.
\eqno(4.8)                                               
$$
This clearly shows that the leading energy-power dependence at $L$-loop
level ( $L\geq 0$ ) is always given by those diagrams with pure GB
self-interactions, i.e., $~~(D_E)_{max}=(D_{E0})_{max}+D_{EL}=2+2L~~$, 
because of  the negative contribution from 
$~~-( {\cal V}_d^{(V)} +{\cal V}_F +2{\cal V}_4^{VVVV}+
   2{\cal V}^{c\bar{c}-\pi} ) ~~$ in (4.8) which includes all 
types of vertices except the pure GB self-interactions.
This conclusion can be directly generalized to all higher order chiral 
Lagrangian operators such as $~{\cal L}_n~$'s in (3.1.2),  and
is easy to  understand since only pure GB 
self-interaction-vertices
contain the highest powers of the momenta in 
each order of the momentum expansion.
The same conclusion holds for pure $V_L$-scattering amplitudes 
since they can be decomposed into the corresponding 
GB-amplitudes plus the $~M_W/E$-suppressed $B$-term  [cf. (5.17)].
We finally conclude that {\it in the EWCL
$~(D_E)_{\max}=2L+2~$ which is  independent of
the number of external lines of a Feynman diagram.} 
To lowest order of EWCL and at the tree level (i.e. $L=0$~),
$~~(D_E)_{\max}=2~~$, which is in accordance with 
the well-known low energy theorem \cite{let}.
For example, by (4.5) and (4.8), the model-independent tree level
contributions to 
$~\pi^{a_1}+\pi^{a_2}\rightarrow\pi^{a_3}+\cdots +\pi^{a_n}~$ and
$~V^{a_1}_T +\pi^{a_2}\rightarrow\pi^{a_3}+\cdots +\pi^{a_n}~$ 
( $n\geq 4$ ) are estimated as
$$
\begin{array}{ll}
T_0[\pi^{a_1},\cdots ,\pi^{a_n}]
=\displaystyle O\left(\frac{E^2}{f^2_\pi}f^{n-4}_\pi\right) ~,~~~~&
B_0^{(0)} = g^2f^{n-4}_\pi ~~;\\[0.4cm]
T_0[V_T^{a_1},\pi^{a_2},\cdots ,\pi^{a_n}]
=\displaystyle O\left(g\frac{E}{f_\pi}f^{n-4}_\pi\right) ~,~~~~&
B_0^{(1)}= O\left( g^2\displaystyle\frac{M_W}{E}f^{n-4}_\pi\right) ~,
\end{array}
\eqno(4.9)        
$$
where $B_0^{(0)}$ and $B_0^{(1)}$ 
are the leading order $B$-terms contained in the
corresponding $V_L$-amplitudes for the above two processes.  

\vspace{0.4cm}
\noindent
{\it 4.2.2. Decoupling scenario (with Higgs Boson)}
\vspace{0.2cm}

For completeness and 
for other possible applications, 
we also generalize Weinberg's power counting method
to another popular effective Lagrangian formalism~\cite{linear} for
the weakly coupled EWSB sector, which is usually called as the 
{\it decoupling scenario.} In this formalism, the lowest order Lagrangian
is just the linear SM with a relatively light Higgs boson and all higher
order new physics effective operators must have dimensions larger than
$4$ and are suppressed by the effective cutoff scale $~\Lambda~$.
Even if a relatively light scalar
is found in future colliders, it remains important to know
whether such a scalar particle trivially serves 
as the SM Higgs boson or originates from a more complicated dynamics.
For instance, the possible new physics effects 
parametrized in (4.10) should be probed 
in details for discriminating the SM Higgs boson
from the non-SM Higgs boson at the LHC and the future linear colliders.  

Following Ref.~\cite{linear}, we can generally write
the $SU(2)_L\otimes U(1)_Y$ linear effective Lagrangian as follows
$$
{\cal L}_{\rm eff}^{\rm linear} ~=~\displaystyle 
{\cal L}_{\rm SM}+\sum_{n} \frac{\ell_n}{\Lambda^{d_n-4}}{\cal O}_n
\eqno(4.10)   
$$
where $~d_n~(\geq 5)~$ is the dimension of the effective operator 
$~{\cal O}_n~$.
In (4.10), the lowest order Lagrangian $~{\cal L}_{\rm SM}~$ 
is just the SM Lagrangian with a relatively light
Higgs boson. The interesting high energy region considered here is
$$
\begin{array}{ccccc}
M_W, m_H, m_t & \ll & E & < & \Lambda 
\end{array}
\eqno(4.11)   
$$
in which $~m_H = \sqrt{2\lambda}f_\pi~$ denotes the Higgs boson mass.

Since the field content of $~{\cal L}_{\rm eff}^{\rm linear}~$ in
(4.10) is the same as that of the SM and the masses of 
all the known fields are much lower than the typical high
energy scale $~E~$ under consideration [cf. eq.~(4.11)], it is clear
that all the essential features of Weinberg's counting method hold
for this linear case.
Following the same reasoning as done in Sec.~4.2.1-4.2.2 [cf. 
eqs.~(4.1)-(4.4)], 
we find that, for a given $S$-matrix element $~T~$, 
the counting formula for the linear case is very similar to Eq.~(4.5):
$$
\begin{array}{l}
T= c_T f_\pi^{D_T}\displaystyle 
\left(\frac{f_\pi}{\Lambda}\right)^{N_{\cal O}}
\left(\frac{E}{f_\pi}\right)^{D_{E0}}
\left(\frac{E}{\Lambda_0}\right)^{D_{EL}}
\left(\frac{M_W}{E}\right)^{e_v} H(\ln E/\mu)~~,\\[0.5cm]
N_{\cal O}=\displaystyle\sum_n (d_n -4)~,~~
D_{E0}=2+\displaystyle\sum_n {\cal V}_n\left(d_n+\frac{1}{2}f_n-2\right)~, 
~~ D_{EL}=2L~,~~ \Lambda_0 =4\pi f_\pi ~, \\
\end{array}
\eqno(4.12)                                                  
$$ 
where the only difference is that $~N_{\cal O}~$ is now determined
by the canonical dimensional counting in (4.1) instead of the naive
dimensional analysis (NDA)~\cite{georgi} for the
non-decoupling scenario discussed in Sec.~4.2.2.

For $~{\cal L}_{\rm SM}~$ (i.e., for $\ell_n=0$ in (4.11)~),
the counting for $~D_E~$ defined in (4.12) or (4.4) 
can be further simplified since we know that the 
total $E$-power dependence of the SM contributions
 will not increase as the loop number $L$ increases 
because of the perturbative unitarity of the light Higgs SM. 
We shall show that, due to the renormalizable feature of 
$~{\cal L}_{\rm SM}~$, the $D_{EL}$ term, $~2L~$, 
in (4.12) will be cancelled by a counter term 
from the vertex-contribution in the $D_{E0}$ term. 

In $~{\cal L}_{\rm SM}~$, there are 
only  $3$-point and $4$-point vertices. 
Due to the renormalizability of the
SM, all the $4$-point vertices do not contain 
partial derivatives, while each $3$-point vertex may contain at most one 
partial derivative. So, we have $d_n=0 \, ~{\rm or}~ \, 1~$. Thus,
$$
\sum_n {\cal V}_n d_n ={\cal V}_d~~,~~~ 
{\cal V}_d \equiv {\cal V}_3^{VVV}+{\cal V}_3^{ssV}+{\cal V}_3^{c\bar{c}V}~~,
\eqno(4.13)
$$                                    
where $~~{\cal V}_d~~$ is the number of all vertices containing one partial
derivative and $~~{\cal V}_n^{\chi_1\cdots\chi_n}~~$ is the total number
of $n$-point  vertices of type $~~\chi_1$-$\chi_2$-$\cdots$-$\chi_n~~$
($\chi$ denotes any possible field in the theory). 
The symbol $~s~$ denotes scalar fields 
( Higgs or GB ), $~c~(\bar{c})~$ denotes (anti-)ghost field and 
$~F~(\bar{F})~$ denotes (anti-)fermion field. 
Furthermore, in $~{\cal L}_{\rm SM}~$,
$$
\begin{array}{l}
{\cal V} = \sum_n {\cal V}_n = {\cal V}_3 + {\cal V}_4  ~~,~~~
\\[0.25cm]
{\cal V}_3 \equiv {\cal V}_d + {\cal V}_F + \bar{\cal V}_3 ~~,~~~
{\cal V}_F\equiv {\cal V}_3^{sF\bar{F}}+{\cal V}_3^{VF\bar{F}} ~~,~~~
\bar{\cal V}_3 \equiv 
  {\cal V}_3^{sVV}+{\cal V}_3^{sc\bar{c}}+{\cal V}_3^{sss}~~.
\\[0.25cm]
{\cal V}_4 \equiv {\cal V}_4^{ssss}+{\cal V}_4^{ssVV}+{\cal V}_4^{VVVV} ~~,
\end{array}
\eqno(4.14)                                       
$$
Substituting (4.13), (4.14) and the SM relation $~{\cal V}_F = 2i_F +e_F~$
into (4.12), we obtain
$$
\begin{array}{ll}
D_E^{\rm SM}~=~ D_{E0}^{\rm SM}+D_{EL}^{\rm SM} 
& =~ 2L+2 -2{\cal V} + {\cal V}_d +{\cal V}_F -e_v \\
    & =~ 2L+2 - ({\cal V}_d +{\cal V}_F +2\bar{\cal V}_3 +2{\cal V}_4)-e_v~~,
\end{array}
\eqno(4.15) 
$$                                              
which, with the aid of another SM relation 
$$
\begin{array}{l}
3{\cal V}_3 + 4{\cal V}_4 = e + 2i  ~~~,~~~~{\rm or},\\[0.25cm]
{\cal V}_d + {\cal V}_F = 4{\cal V}-\bar{\cal V}_3 -2i-e
= 2-2L + 2{\cal V}-\bar{\cal V}_3-e~~~,
\end{array}
\eqno(4.16)                                       
$$                                              
can be further simplified as
$$
D_E^{\rm SM}~  =~ 4 -e -e_v - \bar{\cal V}_3 ~~,
\eqno(4.17)                                           
$$
where $~~\bar{\cal V}_3 \equiv 
  {\cal V}_3^{sVV}+{\cal V}_3^{sc\bar{c}}+{\cal V}_3^{sss}~~$.
Note that the loop-dependence term  $~2L~$ is indeed canceled by 
the counter term from the vertex-contribution 
[cf. (4.15) and (4.17)] as expected. This is 
the unique feature of the renormalizable SM and this feature 
is absent in the EWCL with the derivative expansion which has been fully
studied in Sec.~3. 
In summary, for $~{\cal L}_{\rm SM}~$,
a Feynman diagram with its external lines fixed 
can have the leading energy dependence 
if it does not contains the trilinear vertices 
$~s$-$V$-$V$, $~s$-$c$-$\bar c$~ and
~$s$-$s$-$s$~, and  the $v_\mu$-factor. Equivalently, 
(4.15) shows in another way that at a given $L$-loop level the leading energy 
behavior of a diagram corresponds to the minimal $~~({\cal V}_d +{\cal V}_F 
+2\bar{\cal V}_3 +2{\cal V}_4)~~$ and the vanishing $~e_v~$. 

The NLO linear operators $~{\cal O}_n~$ in (4.10)
have been fully compiled in Ref.~\cite{linear}. Here are a few typical
dimension-$6$ effective operators:
$$
\begin{array}{ll}
{\cal O}_W ~=~ -i4
{\rm Tr}({\bf W}_\mu~^\nu{\bf W}_\nu~^\rho{\bf W}_\rho~^\mu )
~~,~~~~~ & 
{\cal O}_{\partial\phi}~=~ 
\displaystyle{1\over 2}\partial_\mu (\phi^{\dagger}\phi )
                       \partial^\mu (\phi^{\dagger}\phi ) ~~,\\[0.38cm]
{\cal O}_{qq}^{(1,1)}~=~ 
\displaystyle{1\over 2}(\bar{q}\gamma_\mu q)(\bar{q}\gamma^\mu q)
~~,~~~~~ &
{\cal O}_{qq}^{(1,3)}~=
~\displaystyle{1\over 2}(\bar{q}\gamma_\mu\tau^aq)
                        (\bar{q}\gamma^\mu\tau^aq)
~~,\\[0.38cm]
{\cal O}_{\phi W}~=~\displaystyle (\phi^{\dagger}\phi )
{\rm Tr}({\bf W}_{\mu\nu}{\bf W}^{\mu\nu})~~,~~~~~ &
{\cal O}_{\phi B} ~=~ (\phi^{\dagger}{\bf W}_{\mu\nu}\phi )B^{\mu\nu}~~,
\\[0.38cm]
{\cal O}_{\phi}^{(1)}~=~\displaystyle{1\over 2}
(\phi^{\dagger}\phi )(D_\mu\phi^{\dagger}D^{\mu}\phi )~~,~~~~~ &
{\cal O}_{\phi}^{(3)}
~=~(\phi^{\dagger}D^\mu\phi )[(D_\mu\phi)^{\dagger}\phi]~~,\\[0.38cm]
\end{array}
\eqno(4.18))
$$
where $~\phi~$ denotes the Higgs doublet which contains the 
linearly realized Higgs field ($H$) and three would-be Goldstone bosons
($\pi^\pm ,~\pi^0$).

It is straightforward to apply our power counting rule (4.12) 
[and (4.17)]
for estimating various scattering amplitudes contributed by (4.10).
Some typical examples are in order. 
First, we count the model-independent
contributions from $~{\cal L}_{\rm SM}~$ to some $~2\rightarrow 2$~
scattering processes:
$$
\begin{array}{ll}
T[V_T^{a_1}V_T^{a_2}\rightarrow V_T^{a_3}V_T^{a_4}]\displaystyle
~=~O(g^2)+O\left({g^4\over 16\pi^2}\right)~, &\\[0.42cm]
T[\pi^{a_1}\pi^{a_2}\rightarrow \pi^{a_3}\pi^{a_4}]\displaystyle
~=~ O(g^2,\lambda )+O\left({g^4,\lambda^2\over 16\pi^2}\right)~, &
B^{(0)}_0~=~O\left((g^2,\lambda )\frac{M_W^2}{E^2}\right)~,\\[0.35cm]
T[V_T^{a_1}\pi^{a_2}\rightarrow \pi^{a_3}\pi^{a_4}]\displaystyle
~=~O\left((g^2,\lambda )\frac{M_W}{E}\right)
   +O\left(\frac{g^4}{16\pi^2}\right)~~,~~~ & 
B_0^{(1)}~=~O\left(g^2\frac{M_W}{E}\right)~~.
\end{array}
\eqno(4.19)
$$
Second, we count the NLO model-dependent contributions
from (4.18) to some tree-level 
high energy processes at the $~O(1/\Lambda^2)~$:
$$
\begin{array}{l}
T_1[V_T^{a_1}V_T^{a_2}\rightarrow V_T^{a_3}V_T^{a_4}]({\cal O}_W)
~=~\displaystyle O\left(\ell_W \frac{gE^2}{\Lambda^2}\right)~~,\\[0.4cm]
T_1[\pi^{a_1}\pi^{a_2}\rightarrow \pi^{a_3}\pi^{a_4}]({\cal O}_W)
\displaystyle ~=~ 0~~,~~~~~
B^{(0)}_1~=~O\left(\ell_W\frac{gM_W^2}{\Lambda^2}\right)~~,\\[0.42cm]
T_1[\pi^{a_1}\pi^{a_2}\rightarrow HH]({\cal O}_{\partial\phi})~=~
\displaystyle
O\left(\ell_{\partial\phi}\frac{E^2}{\Lambda^2}\right)~~,~~~~~
B^{(0)}_1~=~ O\left(\frac{M_W^2}{\Lambda^2}\right)~~,\\[0.42cm]
\displaystyle
T_1[q\bar{q}\rightarrow q\bar{q}]({\cal O}_{qq}^{(1,1)}) ~=~
O\left(\ell^{(1,1)}_{qq}\frac{E^2}{\Lambda^2}\right)~~.\\[0.4cm]
\end{array}
\eqno(4.20)
$$
The above examples illustrate, 
in the linear effective Lagrangian formalism, 
how to conveniently apply our power counting rule (4.12) 
to determine the high energy behavior of any given
amplitude and estimate its order of magnitude.

\vspace{0.7cm}
\noindent
{\bf 5. Longitudinal-Goldstone Boson Equivalence Theorem:} 
{\it from mathematical formulation to its physical content for probing EWSB}
\vspace{0.3cm}

In this section, we analyze how the ET is deeply rooted in
the underlying Higgs mechanism based upon rigorous Ward-Takahashi (WT) 
identities. Then, we review how the radiative-modification factors to the ET
generally arise at the loop levels and how the 
radiative-modification-free formulation of the ET is constructed
in our convenient renormalization schemes for general $R_\xi$-gauges
including both the 't Hooft-Feynman and Landau gauges.
After this, we focus on analyzing the additive modifications
to the ET, where we discuss the issues of longitudinal-transverse ambiguity
and the kinematic singularity concerning the validity of the ET.
Finally, we formulate the ET as a necessary physical criterion 
for probing the EWSB sector.\\

\vspace{0.5cm}
\noindent
{\bf 5.1.} ET and the Higgs mechanism 
\vspace{0.2cm}

If there were no EWSB, the physical 
degrees of freedom of the weak gauge bosons
$V^a$'s ($V^a$ stands for $W^\pm$ or $Z^0$) would only be their transverse 
components $V^a_T$'s. In the theory with EWSB, from the well-known Higgs
mechanism~\cite{HK}, their longitudinal components ($V^a_L$'s) are 
physical degrees of freedom as well due to 
absorbing the corresponding would-be 
Goldstone bosons ($\pi^a$'s). Thus, the dynamics of $V^a_L$'s reflects 
that of the $\pi^a$'s which are generally described by the 
EWCL in (2.1)-(2.2). Intuitively, we expect that the amplitude of the
$V^a_L$'s is related to that of the $\pi^a$'s, which will be quantitatively 
described by the so-called ET. This makes it possible to probe the EWSB 
mechanism via the experimentally measured processes involving 
$V^a_L$'s. The tree-level form of the ET was first given in 
Ref.~\cite{et-tree} which is
$$
\begin{array}{ll} 
T[V^{a_1}_L,\cdots,V^{a_n}_L;\Phi_{\alpha}]=T[-i\pi^{a_1},\cdots ,
-i\pi^{a_n};\Phi_{\alpha}]+O(M_W/E)\,,
\end{array}
\eqno{(5.1)}
$$                                                      
where $\Phi_{\alpha}$ denotes other possible on-shell physical particles. This
simple relation has been widely used. Since computing the $\pi^a$-amplitude 
is much easier than computing the $V^a_L$-amplitude, eq.(5.1) also provides 
a technical tool for simplifying the calculation of $T(V^{a_1}_L,\cdots,
V^{a_n}_L;\Phi_{\alpha})$. 
In Secs.~5.3-5.4, we further reveal and demonstrate the {\it profound 
physical content of the ET} which serves as a necessary criterion in 
classifying the sensitivities
of probing the EWSB mechanism at high energy colliders.

Up to the quantum loop-level, 
the formulation of the ET becomes much more complicated and
non-trivial. The study of the ET
beyond tree-level was started in Ref.~\cite{et1}. The existence of the
(multiplicative)
radiative-modification factors to the ET was revealed in Ref.~\cite{YY-BS}. 
The precise formulation of the ET to all orders in 
the SM and the EWCL theory was systematically studied in 
Refs.~\cite{et2,et-new} and \cite{et3} 
where both the multiplicative and additive
modification-factors are precisely analyzed.\footnote{
 ~For some other discussions related to the ET, 
see Ref.~\cite{et-extra,etIR} for instance.}  
~~Generally speaking, it is unlikely that the simple 
relation (5.1) can always hold up to loop-level since the physical 
amplitude $T[V^{a_1}_L,\cdots,V^{a_n}_L;\Phi_{\alpha}]$ cannot be simply 
equal to the unphysical one 
$T[-i\pi^{a_1},\cdots,-i\pi^{a_n};\Phi_{\alpha}]$ even if we 
neglect the $~O(M_W/E)~$ terms. 
This is because the wavefunction renormalizations for 
the physical and unphysical fields are different and the latter has
certain arbitrariness (allowed by the WT identities). 
So, we generally expect that {\it there should be certain multiplicative 
modification factors on the R.H.S. of (5.1) which ensure the 
renormalization-scheme- and gauge-parameter- independence of the R.H.S. 
of (5.1).}  To see 
this, we consider a general gauge theory and start from deriving a useful 
Slavnov-Taylor (ST) identity. Let the subscript ``$0$'' denote unrenormalized 
quantities.
We take the general $R_\xi$ gauge 
$$
\begin{array}{ll}
\displaystyle
    {\cal L}_{\rm GF} ~=~ -{1\over2}(F^a_0)^2 ~~,\\[0.3cm]
F^a_0 \equiv 
(\xi^a_0)^{-\frac{1}{2}}\partial_{\mu} V^{a\mu}_0+(\xi^a_0)^{\frac{1}{2}}
\kappa ^a_0\pi^a_0~~\,,
\end{array}
\eqno{(5.2)}                                                
$$
in which we have put a free parameter $~\kappa ^a_0~$ instead 
of taking it to be the mass of $~V^{a\mu}_0~$ for generality.
The invariance of the action under the Becchi-Rouet-Stora-Tyutin (BRST) 
transformation~\cite{BRST} leads to the following ST identity
in the momentum representation~\cite{STID,Higgs-ET}
$$
\begin{array}{l}
<0|F^{a_1}_0(k_1)\cdots F^{a_n}_0(k_n)\Phi |0>=0~.
\end{array}
\eqno{(5.3)}                                               
$$
In (5.3) the external $~\Phi~$ legs have been amputated. Now we give a 
physical analysis of the ET and the Higgs mechanism, starting from (5.3).

First, we consider the case without spontaneous symmetry breaking (SSB). The 
two transverse components of $V^{a\mu}_0$ are physical, while the unphysical 
longitudinal and scalar components are constrained by the gauge fixing 
condition. Let us take the standard covariant gauge for the massless theory, 
i.e. the gauge fixing function (5.2) with $\kappa^a_0=0$. 
The longitudinal and scalar polarization vectors of $V^{a\mu}_0$ can be 
written as 
$$
\begin{array}{l}
\epsilon^\mu_L(k)=(0,{\vec k}/k^0),~~~~\epsilon^\mu_S(k)=(1,{\vec 0})~.
\end{array}
\eqno{(5.4)}                                                     
$$
We then have 
$$
\begin{array}{l}                 
\dis\epsilon^\mu_L(k)+\epsilon^\mu_S(k)=\frac{k^\mu}{k^0}~~,
\end{array}
\eqno{(5.5)}                                                       
$$
so that 
$$
\begin{array}{l}
F^a_0(k)=i(\xi^a_0)^{-\frac{1}{2}}k^0[V^a_{0L}(k)+V^a_{0S}(k)]~~.
\end{array}
\eqno{(5.6)}                                                        
$$
Substituting (5.6) into (5.3) and doing renormalization and $F^a$-leg 
amputation, we directly get the scattering amplitude 
$$
\begin{array}{l}
T[V^{a_1}_L(k_1)+V^{a_1}_S(k_1),\cdots, V^{a_n}_L(k_n)+V^{a_n}_S(k_n);
\Phi_{\alpha}]=0~,
\end{array}
\eqno{(5.7)}                                                         
$$
which is just a quantitative formulation of the $~~V^a_L$-$V^a_S~~$ constraint
mechanism in the physical in/out states for a {\it massless} Abelian
or non-Abelian gauge theory (without Higgs mechanism).

Next, we consider the case with the SSB. 
The gauge fields become massive and the 
longitudinal component $~~V^a_L~~$ is ``released'' to be 
physical by the Higgs mechanism. 
We shall see 
that in the constraint (5.7) $~~V^a_L~~$ will now be replaced by the 
unphysical would-be Goldstone boson field $\pi^a$. Now we can take the 
general $R_\xi$-gauge (5.2) with the arbitrary gauge-parameter $\kappa^a_0$
(which can be {\it either zero or nonzero}).\footnote{ 
 ~The usual choice for
the renormalized $\kappa^a$ to be gauge-boson mass $M_a$ is only for 
cancelling the tree-level $V_\mu^a$-$\pi^a$ mixing and making the propagators
simpler.}
~The longitudinal and scalar polarization vectors for a massive vector 
field with a physical mass $M_a$ can now be written as
$$
\begin{array}{l}
\dis\epsilon^\mu_L(k)=\frac{1}{M_a}(|{\vec k}|, k^0{\vec k}/|{\vec k}|)~,~~~~ 
\epsilon^\mu_S(k)=\frac{k^\mu}{M_a}~.
\end{array}
\eqno{(5.8)}                                                          
$$
Thus 
$$
\begin{array}{l}
F^a_0(k)=i(\xi^a_0)^{-\frac{1}{2}}M_a V^a_{0S}(k)+(\xi^a_0)^{\frac{1}{2}}
\kappa^a_0\pi^a_0(k)~~.
\end{array}
\eqno{(5.9)}                                                          
$$
Repeating the above procedures with care on the 
$~~V^a_{0S}$-$\pi^a_0~~$ mixing and the renormalization, 
we get, corresponding to (5.7),
$$
\begin{array}{l}
T[V^{a_1}_S(k_1)+iC^{a_1}(k^2_1)\pi^{a_1}(k_1),\cdots, 
V^{a_n}_S(k_n)+iC^{a_n}(k^2_n)\pi^{a_n}(k_n);\Phi_{\alpha}]=0~,
\end{array}
\eqno{(5.10)}                                                          
$$
where the exact expression for $~~C^a(k^2)~~$ 
will be derived in Sec.~5.2. 

The rigorous identity (5.10) {\it generally holds for
any gauge theory with SSB via the Higgs mechanism} and is very important.
A few remarks are in order. First, (5.10) is a quantitative formulation
of the $V_S^a$-$\pi^a$ constraint in the physical in/out states, which 
deeply reflects the {\it essence of the underlying Higgs mechanism}: 
after the SSB,
the would-be Goldstone-boson degree of freedom ($\pi^a$) becomes 
``confined'' with the unphysical scalar component of the the massive 
gauge field $V^a_\mu$ so that the longitudinal component of 
$V_\mu^a$ can be ``released'' from the constraint (5.7) 
as the physical degree of freedom. 
Gauge-invariance forces the $\pi^a$'s to be ``confined'' with $V_S^a$'s
via (5.10) (so that no any net effect of them can be observed from the 
physical $S$-matrix elements) and ensures the Higgs mechanism to exactly
work as expected. As to be shown below, 
the identity (5.10) will directly result in the equivalence theorem (ET).
Next, we comment on the new modification-factors [$C^a(k^2)$'s] in (5.10).
Note that in (5.6) and (5.7) 
the $V_L^a$ and $V_S^a$ are two orthogonal projections of the {\it same} 
Lorentz vector field $V^a_\mu$ so that they will not mix and thus 
the amputation and renormalization of the $F^a$-leg are straightforward. 
However, in 
(5.9) and (5.10) the unphysical scalar component $V_S^a$ and would-be GB 
$\pi^a$ do mix with each other: if the gauge-parameter $\kappa^a$ is chosen
to cancel their tree-level mixing as usual, then they will mix again in
all loop-orders. Furthermore, the component $V_S^a$ (as the scalar 
projection of $V^a_\mu$) and the GB ($\pi^a$) are two independent quantum
fields and thus get renormalized in different ways. These are why the
non-trivial modification factors [$C^a(k^2)$'s] generally arise 
in (5.10), in contrast
with the case of massless theory (without SSB via Higgs mechanism).
As a final remark, we note that,
although the Higgs mechanism (and its basic consequence) 
was discovered a long time ago~\cite{HK}, to our knowledge, 
the mathematically quantitative formulation of it {\it at the level
of $S$-matrix elements} [see (5.10)] in comparison with massless case 
(5.7) and its deep connection with the
ET [see (5.16)-(5.17) below] were first revealed very 
recently~\cite{Higgs-ET} and are newly reviewed here in a comprehensive way.
 
Now, we continue our analysis on how the ET arises as a direct consequence
of the Higgs mechanism (5.10). 
By noting that $~M_a~$ is 
the characteristic of the SSB, we infer that (5.5) holds as
$~~ M_a \rightarrow 0~~$. 
Therefore, with $~M_a\neq 0~$, $~~\epsilon^\mu_L(k)+\epsilon 
^\mu_S(k)~~$ must be of the form 
$$
\begin{array}{l}
\dis\epsilon^\mu_L(k)+\epsilon^\mu_S(k)=\frac{k^\mu}{A}
+O\left(\frac{M_a}{E}\right)~,
\end{array}
\eqno{(5.11)}                                                          
$$
with ``$~A~$'' being certain normalization factor. 
Therefore , at high 
energy, $~\epsilon^\mu_L(k)~$ and $~\epsilon^\mu_S(k)~$ are related up to an 
$~O(M_a/E)~$ term. Indeed, if we take the $~M_a\neq 0~$ expressions (5.8),
we see that $~~A=\frac{1}{2}{M_a}~~$, and thus
$$
\begin{array}{l}
\epsilon^\mu_L(k)=\epsilon^\mu_S(k)+O(M_a/E)~.
\end{array}
\eqno{(5.12)}                                                     
$$ 
So we can define
$$
\begin{array}{l}
V_L^a(k) \equiv V_S^a(k) + v^a(k)~,~~~
v^a=v^\mu V^a_\mu ~,~~~ v^\mu = O(M_a/E)~,
\end{array}
\eqno{(5.13)}                                                       
$$
and
$$
\begin{array}{l}
\bar{F}^a\equiv V_S^a+iC^a\pi^a\equiv V_L^a-\bar{Q}^a~~,\\[0.2cm]
\bar{Q}^a\equiv -iC^a\pi^a + v^a = -iC^a\pi^a + O(M_a/E)~~.
\end{array}
\eqno{(5.14)}                                                       
$$
Then (5.10) becomes
$$
\begin{array}{l}
0= T(\bar{F}^{a_1}, \cdots , \bar{F}^{a_n}, \Phi)~~,~~~~~(~n\geq 1~)~~.
\end{array}
\eqno{(5.15)}                                                      
$$
i.e.
$$
\begin{array}{rl}
0= & T(V_L^{a_1}-\bar{Q}^{a_1}, \cdots , V_L^{a_n}-\bar{Q}^{a_n}, \Phi)
\\[0.3cm]
= & T(V_L^{a_1},\cdots , V_L^{a_n};\Phi_{\alpha})
    +(-)^nT(\bar{Q}^{a_1},\cdots, \bar{Q}^{a_n}; \Phi_{\alpha})\\[0.3cm]
  & +\sum^{P_j}_{1\leq j\leq n-1}
T(V_L^{a_{l_1}},\cdots,V_L^{a_{l_j}},\bar{F}^{a_{l_{j+1}}}
-V_L^{a_{l_{j+1}}},\cdots, 
\bar{F}^{a_{l_n}}-V_L^{a_{l_n}}; \Phi_{\alpha}) ~~~~~~~~[~cf.~~(5.14)~~] 
\\[0.3cm]
= & T(V_L^{a_1},\cdots , V_L^{a_n};\Phi_{\alpha})
    +(-)^nT(\bar{Q}^{a_1},\cdots, \bar{Q}^{a_n}; \Phi_{\alpha})\\[0.3cm]
  & +\sum^{P_j}_{1\leq j\leq n-1}
T(V_L^{a_{l_1}},\cdots,V_L^{a_{l_j}},-V_L^{a_{l_{j+1}}},\cdots, 
-V_L^{a_{l_n}}; \Phi_{\alpha})  ~~~~~~~~[~cf.~~(5.15)~] \\[0.3cm]
= & T(V_L^{a_1},\cdots , V_L^{a_n};\Phi_{\alpha})  
    +(-)^nT(\bar{Q}^{a_1},\cdots, \bar{Q}^{a_n}; \Phi_{\alpha}) 
    +\sum^{n-1}_{j=1}C_n^j(-)^{n-j}
    T(V_L^{a_1},\cdots , V_L^{a_n};\Phi_{\alpha})~~.\\[0.5cm]
\end{array}
$$
Using the identity $~~0=(1-1)^n=1+(-)^n+\sum^{n-1}_{j=1}C_n^j(-)^{n-j}~$,~
we have
$$
\begin{array}{l}
 T[V_L^{a_1},\cdots , V_L^{a_n};\Phi_{\alpha}] =
 T[\bar{Q}^{a_1},\cdots, \bar{Q}^{a_n}; \Phi_{\alpha}]~~.
\end{array}
\eqno{(5.16)}                                                        
$$
Substituting (5.14) into (5.16) we get the general formula
$$
T[V^{a_1}_L,\cdots ,V^{a_n}_L;\Phi_{\alpha}]
= C\cdot T[-i\pi^{a_1},\cdots ,-i\pi^{a_n};\Phi_{\alpha}]+ B  ~~,
\eqno(5.17)                                               
$$
$$
\begin{array}{ll}
C & \equiv C^{a_1}_{\rm mod}\cdots C^{a_n}_{\rm mod}
~=~1+O({\rm loop}) ~, \\[0.3cm]
B & \equiv\sum_{l=1}^n (~C^{a_{l+1}}_{\rm mod}\cdots C^{a_n}_{\rm mod}
T[v^{a_1},\cdots ,v^{a_l},-i\pi^{a_{l+1}},\cdots ,
-i\pi^{a_n};\Phi_{\alpha}]  + ~{\rm permutations}~)~~,  \\[0.3cm]
v^a & \equiv v^{\mu}V^a_{\mu} ~, 
~~~~v^{\mu}~\equiv \epsilon^{\mu}_L-k^\mu /M_V = O(M_V/E) ~,
~~~(M_V=M_W,M_Z)~,
\end{array}
\eqno(5.17a,b,c)                                            
$$
where $~~C^a_{mod}\equiv C^a(k^2)|_{k^2=M_a^2}~~$ and will be analyzed
in Sec.~5.2. This is the general 
{\it ET identity} which leads to the precise formulation of the ET when the 
quantitative expression for $C^a_{mod}$ and the condition for neglecting the 
$B$-term are derived (cf. the following subsections). 
To conclude, as rigorously shown above,  
{\it the ET directly results from the $~~V^a_S$-$\phi^a~~$ 
constraint in the SSB theory via the Higgs mechanism [cf.~(5.10)] 
and the high energy relation (5.12) }.
The above general analysis holds for the Higgs mechanisms either with or
without elementary Higgs boson(s).  

\vspace{0.5cm}
\noindent
{\bf 5.2.} ET and its multiplicative modifications
\vspace{0.2cm}

In this subsection, we derive the quantitative expression for the
modification factor $C^a_{\rm mod}$ and study the simplification of it.
For simplicity, we shall first derive our results in the 
$SU(2)_L$ Higgs theory by taking $~g^{\prime}=0~$ in the electroweak
$SU(2)_L\otimes U(1)_Y$ standard model (SM).
The generalizations to the full SM and to
the effective Lagrangian formulations 
are straightforward (though there are some further complications)
and will be given in the later part of this section. 
The field content for the $SU(2)_L$ Higgs theory
consists of the physical fields, $H$, $W^a_\mu$, and $f$($\bar{f}$)
representing the Higgs, the weak gauge bosons and the fermions, 
respectively, and the unphysical
fields $\pi^a$, $c^a$, and $\bar{c}^a$ representing the would-be
Goldstone bosons, the Faddeev-Popov ghosts, and the anti-ghosts
respectively.  
We quantize the theory using the general $R_\xi$-gauge (5.2). For
simplicity, we write (5.2) as
$$
    F^a_0\ ~=~ (\xi_0^a)^{-{1\over2}}\partial_\mu W^{a\mu}_0
        +(\xi_0^a)^{1\over2}\kappa_0^a\pi^a_0 
       ~=~ (\underline{\bf K}_0^a)^T \underline{\bf W}_0^a ~~,\\[0.1cm]
\eqno(5.18a)                                                      
$$
$$
    \underline{\bf K}_0^a \equiv 
    \displaystyle\left( (\xi_0^a)^{-\frac{1}{2}}\partial_{\mu}, 
     -(\xi_0^a)^{\frac{1}{2}}\kappa_0^a \right)^T ~~,~~~
     \underline{\bf W}_0^a \equiv (W_0^{a\mu}, -\pi_0^a)^T ~~, 
\eqno(5.18b)                                                      
$$
For the case of the $SU(2)_L$ theory, we can
take $~\xi_0^a=\xi_0~$, $~\kappa_0^a =\kappa_0~$, for $a=1,2,3$.

As we have seen in Sec.~5.1, the appearance of the modification factor 
$C_{\rm mod}^a$ to the ET is due to the amputation and the renormalization of 
external massive gauge bosons and their corresponding Goldstone-boson fields. 
For the amputation, we need a general ST identity for the propagators
of the gauge boson, Goldstone boson 
and their mixing. The ST identity for the matrix propagator of 
$~\underline{\bf W}_0^a~$ is~\cite{et2}
$$
\underline{\bf K}_0^T \underline{\bf D}_0^{ab} (k)
~=~ - \displaystyle
\left[\underline{\bf X}^{ab}\right]^T (k)
\eqno(5.19)                                                          
$$
with 
$$
\underline{\bf D}_0^{ab}(k) 
~=~ <0|T\underline{\bf W}_0^a (\underline{\bf W}_0^b)^T|0>(k) ~~,~~~~
{\cal S}_0(k)\delta^{ab} ~=~ <0|Tc_0^b\bar{c}_0^a|0>(k)~~,
\eqno(5.20a)                                                         
$$
$$
\underline{\bf X}^{ab}(k)
~\equiv ~\hat{\underline{\bf X}}^{ab}(k){\cal S}_0(k)
~\equiv ~\left( 
\begin{array}{l}
\xi_0^{\frac{1}{2}}<0|T\hat{s}W_0^{b\mu}|0> \\
-\xi_0^{\frac{1}{2}}<0|T\hat{s}\pi_0^b|0> 
\end{array} \right)_{(k)}\cdot {\cal S}_0(k)~~.
\eqno(5.20b)                                                          
$$
To explain how the modification factor $C_{\rm mod}^a$ to the ET arises, 
we start from (5.6) and set $~n=1~$, i.e.,
$$
0=G[F_0^a(k);\Phi_{\alpha}]
 =\underline{\bf K}_0^T G[\underline{\bf W}_0^a(k);\Phi_{\alpha}]
 =- [\underline{\bf X}^{ab}]^T T[\underline{\bf W}_0^a(k);\Phi_{\alpha}] ~~.
\eqno(5.21)                                                            
$$
Here $~G[\cdots ]~$ and $~T[\cdots ]~$ denote the Green function and
the $S$-matrix element, respectively. The identity (5.21) leads directly
to
$$
\frac{k_{\mu}}{M_{W0}}T[W_0^{a\mu}(k);\Phi_{\alpha}] 
= \widehat{C}_0^a(k^2)T[-i\pi_0^a;\Phi_{\alpha}]
\eqno(5.22)                                                        
$$
with $~\widehat{C}^a_0(k^2)~$ defined as
$$
\widehat{C}^a_0(k^2)\equiv{{1+\Delta^a_1(k^2)+\Delta^a_2(k^2)}\over
        {1+\Delta^a_3(k^2)}} ~~,
\eqno(5.23)                                                         
$$
in which the quantities $\Delta^a_i$ are the proper vertices of the
composite operators
$$
\begin{array}{rcl}
        \Delta^a_1(k^2)\delta^{ab} 
    & ~=~&  \displaystyle{g_0\over2M_{W0}}<0|T
        \Dbrack{H^{\phantom b}_0{c}_0^b}|\bar{c}_0^a>(k) ~~,\\[0.4cm]
   \Delta^a_2(k^2)\delta^{ab} 
    & ~=~&   \displaystyle{g_0\over2M_{W0}}\varepsilon^{bcd}<0|T
        \Dbrack{\pi_{0}^{c}{c}_0^d}|\bar{c}_0^{a}>(k) ~~,\\[0.4cm]
    ik^\mu\Delta^a_3(k^2)\delta^{ab} 
     & ~=~&  -\displaystyle{g_0\over2}\varepsilon^{bcd}<0|T
        \Dbrack{W^{\mu b}_0{c}_0^c}|\bar{c}_0^{a}>(k) ~~,
\end{array}
\eqno(5.24)                                                      
$$
where expressions such as $\Dbrack{W_{0}^{\mu b}c_0^c}\!(x)$ indicate
the local composite operator fields formed from $W_{0}^{\mu b}(x)$ and
$c_0^c(x)$~.  The formula (5.23)-(5.24) (valid for the $SU(2)$ Higgs 
theory) were derived in Ref.~\cite{YY-BS}, and the complete expressions
for the realistic $SU(2)_L\otimes U(1)_Y$ SM and EWCL were first given 
in Ref.~\cite{et2,et-new} which are shown to be much more complicated.        
However, we must emphasize that, contrary to all other previous studies,
the essential advantage of our following
approach~\cite{et2,et-new} is that we need {\it not} explicitly calculate
any of these complicated loop level quantities ($\Delta_i$'s) and our
simplifications of these multiplicative modification factors 
[cf.~(5.33-38) below] are based on general WT identities and 
automatically determined by each renormalization scheme itself.

After  renormalization, (5.22) becomes
$$
\frac{k_{\mu}}{M_{W}}T[W^{a\mu}(k);\Phi_{\alpha}] 
= \widehat{C}^a(k^2)T[-i\pi^a;\Phi_{\alpha}]
\eqno(5.25)                                                        
$$
with the  finite renormalized coefficient
$$
\widehat{C}^a(k^2) = Z_{M_W}\left(\frac{Z_W}{Z_\pi}\right)^{\frac{1}{2}}
\widehat{C}^a_0(k^2)  ~~.
\eqno(5.26)                                                          
$$
The renormalization constants are defined as
$~W^{a\mu}_0 =Z_W^{1\over 2} W^{a\mu}~$,~
$\pi_0^a = Z_\pi^{1\over 2}\pi^a~$,
and $~M_{W0}=Z_{M_W}M_{W}~$. 
The modification factor to the ET is precisely
the value of this finite renormalized coefficient $~\widehat{C}^a(k^2)~$
on the gauge boson mass-shell:
$$
C^a_{\rm mod} =\displaystyle \frac{M_W}{M^{phys}_W}\left.\widehat{C}^a(k^2)
\right|_{k^2=(M^{phys}_W)^2}~~,
\eqno(5.27)                                                          
$$
where $M^{phys}_W$ is the physical  mass of $M_W$ which is
equal to $M_W$ if the usual on-shell subtraction for $M_W$ is adopted.

We see that the appearance of the factor $C\equiv C^{a_1}_{\rm mod}\cdots
C^{a_n}_{\rm mod}$ on the R.H.S. of (5.17) is due to the amputation and 
renormalization of external $W^{a\mu}$ and $\pi^a$ lines by using the ST 
identity (5.19). Thus it is  natural that the factor $C^a_{\rm mod}$ contains
$W^{a\mu}$-ghost, $\pi^a$-ghost and Higgs-ghost interactions expressed
in terms of these $\Delta^a_i$-quantities.
In general, the loop-level $\Delta_i^a$-quantities are 
non-vanishing and make $~\widehat{C}_0(k^2)\neq 1~$ and $~C^a_{\rm mod}\neq 1~$
order by order. In the Landau gauge, these $\Delta_i^a$-quantities can be
partially simplified, especially at the one-loop order, because the tree-level
Higgs-ghost and $\pi^a$-ghost vertices vanish. This
makes $~\Delta_{1,2}^a=0~$ at one loop.\footnote{ 
 ~We note that, in the non-Abelian case,
the statement that $~\Delta^a_{1,2}=0~$ for Landau gauge
in Refs.~\cite{YY-BS,et2} is valid at the one-loop order.}~
In general~\cite{et-new},
$$
\Delta_1^a=\Delta_2^a = 0+O(2~{\rm loop})~,
~~~\Delta_3^a =O(1~{\rm loop})~,~~~~~
(~{\rm in~Landau~gauge}~)~.
\eqno(5.28)                                                     
$$
Beyond the one-loop order, $~\Delta^a_{1,2}\neq 0~$ since the 
Higgs and Goldstone-boson fields can still indirectly couple to the ghosts
via loop diagrams containing internal gauge fields.

The explicit expressions for the $\Delta^a_i$'s will be further greatly
complicated in the full $SU(2)_L\otimes U(1)_Y$ theory due to various
mixing effects~\cite{et2},  as mentioned above.
Since the calculation of these $\Delta^a_i$'s is very cumbersome,
the use of the general loop-level form of the ET [cf. (5.40)] from the
general identity (5.17) is much more
inconvenient than that of the tree-level form (5.1). However, as we
have already mentioned, both the modification factor $C\equiv 
C^{a_1}_{\rm mod}\cdots C^{a_n}_{\rm mod}$ and the GB-amplitude 
$T[-i\pi^{a_1},\cdots,-i\pi^{a_n};\Phi_{\alpha}]$ 
are unphysical quantities which are renormalization-scheme and 
gauge-parameter dependent [the unphysical parts of
these two quantities cancel each other to make their product physical
and be equal to the physical amplitude 
$T[V^{a_1}_L,\cdots,V^{a_n}_L;\Phi_{\alpha}]$ 
on the L.H.S. of (5.17)]. So that it is 
possible to choose certain renormalization schemes 
to simplify the expression of $C^a_{\rm mod}$. 
Especially, if we can find certain schemes to exactly make
$C^a_{\rm mod}=1$, (5.17) will be as convenient as (5.1) in these schemes.
In the following, we shall deal with this simplification which concerns
only the WT identities for the propagators and has nothing to do with
the explicit calculation of the $\Delta^a_i$'s, so it really makes the
use of (5.17) as simple as the tree-level case.

For this purpose, we consider the WT identities for the $W^{a\mu}$, $\pi^a$
and the ghost-field propagators which, in the momentum representation,
take the form \cite{et2}
$$
\begin{array}{ll}
i{\cal D}^{-1}_{\mu\nu,ab}(k)
&=\left( g_{\mu\nu}-\displaystyle
\frac{k_\mu k_\nu}{k^2}\right) [-k^2+M^2_{W}-\Pi_{WW}(k^2)]+
\displaystyle\frac{k_\mu k_\nu}
{k^2}[-\xi^{-1} k^2+M^2_{W}-\tilde{\Pi}_{WW}(k^2)]\delta_{ab}~,\\[0.5cm]
i{\cal D}^{-1}_{\pi\mu,ab}(k)
& =ik_\mu[M_{W}-\kappa-\tilde{\Pi}_{W\pi}(k^2)]\delta_{ab}~,\\[0.35cm]
i{\cal D}^{-1}_{\pi\pi,ab}(k)
& =k^2-\xi\kappa^2-\tilde{\Pi}_{\pi\pi}(k^2)\delta_{ab}~,\\[0.35cm]
iS^{-1}_{ab}(k)
& =k^2-\xi\kappa M_{W}-\tilde{\Pi}_{c\bar{c}}(k^2)\delta_{ab}~,
\end{array}
\eqno(5.29)                                                      
$$
where $\Pi_{WW}$ is the proper self-energy of the physical part of
the gauge boson, and $\tilde{\Pi}^a_{ij}$'s are unphysical self-energies.
these identities enables us to express $\widehat{C}^a(k^2)$ in (5.26) in
terms of the self-energies as \cite{et2}
$$
\begin{array}{ll}
\widehat{C}^a(k^2)=
\displaystyle\frac{ M^2_W-\tilde{\Pi}_{WW}+(\Omega^{-1}_\xi-1)\xi^{-1}k^2}
{M^2_W-M_W\tilde{\Pi}_{W\pi}(k^2)+M_W \kappa(\Omega^{-1}_\xi\Omega_\kappa-1)}=
\displaystyle\frac{k^2}{M_W}
\displaystyle\frac{M_W-\tilde{\Pi}_{W\pi}+(\Omega^{-1}_\xi 
\Omega_\kappa-1)\kappa }
{k^2-\tilde{\Pi}_{\pi\pi}+(\Omega^{-1}_\xi\Omega^2_\kappa-1) 
\xi\kappa^2}  ~,
\end{array}
\eqno(5.30)                                                          
$$
where $\Omega_\xi~{\rm and}~\Omega_\kappa$ are finite constants
appearing in the relations between the renormalization constants
constrained by the WT identities (5.29) which are \cite{et2}
$$
\begin{array}{l}
Z_\xi=\Omega_\xi Z_W\,,   ~~~~~~~~~~~~~~~~Z_\kappa=\Omega_\kappa Z^{1/2}_W
Z^{-1/2}_\pi Z^{-1}_\xi\,,~~~~~~~~~~~~~~~~{\rm etc.}\,\,\,,
\end{array}
\eqno(5.31)                                                          
$$
and finite quantities $~\Omega_{\xi ,\kappa}(=1+O({\rm loop}))~$ 
are to be determined by the
normalization conditions at the subtraction point.
From (5.30) we see that the modification factor $C^a_{\rm mod}=
\widehat{C}^a(k^2)|_{k^2=(M^{phys}_W)^2}$ can be simplified by taking certain 
on-shell renormalization schemes in which the proper self-energies
vanish at the subtraction point (mass-shell). For the physical sector,
we take the standard on-shell scheme. For the unphysical sector, the 
prescription of the on-shell renormalization is arbitrary since changing the 
determination of the renormalization constants for the unphysical 
fields and parameters does not affect 
the physical observables. We can thus take this 
arbitrariness to construct certain on-shell renormalization schemes
in the unphysical sector such that the on-shell conditions (for 
tree-mass-pole)
$$
\tilde{\Pi}_{WW}(\xi\kappa M_W)=\tilde{\Pi}_{W\pi}(\xi\kappa M_W)
=\tilde{\Pi}_{\pi\pi}(\xi\kappa M_W)=\tilde{\Pi}_{c\bar{c}}(\xi\kappa M_W)
=0
\eqno(5.32)                                                         
$$
are satisfied and the $C^a_{\rm mod}$-factor is rigorously simplified 
according to our general identity (5.30). We have
developed four kinds of renormalization schemes in which $C^a_{\rm mod}$
is essentially simplified and no extra explicit calculations of
$C^a_{\rm mod}$'s are needed~\cite{et2}\cite{et3}\cite{et-new}. \\
\null\noindent
{\bf a.} {\it Scheme-I~}\cite{et2}\\
This is closest to the conventional on-shell scheme. Take $\kappa=M_W$ and 
$\xi=1$. Determine $\Omega_\xi$ and $\Omega_\kappa$ by the requirement (5.32), 
and determine other renormalization constants by the usual normalization 
conditions requiring the residues of the propagators to be unity at 
$k^2=\xi\kappa M_W=M^2_W$. In this scheme, the modification factor is 
simplified as a {\it single known quantity~:}
$$
C^a_{\rm mod}=\Omega_\kappa^{-1}\,.
\eqno(5.33)                                                          
$$
Note that $\Omega_\kappa$ is already determined by (5.32) during the
renormalization, so that this simplification has nothing to do with the
explicit calculation of the $\Delta^a_i$'s in (5.24).\\\\
\null\noindent
{\bf b.} {\it Scheme-II~}\cite{et2}\\
This scheme is valid for gauges with $\xi\neq 0$
and particularly convenient in the 't Hooft-Feynman gauge ($\xi=1$). 
We take $~\kappa =\xi^{-1}M_W~$ with arbitrary nonzero $~\xi$~. 
Set $~\Omega_\kappa=1$,~
and determine $\Omega_\xi$ and $Z_\pi$ by the requirement (5.32). Other
renormalization constants are determined by the usual normalization conditions.
In this scheme, the modification factor is completely removed:
$$
C^a_{\rm mod}=1.
\eqno(5.34)                                                         
$$
Note that the determination of $Z_\pi$ in this scheme is different from
the usual way. Instead of requiring the residue of ${\cal D}_{\pi\pi}$
to be unity, we require directly $\tilde{\Pi}_{\pi\pi}=0$ at 
$k^2=\xi\kappa M_W=M^2_W$. This renormalization procedure is as
convenient as the usual one for practical applications.\\\\
\null\noindent
{\bf c.} {\it Scheme-III~}\cite{et3}\\
This scheme is specially prescribed for studying the pure $V^a_L$-scatterings 
to the precision of neglecting the whole $B$-term in (5.17) which corresponds 
to keeping only terms independent of the gauge and Yukawa coupling constants 
$g$, $e$, and $y_f$. In this case, $C^a_{\rm mod}$ can be completely 
simplified to unity [as (5.34)] by choosing the renormalization constant 
$Z^a_\pi$ to be
$$
\displaystyle
Z^a_\pi=\left[\left(\frac{M_a}{M^{phys}_a}\right)^2
Z_{V^a}Z^2_{M_a}\right]_{g,e,y_f=0}~~.
\eqno(5.35)                                                         
$$\\\\
\null\noindent
{\bf d.} {\it Scheme-IV~}\cite{et-new}\\
This scheme is prescribed for the $R_\xi$ gauge 
with arbitrary finite $\xi$ (including 
$\xi=0$) and is specially convenient for the Landau gauge ($\xi=0$). We take
$\kappa=M_W$, and $\Omega_\xi=\Omega_\kappa=1$. Other renormalization
constants other than $Z_\pi$ are determined in the usual
normalization conditions at the subtraction point. Only $Z_\pi$ is
specially chosen as
$$
\displaystyle
Z_\pi=Z_W\frac{Z^2_{M_W}M^2_W-\tilde{\Pi}_{WW,0}(M^2_W)}
{M^2_W-\tilde{\Pi}_{\pi\pi,0}(M^2_W)}
\eqno(5.36)                                                          
$$
for simplifying $C^a_{\rm mod}$ to be exactly unity [as (5.34)]. The choice
of $\Omega_\xi=\Omega_\kappa=1$ implies that
$$
F^a_0=F^a\,,
\eqno(5.37)                                                          
$$
i.e. the gauge fixing function is unchanged after the renormalization in
this scheme. 
We finally remark that all the above {\it Scheme I-IV} are also valid for the 
$1/{\cal N}$-expansion~\cite{1/N} since our formulation is based upon 
the general WT identities (for self-energies) which take the {\it same} form 
in any perturbative expansion.

 The generalization of these four renormalization schemes to the full
$SU(2)\otimes U(1)$ SM is straightforward but complicated. 
But our final results are extremely simple and useful, which can be summarized
as~\cite{et2,et3,et-new}
$$
C^W_{\rm mod}=1/\Omega^{WW}_\kappa~,~~~~~~
C^Z_{\rm mod}=1/\Omega^{ZZ}_\kappa~,~~~~~~~
({\rm in}~Scheme~I)\,,
\eqno(5.38a)                                                         
$$
and 
$$
C^W_{\rm mod}=1~,~~~~~~C^Z_{\rm mod}=1,
~~~~~~~~~~~~~~~~({\rm in}~Scheme~II-IV)\,.
\eqno(5.38b)                                                          
$$

So far, we have completed the analysis of the multiplicative 
modification factor $C\equiv
C^{a_1}_{\rm mod}\cdots C^{a_n}_{\rm mod}$ in (5.17). The remaining task
for obtaining the precise formulation of the ET is to make clear the
condition for neglecting the additive $B$-term in (5.17), 
which will be given in Sec.~5.3.

\vspace{0.5cm}
\noindent
{\bf 5.3.} ET and its additive modifications
\vspace{0.25cm}

\noindent
{\it 5.3.1. The longitudinal-transverse ambiguity and the precise
formulation of the ET}
\vspace{0.2cm}

Here we consider the condition for neglecting the $B$-term in (5.17). 
We see from (5.17) that the Lorentz invariant (LI) amplitude
$T[V^{a_1}_L\cdots V^{a_n}_L;\Phi_{\alpha}]$  can be decomposed 
into two parts: the 1st part is 
$~C\cdot T[-i\pi^{a_1}\cdots -i\pi^{a_n};\Phi_{\alpha}]~$ which is LI; the 2nd
part is the $~v_\mu$-suppressed $B$-term which is  Lorentz
non-invariant (LNI) because it contains the external {\hbox{spin-1}} 
$V_{\mu}$-field(s).  
(Without losing generality~\cite{et3}, 
here we have assumed that $~\Phi_\alpha~$ contains 
possible physical scalars, photons and light fermions.). Since the size
of the LNI $B$-term is Lorentz frame dependent, we cannot talk about the 
condition for neglecting the $B$-term relative to the LI part 
$C\cdot T[-i\pi^{a_1}\cdots -i\pi^{a_n};\Phi_{\alpha}]$ in general Lorentz 
frames. It makes sense only in a group of Lorentz frames within which
Lorentz transformations do not significantly change $B$ (i.e. the $V_L$-$V_T$
mixing effect is negligible). We call such frames {\it safe frames}. By examining
the Lorentz transformation of $B$, we can generally estimate $B$ as~\cite{et3}
$$
B = O\left(\frac{M_W^2}{E_j^2}\right)
  T[ -i\pi^{a_1},\cdots , -i\pi^{a_n}; \Phi_{\alpha}] +   
  O\left(\frac{M_W}{E_j}\right)T[ V_{T_j} ^{a_{r_1}}, -i\pi^{a_{r_2}},
                      \cdots , -i\pi^{a_{r_n}}; \Phi_{\alpha}]~~,
\eqno(5.39)                                                   
$$
where $E_j$ is the energy of the $j$-th external $V^a_L$-line. From
(5.39) we see that the condition for a frame to be {\it safe} is ~~
$E_j\sim k_j\gg M_W$~($j=1,~2,\cdots ,~n$). For a given process, $E_j$
can be easily obtained from the kinematics, so that this condition is
easy to implement in practice. 
Note that the $B$-term 
is only $~O(M_W/E_j)$-suppressed {\it relative} to the
leading contributions in the GB-amplitude and is therefore {\it not
necessarily} of the $O(M_W/E_j)$ in magnitude. (5.39) explicitly shows that 
the magnitude of the $B$-term depends on the size of the amplitudes 
$T[-i\pi^{a_1},\cdots ]$ and $T[V_{T_j}^{a_{r_1}},-i\pi^{a_{r_2}},\cdots ]$ 
so that it can be {\it either larger  or smaller than $O(M_W/E_j)$} 
[cf. Eq.~(5.46) below]. 

 With this consideration, we can {\it precisely 
formulate the ET} as
$$                                                                             
T[V^{a_1}_L,\cdots ,V^{a_n}_L;\Phi_{\alpha}]                                   
= C\cdot T[-i\pi^{a_1},\cdots ,-i\pi^{a_n};\Phi_{\alpha}]+ 
O(M_W/E_j){\rm -suppressed} ~~,
\eqno(5.40)                                               
$$                                                                             
$$                                                                             
E_j \sim k_j  \gg  M_W , ~~~~~(~ j=1,2,\cdots ,n ~)~~,           
\eqno(5.40a)                                              
$$
$$
C\cdot T[-i\pi^{a_1},\cdots ,-i\pi^{a_n};\Phi_{\alpha}]\gg B ~~,
\eqno(5.40b)                                              
$$                                                                             
which is in general different from the naive tree-level form (5.1). In the 
renormalization {\it Scheme II-IV} the modification 
factor $C$ in (5.40) is exactly simplified to $C=1$. Eqs.~(5.40a,b) 
are the precise conditions for the validity of the 
{\it longitudinal-Goldstone boson equivalence} in (5.40), 
i.e. the validity of the ET. Note that (5.40a) is 
stronger than the usual requirement $E\gg M_W$ (E stands for the center-of-
mass energy) for the validity of the ET. The difference between (5.40a)
and $E\gg M_W$ has physical significance which has been explicitly shown in 
Ref.~\cite{et3}.

We emphasize that, in principle, the complete set 
of diagrams (including those with internal gauge boson lines) has to be 
considered when calculating $~T[-i\pi^{a_1},\cdots ,-i\pi^{a_n};
\Phi_{\alpha}]~$, as already implied in (5.40). If not, this equivalence might 
not manifest for scattering processes involving $t$- or $u$- channel diagram
in either forward or backward direction.
A detailed discussion on this point is given in Sec.~5.3.2. 
Furthermore, the ET (5.40) and its high energy condition (5.40a) 
indicate the absence of infrared (IR) power divergences (like 
$~\left(\frac{E_j}{M_W}\right)^r,~r>0~$) in the $~M_W\rightarrow 0~$ limit for 
fixed energy $~E_j\sim k_j~$~\cite{etIR}. This can be understood by noting 
that the limit $~M_W\rightarrow 0~$ implies $~g\rightarrow 0~$ after fixing 
the physical vacuum expectation value (VEV) at $~f_\pi =246$~GeV. Taking 
$~g\rightarrow 0~$ limit leads to the well-defined un-gauged linear or non-
linear sigma-model which suggests turning off the gauge coupling to be a 
smooth procedure. The smoothness of the $~g\rightarrow 0~$ limit indicates the 
absence of IR power divergences for $~M_W\rightarrow 0~$. 
In our present formalism, we shall fix the gauge boson mass $M_W(M_Z)$ at its 
experimental value. 

Finally, we remark that the condition (5.40b) for ignoring the additive 
$B$-term can be technically relaxed by a new prescription, called 
`` Divided Equivalence Theorem '' (DET)~\cite{et-new}, based upon our 
(multiplicative) modification-free formulations in Sec.~5.2 
(cf. {\it Scheme II-IV}). The basic observation is that, within 
{\it Scheme II-IV}, we have $~C^a_{\rm mod}=1~$ so that the 
{\it equivalence} in (5.40) can be conveniently
divided order by order in a given
perturbative expansion: $~T=\sum_{\ell =0}^N T_{\ell}~$ and 
$~B=\sum_{\ell =0}^N B_{\ell}~$.  I.e., the ET (5.40) can be expanded
as
$$                                                                             
T_{\ell}[V^{a_1}_L,\cdots ,V^{a_n}_L;\Phi_{\alpha}]                      
= T_{\ell}[-i\pi^{a_1},\cdots ,-i\pi^{a_n};\Phi_{\alpha}]+ 
B_{\ell} ~~,
\eqno(5.41)                                               
$$        
and the conditions (5.40a,b) become, at the $\ell$-th order,
$$                                                                             
E_j \sim k_j  \gg  M_W , ~~~~~(~ j=1,2,\cdots ,n ~)~,           
\eqno(5.41a)                                              
$$
$$
T_{\ell}[-i\pi^{a_1},\cdots ,-i\pi^{a_n};\Phi_{\alpha}]\gg B_{\ell} ~~,~~~~~
(~\ell =0,1,2,\cdots ~)~.
\eqno(5.41b)                                              
$$                                                                
The $\ell$-th order $B$-term is deduced from (5.39):
$$
B_{\ell} = O\left(\frac{M_W^2}{E_j^2}\right)
  T_{\ell}[ -i\pi^{a_1},\cdots , -i\pi^{a_n}; \Phi_{\alpha}] +   
  O\left(\frac{M_W}{E_j}\right)
  T_{\ell}[ V_{T_j} ^{a_{r_1}}, -i\pi^{a_{r_2}},
                      \cdots , -i\pi^{a_{r_n}}; \Phi_{\alpha}]~~.
\eqno(5.42)                         
$$ 
When the NLO ($\ell = 1$) contributions 
(containing possible new physics effects)
are included, the main limitation
on the predication of the ET for the $V_L$-amplitude via computing 
the GB-amplitude 
is due to ignoring the {\it leading order $B_0$-term.}
This leading order $B_0$-term
is of $O(g^2)$~\cite{et3} in the heavy Higgs SM and the EWCL and 
cannot always be neglected in comparison with the NLO GB-amplitude
$~T_1~$ though we usually have $~T_0\gg B_0~$ and 
$~T_1\gg B_1~$ respectively (cf. Ref.~\cite{et3,global1} and Sec.~6 below) 
because of (5.42).
Based upon the above new equations (5.41)-(5.42), we can precisely 
formulate the ET at {\it each given order-$\ell$} 
where only $~B_{\ell}~$, {\it but not the whole $B$-term}, will be
ignored to build the longitudinal-Goldstone boson equivalence.
Hence, {\it the equivalence is divided order by order} in the perturbative
expansion. The condition for this divided equivalence is
$~~T_{\ell} \gg B_{\ell}~~$ (at the $\ell$-th order)
which is much weaker than $~~T_{\ell}\gg B_0~~$ [deduced from (5.40b)]
by a factor of $~B_{\ell}/B_0~(<1)~$  for $~\ell\geq 1~$. 
Therefore, to improve the numerical prediction of $V_L$-amplitude for 
most of the NLO contributions (in $~T_1~$) 
by using the ET, we propose the following simple new prescription
(called the DET):
\begin{description}
 \item[{\bf (i).}] Perform a direct and precise unitary gauge
calculation for the tree-level $V_L$-amplitude $~T_0[V_L]~$ 
which is quite simple.
\item[{\bf (ii).}]  
Make use of the DET (5.41) and deduce $~T_1[V_L]~$ from the
Goldstone boson amplitude $~T_1[GB]~$, by ignoring $~B_1~$ only. 
\end{description}
The direct tree-level unitary gauge calculation of 
$V_L$-amplitude avoids any approximation at this order and 
is shown~\cite{et-new} 
to be much simpler than computing the corresponding GB-amplitude
plus the very complicated $B_0$-term in the $R_\xi$-gauge (as adopted
in Ref.~\cite{gk} before).  
In Ref.~\cite{et-new}, we further demonstrated that, up to NLO of 
the EWCL, the precision of the DET (5.40-42) is typically
increased by a factor of ~$B_0/B_1\simeq \Lambda^2/E^2~$
($\sim 10$~ at the TeV scale) relative to the usual
prescription of the ET [cf.~(5.40a,b)] as ignoring the $B$-term is
concerned. But, we need to clarify that, though the DET technically
improves the precision of predicating the $V_L$-amplitude, it does 
{\it not} increase the ``equivalence'' between the {\it whole}
$V_L$ and GB amplitudes [which is still limited by the condition
{\it (5.40b)} besides (5.40a)]. The Lorentz non-invariant
leading $B_0$-term is always there to 
modify the overall ``equivalence'' in (5.40) and sets up
a necessary {\it physical criterion} for the sensitivity on probing the 
Goldstone dynamics from measuring the $V_L$-amplitude, in spite that
we can numerically include $B_0$ via the DET. This important
physical content of the ET will be further analyzed in Sec.~5.4.

\vspace{0.4cm}
\noindent
{\it 5.3.2. On the kinematic singularity}
\vspace{0.25cm}

Here we examine the validity of the ET in some special kinematic regions
and its physical implication in probing the EWSB, which often cause
confusion in the literature.
It is known that there are kinematic regions
in which the Mandelstam variables 
~$t$~ or ~$u$~ is small or even vanishing
despite the fact that $~\sqrt{s}\gg M_W~$ for high energy scatterings.
Therefore, the amplitude that contains a~$t$- or $u$-channel diagram 
with massless photon field can generate
a kinematic singularity when the scattering angle $\theta$ 
approaches to $0^{\circ}$ or $180^{\circ}$.
In the following, we study in such special kinematic regions
whether the $B$-term [cf. (5.17)] can be safely ignored to validate
the ET and its physical consequence to probing the EWSB sector.

For illustration, let us consider the tree level 
$~W^+_LW^-_L\rightarrow W^+_LW^-_L~$ scattering
in the chiral Lagrangian formalism.
Generalization to loop orders is obvious since the kinematic problem
 analyzed here only concerns the one-particle-reducible (1PR) internal
$W$, $Z$ or photon line in the $t$-channel (or $u$-channel) diagram.
Both the tree level $~W^+_LW^-_L\rightarrow W^+_LW^-_L~$ and
$~\pi^+\pi^-\rightarrow \pi^+\pi^-~$ 
amplitudes in the chiral Lagrangian formalism 
contain contact diagrams, $s$-channel $Z$-exchange and
photon-exchange diagrams, and $t$-channel $Z$-exchange and
photon-exchange diagrams. 
In the C.M. frame, the precise tree-level amplitudes 
$~T[W_L]~$ and $~T[{\rm GB}]~$ are:
$$
\begin{array}{l}
T[W_L]=
 ig^2\left[ -(1+\kappa )^2 \sin^2\theta 
+ 2\kappa (1+\kappa )(3\cos\theta -1)
-{\rm c}_{\rm w}^2
   \displaystyle\frac{4\kappa (2\kappa +3)^2\cos\theta}{4\kappa +3
  -{\rm s}_{\rm w}^2{\rm c}_{\rm w}^{-2}}\right.\\[0.4cm]
  \left.  +{\rm c}_{\rm w}^2
   \displaystyle\frac{8\kappa (1+\kappa )(1-\cos\theta )(1+3\cos\theta )
   +2[(3+\cos\theta )\kappa +2][(1-\cos\theta )\kappa -\cos\theta ]^2}
   {2\kappa (1-\cos\theta )+{\rm c}_{\rm w}^{-2}} \right]\\[0.4cm]
 +ie^2 \left[ -\displaystyle\frac{\kappa (2\kappa +3)^2\cos\theta}
    {\kappa +1} +4(1+\kappa )(1+3\cos\theta )+
   \displaystyle\frac{[(3+\cos\theta )\kappa +2][(1-\cos\theta )\kappa 
       -\cos\theta ]^2}{\kappa (1-\cos\theta )} \right]~~,
\end{array}
\eqno(5.43a)                                                   
$$  \\
$$
\begin{array}{l}
T[{\rm GB}] = 
ig^2 \left[\displaystyle\frac{(1+\cos\theta )}{2}\kappa 
           +\frac{1}{3}
+\displaystyle\frac{({\rm c}_{\rm w}^2-{\rm s}_{\rm w}^2)^2}
{2{\rm c}_{\rm w}^2}\left(-\displaystyle\frac{2\kappa\cos\theta}
{4\kappa +3-{\rm s}_{\rm w}^2{\rm c}_{\rm w}^{-2}}
+\displaystyle\frac{(3+\cos\theta )\kappa +2}{2(1-\cos\theta )\kappa
+ {\rm c}_{\rm w}^{-2}}\right)\right]\\[0.4cm]
+ie^2\left[ -\displaystyle\frac{4\kappa\cos\theta}{4\kappa +1}
+\displaystyle\frac{(3+\cos\theta )
   \kappa +2}{(1-\cos\theta )\kappa}\right]~~, \\[0.4cm]
\end{array}
\eqno(5.43b)                                                 
$$\\[0.10cm]
where $~\kappa \equiv p^2/M_W^2~$ with $~p~$ equal to the C.M. momentum;
$~{\rm s}_{\rm w}\equiv \sin\theta_{\rm W}~$,
      $~{\rm c}_{\rm w}\equiv \cos\theta_{\rm W}~$ with $\theta_{\rm W}$
equal to the weak mixing angle; 
and $\theta$ is the scattering angle.
In (5.43a) and (5.43b) the terms without a momentum factor in the denominator
come from contact diagrams, terms with denominator independent of
scattering angle come from $s$-channel diagrams and terms with denominator
containing a factor $~1-\cos\theta~$ 
are contributed by $t$-channel diagrams.
Let us consider two special kinematic regions defined below.

\noindent
(i). In the limit of ~$\theta \rightarrow 0^{\circ}$:

As $~\theta\rightarrow 0^{\circ}~$, the $t$-channel photon propagator 
has a kinematic pole, but  
both $W_L$ and GB amplitudes have the {\it same} pole structure, i.e.
$$
\begin{array}{ll}
(T[W_L]-T[{\rm GB}])_{\rm pole~term}
& =-ie^2\left[\displaystyle
\frac{[(3+\cos\theta )\kappa +2]\cos^2\theta}{(1-\cos\theta )\kappa} -
\frac{(3+\cos\theta )\kappa +2}{(1-\cos\theta )\kappa}\right]\\[0.45cm] 
& =-ie^2(1+\cos\theta )(3+\cos\theta +2\kappa^{-1}) = O(g^2)~~,
\end{array}
\eqno(5.44)                                                
$$
which is finite.\footnote{
 ~This conclusion can be directly generalized to other $t$ or $u$
channel processes.}~~ 
Hence, {\it the B-term, 
which is defined as the difference $~T[W_L]-T[{\rm GB}]~$,
is finite at $~\theta =0^{\circ}~$, and is of $~O(g^2)~$.}
This means that when $~\theta~$ is close to the $t$-channel photon pole,
the $B$-term is negligibly small relative to the GB-amplitude
so that (5.40b) is satisfied and
the ET works. More explicitly,  in the limit of 
$~\theta =0^{\circ}~$ (i.e. $~t=0~$), and from (5.43a,b), 
the $W_L$ and GB amplitudes are
$$
\begin{array}{ll}
T[W_L] & = i\left[ 4(3-8{\rm c}_{\rm w}^2+8{\rm c}_{\rm w}^4)
\displaystyle\frac{p^2}{f_\pi^2}
+2e^2\left( 2+\displaystyle\frac{M_W^2}{p^2}\right)\frac{1}{1-c_0}\right]
 +O(g^2)~~,\\[0.45cm]
T[{\rm GB}] & = i\left[ 4(3-8{\rm c}_{\rm w}^2+8{\rm c}_{\rm w}^4)
\displaystyle\frac{p^2}{f_\pi^2}
+2e^2\left( 2+\displaystyle\frac{M_W^2}{p^2}\right)
\displaystyle\frac{1}{1-c_0}\right]
 +O(g^2)~~,\\[0.4cm]
T[W_L] & = T[{\rm GB}] + O(g^2) ~~,
\end{array}
\eqno(5.45)                                               
$$
where $~~c_0\equiv \lim_{\theta\rightarrow 0}\cos\theta~$.~
Notice that in this case one cannot make 
the $M_W^2/t~$ expansion\footnote{ ~This
expansion is {\it unnecessary} for the validity of the ET, 
cf. (5.40) and (5.40a,b).}~~ because ~$t$ vanishes identically. Since 
both $W_L$ and GB amplitudes have exactly the same kinematic
singularity and the $B$-term is much smaller than $T[{\rm GB}]$,
{\it the ET still holds} in this special kinematic region.
We also emphasize that {\it in the 
kinematic regions where $t$ or $u$ is not much larger than $M_W^2$,
the $t$-channel or $u$-channel 
internal gauge boson lines must be included
according to the precise formulation of the ET} 
[cf. (5.40) and (5.40a,b)]. 
This does not imply, in any sense, a violation of the ET 
since the ET, cf. (5.40) and (5.40a,b), does not require
either $t \gg M_W$ or $u \gg M_W$.

\noindent
(ii). In the limit of ~$\theta \rightarrow 180^{\circ}$: 

In the kinematic region with $~s,~t\gg M_W^2~$, (5.41a) and (5.41b) yield
$$
\begin{array}{ll}
T[W_L] & =i\displaystyle\left[ 2(1+\cos\theta )\frac{p^2}{f_\pi^2}
                   +O(g^2)\right] ~~,\\[0.4cm]
T[{\rm GB}] & = i\displaystyle\left[ 2(1+\cos\theta )\frac{p^2}{f_\pi^2}
                   +O(g^2)\right]  ~~,\\[0.55cm]
T[W_L] & = T[{\rm GB}] + O(g^2) ~~,
\end{array} 
\eqno(5.46)                                                  
$$
where the $~O(g^2)~$ term is the largest term 
we ignored which denotes the
order of the $B$-term [cf. (5.48)]; all other terms we ignored 
in $(5.44)$ are of $~O(M_W^2/p^2)~$ or $~O(e^2)~$
which are smaller than $O(g^2)$  and thus will not affect 
the order of magnitude estimate of the $B$-term.
For $~s,~t\gg M_W^2~$, the $W_L$ and GB amplitudes are dominated
by the $p^2$-term in (5.46), which is actually proportional to $u$ for
this process.  
When the scattering angle $\theta$ is close to $180^{\circ}$,
$u$ becomes small and 
thus this leading $p^2$ term is largely suppressed
so that both the $W_L$ and GB amplitudes can be 
as small as the $B$-term, i.e.
of  $O(g^2)$. In this case our condition (5.40a) is satisfied
while (5.40b) is not, which means that the EWSB sector
cannot be sensitively probed for this kinematic region. Since
the total cross section of this process 
is not dominated by this special kinematic region and
is mainly determined by the un-suppressed
leading large $p^2$-term, 
so {\it the kinematic dependence of the amplitude
will not affect the order of magnitude of the total cross section}.
Hence, {\it our application of the power counting analysis in Sec.~7 for 
computing the total event rates remains valid even
though we have ignored the angular dependence in
estimating the magnitude of the scattering amplitudes.}
Neglecting the angular dependence in the amplitude may 
cause a small difference in the event rate 
as compared to that from a precise calculation.
For the processes such as
$~W^\pm_L W^\pm_L\rightarrow W^\pm_L W^\pm_L ~$ and
$~W_L^+ W^-_L \rightarrow Z_LZ_L~$, the leading $p^2$-term
is proportional to $~s/f_\pi^2~$ with no angular dependence, so that
the angular integration causes {\it no difference} between our power
counting analysis and the exact calculation for the leading 
$p^2$-term contribution.\footnote{
 ~~The small difference (a factor of 1.4) in Fig.~4 
mainly comes from  neglecting  the tree level sub-leading terms 
in our order of magnitude estimate for the amplitudes.}~~
In the above example for $~W^+_L W^-_L\rightarrow W^+_L W^-_L ~$ channel
[cf. (5.46)], the leading amplitude is proportional to $~-u/f_\pi^2~$.
When applying the power counting method, 
we ignore the $\theta$-dependence and estimate it as $~s/f_\pi^2~$.
In computing the total rate, we integrate out the scattering angle.
This generates a difference from the precise one:
$$
{ {\int_{-1}^{1} u^2 \, {\rm d}\! \cos \theta} \over  
{\int_{-1}^{1} s^2 \,   {\rm d}\! \cos \theta} } = \frac{1}{3} ~~,
$$
which, as expected, is only a factor of $3$ and does not affect our order
of magnitude estimates.

Finally, we make a precise numerical analysis on the equivalence between
the $W_L$ and the GB amplitudes to show how well the ET works in different
kinematic regions and its implication to probing the EWSB sector.
We use the full expressions (5.43a,b) for $W_L$ and GB amplitudes 
as required by the ET, cf. (5.40) and (5.40a,b).
In Fig.~1a, we plot the ratio $~|B/g^2|~$ for scattering angle
$~\theta = 2^{\circ}, 10^{\circ}, 45^{\circ}, 
90^{\circ}, 100^{\circ}, 120^{\circ}, 135^{\circ}, 
150^{\circ}, 180^{\circ}~$.
It shows that the LNI $B$-term is {\it always of $~O(g^2)~$
in the whole kinematic region,} and thus is irrelevant to
the EWSB sector,  in accordance
 with our general physical analysis in Sec.~5.4 below. 
Hence, to have a sensitive probe of the EWSB mechanism, condition
(5.40b) [or (5.47)] must be satisfied.
Fig.~1b shows that
for $~0^{\circ}\leq \theta \leq 100^{\circ}~$, 
the ratio $~~|B/T[W_L]|\leq 10\% ~~$ when
$~M_{WW}\geq 500$\,GeV. 
For $~\theta \geq 120^{\circ}~$, this ratio becomes
large and reaches $~O(1)~$ when $~\theta~$ is close to $180^{\circ}$.
This is because the kinematic factor
$(1+\cos\theta )$, associated with the leading $p^2$ term [cf. (5.46)],
becomes small. This, however, will
not alter the conclusion that for $4W_L$-scattering the total cross
section from $T[{\rm GB}]$ is much larger than that from the $B$-term
as $~M_{WW}\geq 500$\,GeV.\footnote{  ~~In practice, 
this is even more true after applying the necessary kinematic
cuts to require the final state $W$-bosons to be in the central rapidity 
region of the detector for detecting the signal event~\cite{wwlhc}, so that
the $\theta$ angle cannot be close to either 
$180^{\circ}$ or $0^{\circ}$.}~~  
Note that in Fig.~1b, for  $~\theta \leq 10^{\circ}~$, i.e. close to the
$t$-channel photon pole,
 the ratio $~|B/T[W_L]|~$ is below $~1\% ~$ and thus the ET holds
very well.
In Fig.~1c, we plot both the $W_L$ and GB amplitudes for
$~\theta = 10^{\circ},45^{\circ}, 100^{\circ}, 
150^{\circ}~$. The solid lines denote the 
complete $W_L$ amplitude and the dotted lines denote the GB amplitude.
We find that when $~\theta \leq 100^{\circ}~$, the GB amplitude is almost
indistinguishable from the $W_L$ amplitude. For $~\theta =150^{\circ}~$,
the $W_L$ amplitude is of the same order as the $B$-term, i.e. of 
$~O(g^2)~$, when $~M_{WW}< 1$\,TeV. In this case the $W_L$ or GB
amplitude is
too small and the strongly coupled EWSB sector cannot be sensitively
probed. As the energy $E$ increases, we see that the $W_L$ and GB
amplitudes rapidly dominate over the $B$-term and agree 
better and better even for large scattering angles.
   Finally, using the effective-$W$ method~\cite{effective-W},
we compare the LHC production rates in Figs.~2a and 2b for 
the invariant mass ($M_{WW}$) and the polar angle ($\cos\theta$) 
distributions, respectively. 
To avoid the $t$-channel photon singularity (at
$~\theta =0^{\circ}~$) in the phase space integration, 
we add an angular cut $~-1.0\leq \cos\theta  \leq 0.8~$
(i.e., $~36.9^{\circ}\leq \theta\leq 180^{\circ}~$).
Fig.~2a shows that the total cross sections computed from the
$W_L$ and 
the $GB$ amplitudes [cf. eq.~$(5.41)$] indeed agree with each other
very well. From Fig.~2b, we see that the difference clearly appears 
only for the region of large scattering angle (i.e., $\cos\theta <-0.6~$ or
$~\theta > 127^{\circ}~$) where both the leading 
$W_L$ and GB amplitudes are suppressed by the kinematic factor 
$~1+\cos\theta~$ and thus the event rates are too low to be
sensitive to the EWSB sector. Hence, the difference from the large
$\theta$ region has only {\it negligible} effects on the total 
cross sections, as clearly shown in Fig.~2a. This also agrees with
our conclusion from Figs.~1a-c.
We have also made the comparison with symmetric angular cuts (such as
$~|\cos\theta | \leq 0.8~$) and found similar good agreement to
that in Fig.~10d. This is clear since in the large $\theta$ region 
the event rates become much lower and are close to their 
difference, i.e., of $~O(|R_B|)~$.

The above conclusions hold for the tree level contributions from
the lowest order operators in 
$~{\cal L}_{\rm G} +{\cal L}^{(2)}+{\cal L}_{\rm F}~$, cf. (5.2).
However, independent of the kinematic region considered,
not all the contributions from the NLO effective operators
can dominate the $B$-term and satisfy the condition (5.40b) [or (5.47)]. 
This is why the condition (5.40b) [or (5.47)] can serve as the 
criterion for classifying the sensitivities of these NLO operators 
in probing the EWSB sector for a given scattering process.

We therefore conclude that 
for the process considered here {\it the B-term, as 
defined in (5.17), can be at most of $O(g^2)$ 
for all kinematic regions} (cf. Fig.~1a), 
and is insensitive to the EWSB 
mechanism, in accordance with our general analysis in Sec.~5.4.
{\it When ~$t$ or ~$u$
is not large, the ~$t$- or ~$u$-channel 
internal  lines must be included}.
We find that in certain kinematic region {\it even  $t$ (or $u$) is close to 
zero, the ET still works well} [cf. Eq.~(5.45) and Fig.~1b].
This is because
the validity of the ET does not require either
$~t \gg M_W^2$ or $~u \gg M_W^2$ [cf. (5.40) and (5.40a,b)]. 
 For some scattering processes,
there may be special kinematic regions in which the GB and the $W_L$
amplitudes are largely suppressed\footnote{ ~~This large suppression can also 
arise from the polarization effects of the in/out states.}~~
 so that the EWSB sector cannot be sensitively  probed 
in these special kinematic regions (cf. Figs.~1b,c and 2b). 
But, as shown in this work, 
measuring the total event rates from these processes
can still be used to sensitively probe the EWSB sector 
(cf. Fig.~2a and Figs.~4-12).  

\vspace{0.5cm}
\noindent
{\bf 5.4.} Formulating the ET as a Criterion for Probing the EWSB
\vspace{0.3cm}

In Sec.~5.3.1, we decomposed the amplitude $T[V^{a_1}_L,\cdots,
V^{a_n}_L;\Phi_{\alpha}]$ into the LI part and the LNI part. Such a 
decomposition shows the {\it essential difference} between the $V_L$- and 
the $V_T$-amplitudes: the former contains a LI GB-amplitude that can 
yield a large $V_L$-amplitude in the case of
strongly coupled EWSB sector, but the latter does not.
We note that only the LI part (the GB-amplitude) of the $V_L$-amplitude 
is {\it sensitive} to probing the EWSB sector, 
while its LNI part, containing a
significant {\it Lorentz-frame-dependent} $~B$-term (which is related to
the $V_L$-$V_T$ mixing effects under proper Lorentz transformations),
 is {\it insensitive} to the EWSB mechanism.
Thus the $B$-term serves as an intrinsic background to the probe of the
EWSB mechanism, and for a {\it sensitive and unambiguous} probe of the EWSB,
the LI GB-amplitude should dominate the $V_L$-amplitude and the LNI $B$-term 
should be negligible.
It is obvious that one can technically improve the prediction for the 
$V_L$-amplitude from the right-hand side (RHS) of (5.17) 
by including the complicated $B$-term 
( or part of $B$ )~\cite{gk} or even directly calculate its
left-hand side (LHS) of (5.17) despite the complexity.
However, this is {\it not} an improvement of the 
longitudinal-Goldstone boson {\it equivalence} and thus the sensitivity 
of probing the EWSB mechanism via  $V_L$-scattering experiments.
 {\it The physical content of the ET is essentially
independent of how to numerically compute the $V_L$-amplitude.}

Now we study the precise meaning of the condition (5.40b) for
neglecting the $B$-term, i.e. the condition for sensitively probing the
EWSB mechanism. 
The amplitude $T$, to a finite order, can be written as 
$~~T= \sum_{\ell=0}^N T_\ell ~$ in the perturbative calculation.
Let $~~ T_0 > T_1,\cdots, T_N \geq T_{\min}~$, where
$~T_{\min}= \{ T_0, \cdots , T_N \}_{\min}~$, then condition (5.40b)
and eq.~(5.39) imply
$$
\begin{array}{l}
~~~~~~~~~~
T_{\min}[-i\pi^{a_1},\cdots ,-i\pi^{a_n};\Phi_{\alpha}] ~\gg ~\\[0.35cm]
O\left(\frac{M_W^2}{E_j^2}\right) \,
T_0[ -i\pi^{a_1},\cdots , -i\pi^{a_n}; 
   \Phi_{\alpha}] + 
  O\left(\frac{M_W}{E_j}\right) \,T_0[ V_{T_j}^{a_{r_1}}, -i\pi^{a_{r_2}},
                      \cdots , -i\pi^{a_{r_n}}; \Phi_{\alpha}] ~~.
\end{array}                                                                    
\eqno(5.47)                                                 
$$                                                                             
Note that the above formulation of the ET discriminates processes
which are insensitive to probing the EWSB sector when either
(5.40a) or (5.40b) fails.
Furthermore, {\it as a necessary criterion, condition (5.47) determines 
whether or not the corresponding $~V_L$-scattering process 
in (5.40) is sensitive to probing 
the EWSB sector to the desired precision in perturbative calculations.}

From (5.39) or the RHS of (5.47) and the precise electroweak power counting
rule (cf. Sec.3.2), we can directly estimate the 
{\it largest} model-independent $B$-term to be
$~~~B_{\max} = O(g^2)f_{\pi}^{4-n}~~$ in the EWCL 
formalism, which comes from the $n$-particle pure $V_L$-amplitude.
(This conclusion also holds for the heavy Higgs SM.)
It is crucial to note that $B_{\max}$ is of the same order 
of magnitude as the leading $V_T$-amplitude:
$$
B_{\max} \approx T_0[V^{a_1}_T,\cdots ,V^{a_n}_T] 
= O(g^2)f_{\pi}^{4-n}~~.
\eqno(5.48)                                                   
$$
Since both the largest $B$-term and the leading $V_T$-amplitude are of 
$O(g^2)$~, they are therefore irrelevant to the EWSB mechanism 
as pointed out in the above discussion. Thus, {\it  (5.47) provides a 
useful criterion for discriminating physical processes 
which are sensitive, marginally sensitive, or insensitive to 
the EWSB sector.}

In conclusion, our formulation of the ET 
provides a necessary criterion for probing the EWSB sector as follows. 
If the ET is valid for a scattering process to the order of
$~T_{\min}~$ [i.e. $~T_{\min}\gg B_0~$, cf.~(5.47),
where $B_0$ is the tree-level leading contribution to $B$.], 
this process is classified to be {\it sensitive} to probing $~T_{\min}~$.
Otherwise, we classify this process to be either 
{\it marginally sensitive} (for $~T_{\min}> B_0$ but 
$T_{\min}\not\gg B_0$~) or {\it insensitive} (for $~T_{\min}\leq B_0~$)
to testing $~T_{\min}~$. 
This classification is given at the level of the 
$S$-matrix elements. Up to the next-to-leading order, 
$~T_{\min}=T_1~$. For this case, 
by simply squaring both side of the condition (5.47)
and integrating over the phase space, 
we can easily derive the corresponding condition
(criterion) at the level of the constituent cross sections for 
$~\hat{\sigma}_{1}\simeq \int_{\rm phase} 2T_0T_{1}~$ and 
$~\hat{\sigma}_{B}\simeq \int_{\rm phase} 2T_0B_0~$, 
where $~\int_{\rm phase}~$ denotes the phase space integration. 
The constituent cross sections ($\hat{\sigma}$)
are functions of the invariant mass 
( $\sqrt{\hat{s}}$ ) of the final state weak bosons.
Defining the differential parton luminosity (for either 
the incoming light fermion or the weak boson) as
$~\frac{{\rm d} \,{\cal L}_{\rm partons}}{{\rm d} \,{\hat{s}}} ~$,
the total cross section is thus given by
$$  
\sigma  ~=~ \int {\rm d} \, {\hat{s}}
\frac{{\rm d} \, {\cal L}_{\rm partons}}{{\rm d} \, {\hat{s}}} 
\hat{\sigma}
({\hat{s}}) ~~.
\eqno(5.49)                                         
$$ 
Using (5.49) we can further derive the corresponding 
conditions for total event rates $R_1$ 
(calculated from $\hat{\sigma}_1$) and
$R_B$ (calculated from $\hat{\sigma}_B$), 
and then define the corresponding criterion for 
testing the sensitivities of various 
operators and processes at the level of event rates.
They are:
(i). Sensitive, if $~R_1 \gg R_B~$; (ii). Marginally sensitive,
  if $~R_1 > R_B~$, but $~R_1\not\gg R_B~$; (iii). Insensitive,
  if $~R_1 \leq R_B~$. 
A specific application to the LHC physics is given  in Sec.~7.
We note that, at the event rate level, the above criterion is 
{\it necessary} but not sufficient since the leading $B$-term, of the
same order as the LNI $V_L$-$V_T$ mixing effects [cf. (5.39)] and as
an intrinsic background to any strong $V_L$-$V_L$ scattering process,
denotes a universal part of the full backgrounds~\cite{mike-EWA}.
The sufficiency will of course require detailed numerical analyses
on the detection efficiency for suppressing the full backgrounds to 
observe the specific decay mode of the final state (cf.~Ref.~\cite{wwlhc}). 
This is beyond our present first step theoretical 
global study. Some general discussions on the experimental detections
will be presented in Sec.~8.

Before concluding this section, we note that
in our power counting analysis
(cf. Sec.4), 
{\it both the GB-amplitude and the $B$-term are explicitly estimated}
(cf. Tables 1-4 in Sec.4).
The issue of numerically
including/ignoring $B$ in an explicit calculation
is {\it essentially  irrelevant} here. If $~T_1 \leq B~$, this
means that the sensitivity is poor so that the probe of $T_1$ is
experimentally harder and requires a higher experimental 
precision of at least the order of $B$ to test $T_1$.

\vspace{0.7cm}
\noindent
{\bf 6. Global Classifications for $S$-matrix Elements}
\vspace{0.3cm}

Armed with the above counting rule (4.5), 
we can conveniently estimate  contributions 
from all effective operators in the EWCL to  
any high energy scattering process.  
In the literature (cf. Ref.~\cite{wwlhc}), what usually done 
was to study
only a small subset of all effective operators for simplicity. But,
to discriminate different underlying theories for a complete test of the
EWSB mechanism, it is necessary to measure all these
operators via various high energy processes.
As the first step global study,
our electroweak power counting analysis 
makes it possible to quickly grasp the overall physical picture 
which provides a useful guideline
for selecting relevant operators and scattering processes to perform
further detailed numerical studies. 
In this and the next sections, we shall systematically classify all
possible NLO effective operators for both the
$S$-matrix elements and  event rates at the LHC and the LC.

\vspace{0.5cm}
\noindent
{\bf 6.1.} Power Counting Hierarchy
\vspace{0.25cm}

We concentrate on the high energy weak-boson fusion and 
quark-anti-quark annihilation processes.
As shown in Refs.~\cite{wwlhc,mike-ww}, for the non-resonance case, 
the most important fusion process 
for probing the EWSB sector is the same-charged channel: 
$~W^{\pm}W^{\pm}\rightarrow W^{\pm}W^{\pm}~$, which gets 
dominant contributions from the 4-GB vertices in the EWCL.
In Tables~Ia and 1b we estimate the contributions from the lowest order 
(model-independent) operators in 
$~{\cal L}_{\rm MI}\equiv 
{\cal L}_{\rm G} +{\cal L}_{\rm F} +{\cal L}^{(2)}~$ up to one-loop 
and from all the 
NLO (model-dependent) bosonic operators in (2.2) at the tree-level
for $~W^{\pm}W^{\pm}\rightarrow W^{\pm}W^{\pm}~$.
The contributions of different operators to a given amplitude are
different due to their different structures.
For instance, the commonly discussed operators $~{\cal L}_{4,5}~$
contribute the model-dependent leading term of
$~~O\left(\frac{E^2}{f_\pi^2}\frac{E^2}{\Lambda^2}\right)~~$ 
to the $~T[4W_L]~$ amplitude,
and the sub-leading term of 
$~~O\left(g\frac{E}{f_\pi}\frac{E^2}{\Lambda^2}\right)~~$ to 
the $~T[3W_L,W_T]~$ amplitude, while $~{\cal L}_{3,9}~$ give their largest
contributions to $~T[3W_L,W_T]~$ rather than $~T[4W_L]~$ at high energies.
The model-independent
and model-dependent contributions to various $B$-terms 
are summarized in Tables~IIa and IIb, in which
$B^{(i)}_{\ell}$ ($i=0,\cdots ,3;~\ell =0,1,\cdots$)
denotes the $B$-term from $V_L$-amplitudes containing $i$ external 
$V_T$-lines with $B^{(i)}_{0}$ obtained from the leading order and
$B^{(i)}_1$ from the NLO calculations.  
We see that the largest $B$-term is  
$B^{(0)}_0$ from the $4W_L$ amplitudes, as given in (5.48).
The term $B^{(0)}_0$ [of $O(g^2)$], is 
a model-independent constant containing only 
the SM gauge coupling constants.
All the other $B$-terms are further suppressed 
by a factor of $M_W/E$ or $(E/\Lambda )^2$, or their product.

For all the $~q\bar{q}^{(\prime )}\rightarrow V^aV^b~$ 
processes (with $q$ or ${q}^{(\prime)}$ being light quarks except
the top),  which get dominant contributions from $s$-channel diagrams 
(containing the $~V_T$-GB-GB vertices), the
model-independent and the model-dependent contributions are estimated
in Tables III and IV, respectively. 
Note that the tree-level $~q\bar{q}\rightarrow ZZ~$ 
annihilation process has no model-dependent NLO contribution, 
therefore to probe new physics in the EWSB sector, we have to study
$~q\bar{q}^{(\prime )}\rightarrow W^+W^-,~W^\pm Z~$ annihilations.
As shown in Tables~IVa and IVb, the operators $~{\cal L}_{2,3,9}~$
give the leading contributions, 
of $~O\left(g^2\frac{E^2}{\Lambda^2}\right)~$, to 
$~q\bar{q}\rightarrow W^+W^-~$ via
$~T_1[q\bar{q};W^+_LW^-_L]~$ channel\footnote{ ~~We note that
the contributions from $~{\cal L}_2~$ 
(and also $~{\cal L}_{1,13}~$)  are always associated
with a suppressing factor $~\sin^2\theta_W \sim {1\over 4}$~.},~
and the operators $~{\cal L}_{3,11,12}~$
give the same leading contributions 
to $~q\bar{q}^{\prime}\rightarrow W^\pm Z~$
via $~T_1[q\bar{q}^{\prime};W^\pm_L Z_L]~$ channel.
But $~{\cal L}^{(2)\prime}~$ does not 
contribute any positive $E$-power term to any of the
 $~V^aV^b$ final states via
tree-level quark-anti-quark annihilations.
 Tables~III and IV also show that the largest $B$-term is
$~B=B_0^{(1)}=O\left(g^2\frac{M_W}{E}\right)~$ which is model-independent
and comes from $~T_0[q\bar{q}^{(\prime )};V_T,v]~$, a part of 
the $~T_0[q\bar{q}^{(\prime )};V_T,V_L]~$ amplitude (cf. Table~IIIa).
All model-dependent $B$-terms, as listed in Tables~IVa and IVb,
are either constant terms of 
$~O\left((g^4,e^2g^2)\frac{f_\pi^2}{\Lambda^2}\right)~$ or further
suppressed by negative $E$-power(s) and are thus negligibly small.

From Tables I-II, we  further classify in Table~V
the sensitivities to all the bosonic  operators for probing
the EWSB sector either
directly (from pure GB interactions) or indirectly (from interactions
suppressed by the SM gauge coupling constants).
The same classification for 
all $~q\bar{q}^{(\prime )}\rightarrow V^aV^b~$
annihilation processes is separately given in Table~VI.
The classifications in Tables~V and VI 
are based upon the following hierarchy in the power counting:
$$ \dis
\frac{E^2}{f_\pi^2}\gg 
\left[\frac{E^2}{f_\pi^2}\frac{E^2}{\Lambda^2},~g\frac{E}{f_\pi}\right] \gg 
\left[g\frac{E}{f_\pi}\frac{E^2}{\Lambda^2}, ~g^2\right] \gg 
\left[g^2\frac{E^2}{\Lambda^2}, ~g^3\frac{f_\pi}{E}\right] \gg 
\left[g^3\frac{Ef_\pi}{\Lambda^2},~g^4\frac{f^2_\pi}{E^2}\right]\gg 
g^4\frac{f_\pi^2}{\Lambda^2}~~.
\eqno(6.1)                               
$$
In the typical TeV region, for 
$~~E\in (750\,{\rm GeV},~1.5\,{\rm TeV})$,
this gives:
$$
\begin{array}{c}
(9.3,37)\gg \left[(0.55,8.8),(2.0,4.0)\right]\gg 
\left[ (0.12,0.93),(0.42,0.42)\right] \gg \\
\left[ (0.025,0.099),(0.089,0.045)\right]\gg 
\left[ (5.3,10.5),(19.0,4.7)\right]\times 10^{-3}\gg 
(1.1,1.1)\times 10^{-3} ~,
\end{array}
\eqno(6.2)                             
$$
where $E$ is taken to be the invariant mass of the $VV$ pair.
The numerical values in (6.2) convincingly show the existence of 
the power counting hierarchy in (6.1). 
This governs the order of magnitude
of the results from  detailed numerical calculations. 
This  hierarchy makes it possible to conveniently and globally 
classify the sensitivities of various scattering processes to
the complete set of the effective operators in the EWCL. 
The construction of this power counting hierarchy is based upon the
property of the chiral perturbation expansion and  
can be understood as follows.
The leading term $~\frac{E^2}{f_\pi^2}~$ in (6.1) comes from the 
model-independent lowest order $4V_L$ ($\neq 4Z_L$) scatterings.
 Starting from this leading term, 
(6.1) is built up by {\it increasing either
the number of derivatives 
(i.e. the power of $E/\Lambda$) or the number of
external transverse gauge bosons 
(i.e. the power of gauge coupling constants).}
The NLO contributions from the derivative expansion are 
always suppressed
by $~E^2/\Lambda^2~$ relative to the model-independent leading term.
Also, for each given process, when an external $V_L$-line is replaced 
by a corresponding $V_T$-line, 
a factor $\frac{E}{f_\pi}$ in the amplitude would be replaced by 
a gauge coupling $g$ (or $g^\prime$).\footnote{
 ~~The counting on the amplitudes $T_0[4W_T]$ and 
$T_0[q\bar{q}^{(\prime )};V_TV_T]$ are exceptions of this rule
since they have a contribution from the tree-level pure
Yang-Mills gauge term $~{\cal L}_{\rm G}~$.
These two similar exceptions can be found at the second line of 
Tables~Ia and IIIa, respectively.}
~~This explains why the power counting hierarchy takes the form of (6.1).

\vspace{0.5cm}
\noindent
{\bf 6.2.} Classifications for $VV$-fusions and $f\bar{f}'$-annihilations
\vspace{0.25cm}

Tables~V and VI are organized in accordance 
with the power counting hierarchy 
given in (6.1) for all $VV$-fusion and 
$q\bar{q}^{(\prime )}$-annihilation amplitudes.
It shows the {\it relevant} effective new physics operators and the 
corresponding physical processes for probing the EWSB sector
when calculating the scattering amplitudes to the required precision.
For instance, according to the classification of  Table~V, 
the model-independent operator $~{\cal L}_{\rm MI}~$ can be probed
via studying the leading tree-level scattering amplitude
$~T_0[4V_L]~(\neq T_0[4Z_L])~$ which is of 
$~O\left(\frac{E^2}{f_\pi^2}\right)~$. A sensitive probe of 
$~{\cal L}_{\rm MI}~$ via this amplitude requires $~T_0\gg B_0~$, i.e.,
$~O\left(\frac{E^2}{f_\pi^2}\right)\gg O(g^2)~$ or 
$~~\left(2M_W/E\right)^2\ll 1~~$ 
which can be well satisfied in the high energy
region $~E\geq 500$~GeV. We note that the test of the leading order 
operator ${\cal L}_{\rm MI}$ will first distinguish
the strongly interacting EWSB sector from the weakly interacting one.
To test the model-dependent operators $~{\cal L}_{4,5,6,7,10}~$ demands
a higher precision than the leading tree level contribution by a factor
of $~\frac{E^2}{\Lambda^2}~$. As an example, 
in order to sensitively test the 
$~{\cal L}_{4,5}$ operators with coefficients of $O(1)$ 
via the $4V_L$-processes, the criterion (5.47) requires
$~~O\left(\frac{E^2}{f_\pi^2}\frac{E^2}{\Lambda^2}\right)\gg O(g^2)~$,~
or, $~~\left(0.7{\rm TeV}/E\right)^4\ll 1~~$. 
This indicates that sensitively
probing $~{\cal L}_{4,5}~$ via the $4V_L^\pm$-scatterings
requires $~E\geq 1$~TeV. Thus, we find that,
in the TeV region, the $~4V_L$ scatterings can 
sensitively probe $~{\cal L}_{4,5}~$; 
while, similarily, $~{\cal L}_{6,7}~$ can
be probed via $~2W_L+2Z_L~$ or $~4Z_L~$ scattering and $~{\cal L}_{10}~$ 
can only be tested  via $~4Z_L~$ scattering. 
As shown in Table~III, to probe the operators
$~{\cal L}_{2,3,9,11;12}~$, one has to detect the $~3V_L+V_T$
scatterings, which are further suppressed by a factor $~\frac{M_W}{E}~$
relative to the leading model-dependent contributions from the
$~{\cal L}_{4,5}~$ and $~{\cal L}_{6,7,10}~$ via $4V_L$ processes.
Since the model-independent
leading order $2V_T+2V_L$ and $4V_T$ amplitudes 
(from ${\cal L}_{\rm MI}$) and the largest constant
$B$-term $\left( B^{(0)}_0\right)$  are all around of the same order, 
i.e. $~O\left(g\frac{E}{f_\pi}\frac{E^2}{\Lambda^2},g^2\right)~$ 
[cf. (6.2)],\footnote{
 ~~They can in principle be separated if the polarization of
the external $V$-lines are identified.
For the final state $V$'s, one can study the angular 
distribution of the leptons from $V$-decay. 
For the incoming $V$'s, one can use forward-jet
tagging and central-jet vetoing to select 
longitudinal $V$'s \cite{wwww}.} 
~~it requires  a significantly higher precision 
to sensitively probe the operators $~{\cal L}_{2,3,9,11;12}~$ 
which can only contribute
the $g$-suppressed indirect EWSB information
and therefore are more difficult to be tested. Here the ratio
$~B_0/T_1\sim g^2/\left[g\frac{E}{f_\pi}\frac{E^2}{\Lambda^2}\right]~$ 
gives 
$~\left(1.15{\rm TeV}/E\right)^3\simeq 0.45\not\ll 1~$ for $E=1.5$~TeV,
which shows the probe of these operators is at most marginally
sensitive when their coefficients $~\ell_n=O(1)~$.
Finally, the operators $~{\cal L}_{1,8;13,14}~$ can be probed 
via the amplitude
$~T_1[2V_L,2V_T] ~(\neq T_1[2Z_L,2Z_T])~$ which is of 
$~O\left(g^2\frac{E^2}{\Lambda^2},g^3\frac{f_\pi}{E}\right)~$ 
and numerically much smaller [cf. (6.2)] in comparison with the
leading $B$-term in (5.48). 
Therefore, $~{\cal L}_{1,8;13,14}~$ should be effectively probed 
via scattering processes other than the $VV$-fusions.

We then look at Table~VI for $q\bar{q}^{(\prime )}$-annihilations.
For the lowest order Lagrangian 
$~{\cal L}_{\rm MI}= 
  {\cal L}_{\rm G}+{\cal L}^{(2)}+{\cal L}_{\rm F}~$,
the model-independent operators 
$~{\cal L}^{(2)}~$ and $~{\cal L}_{\rm G}~$
can be probed via tree-level amplitudes [of $~O(g^2)~$]
with $~V_LV_L~$ and ~$V_TV_T$~ final states, respectively.\footnote{
 ~~$~{\cal L}^{(2)}~$ just gives the low energy theorem results
and thus denotes the model-independent part of the EWSB sector, while
$~{\cal L}_{\rm G}~$ is the standard tree-level Yang-Mills gauge term
which is irrelevant to the EWSB mechanism.}
~~Thus, the contribution of $~{\cal L}^{(2)}~$ to $V_LV_L$ final state is
not enhanced by any $E$-power in the high energy region, in contrast to
the case of $VV$-fusions (cf. Table~Ia). 
So, the leading order $~T_0[q\bar{q}^{(\prime )};V_LV_L]~$ amplitude,
similar to the $~T_0[q\bar{q}^{(\prime )};V_TV_T]~$ amplitude,
is not sensitive to the strongly coupled EWSB sector. 
We then discuss the contributions of model-dependent 
NLO operators to the $q\bar{q}^{(\prime )}$-annihilations.
We first note that the operators $~{\cal L}_{4,5,6,7,10}~$ cannot
contribute to $q\bar{q}^{(\prime )}$-annihilations at $~1/\Lambda^2$-order
and thus should be best probed via $VV$-fusions (cf. Table~V).
Among all other NLO operators, 
the probe of $~{\cal L}_{2,3,9}~$ 
are most sensitive via $~q\bar{q}\rightarrow W^+_LW^-_L$ amplitude
and the probe of $~{\cal L}_{3,11,12}~$ are best via
$~q\bar{q}^{\prime}\rightarrow W^\pm_LZ_L$ amplitude. For operators 
$~{\cal L}_{1,8;13,14}~$, the largest amplitudes are
$~T_1[q\bar{q}^{(\prime )};W^+_LW^-_T/W^+_TW^-_L]~$ and
$~T_1[q\bar{q}^{\prime};W^\pm_LZ_T/W^\pm_TZ_L]~$, which are
at most of $~O\left(g^3\frac{Ef_\pi}{\Lambda^2}\right)~$ and are
suppressed by a factor $~\frac{f_\pi}{E}~$ relative to the 
model-dependent leading amplitudes of 
$~O\left(g^2\frac{E^2}{\Lambda^2}\right)~$.

In summary, applying the power counting technique allows us to
conveniently estimate contributions of various operators 
to any scattering amplitude.
For a given scattering process, this result tells us which 
operators can be sensitively probed. Similarly, the same result can 
also tell us which process would be most sensitive for probing 
new physics via a given effective operator.
In the next section, we shall examine the important
$W^\pm W^\pm \rightarrow W^\pm W^\pm$ fusion and 
$~q\bar{q}^{\prime}\rightarrow W^\pm Z~$ annihilation
processes at the LHC to illustrate how to use 
the electroweak power counting method
to estimate the event rates and 
how to use the ET as a theoretical criterion
to classify the sensitivities of these typical scattering processes
to the NLO bosonic operators in the EWCL.\\

\vspace{0.5cm}
\noindent
{\bf 6.3.} A comparative analysis for the LHC and LCs
\vspace{0.25cm}

The values of coefficients ($\ell_n$'s) of the 15 NLO operators 
in the EWCL (2.2) depend on the details of the underlying dynamics 
and reflect the possible new physics.
Among the 15 NLO coefficients,
$\ell_1$, $\ell_0$ and $\ell_8$ correspond to 
${\rm S}$, ${\rm T}$ and ${\rm U}$ parameters~\cite{STU}. 
They have been measured 
from the current low energy LEP/SLC data 
and will be further improved at LEPII and upgraded Tevatron.
To distinguish different models of the EWSB, the rest of the $\ell_n$'s  
has to be measured by studying the scattering processes
involving weak gauge bosons. 
What is usually done in the literature is to consider only 
a small subset of these operators at a time. For instance, 
in Ref.~\cite{wwlhc}, a non-resonant model (called Delay-K model) was 
studied which includes ${\cal L}^{(2)}$ as well as
the NLO operators ~${\cal L}_4$ and ${\cal L}_5$~. It was found that 
for the gold-plated mode (i.e. pure leptonic decay mode) 
of $W^\pm W^\pm$, a total number of about 10 signal events is expected
 at the LHC with a 100\,$\ifb$ luminosity after 
imposing relevant kinematic cuts to suppress backgrounds.
In the end of the analysis the ratio of signal to background is about 1.
Another non-resonant model (called LET-CG model), which contains 
only the model-independent operator ${\cal L}^{(2)}$,
was also studied in that paper.
The difference between the predictions of these two models
signals the effects from the NLO operators ${\cal L}_{4,5}$~.
With just a handful events, it requires
higher integrated luminosities to probe these NLO operators and
compare with the model-independent contributions from ${\cal L}^{(2)}$~.
Generally speaking, if one combines measurements from various 
$VV$-modes, it is possible (although not easy) to distinguish 
models of EWSB which effectively include different subsets
of the 15 NLO operators and the 
model-independent operator ${\cal L}^{(2)}$~.  

The important question to ask is:
`` How and to what extent can one measure 
{\it all}~ the NLO coefficients $\ell_n$  at future colliders  
to {\it fully} explore the EWSB sector? ''
To answer this question, as the first step, one should 
{\bf (i)}. find out, for each given NLO operator, whether 
it can be measured via leading and/or sub-leading amplitudes 
of relevant processes at each collider;
{\bf (ii)}. determine whether a given NLO 
operator can be sensitively (or marginally sensitively) probed 
through its contributions to the leading (or sub-leading) 
amplitudes of the relevant scattering process at each given collider;
{\bf (iii)}. determine whether carrying out the above study for various 
high energy colliders can {\it complementarily} cover all 
the 15 NLO operators to probe the strongly interacting EWSB sector.
For abbreviation, the above requirements {\bf (i)}-{\bf (iii)} 
will be referred hereafter as the `` {\it Minimal Requirements} ''.

The minimal requirements (i)-(ii) have been analyzed for $V$-$V$ fusions
and $q$-$\bar{q}^{(\prime )}$ annihilations in the last sub-section. 
This sub-section is mainly devoted to discuss our 
{\it Minimal Requirement}-{\bf (iii)} for different high energy colliders
besides including more channels.
It is understood that the actual sensitivity
of a collider to probe the NLO operators depends
not only on the luminosities of the active partons (including 
weak-gauge bosons) inside hadrons or electrons,
but also on the detection efficiency of the signal events after
applying background-suppressing kinematic cuts
to observe the specific decay mode of the final state
weak-bosons (cf. Refs.~\cite{wwlhc,wwnlc} and Sec.~8 below). 
However, all of these will only add fine structures 
to the sub-leading contributions listed in Table~II but not 
affect our conclusions about the leading contributions
as long as there are enough signal events produced.
In this global analysis, we shall not perform a detailed numerical study 
like Refs.~\cite{wwlhc,wwnlc}, 
but only give a first-step qualitative global power counting 
analysis which serves as a useful guideline for further elaborating
numerical calculations.

After examining all the relevant 
$ 2 \rightarrow 2$ and $2 \rightarrow 3$ hard scattering
processes, we summarize in Table~VII our global classification for 
the sensitivities of various 
future high energy colliders to probing the 15 
model-dependent NLO bosonic operators.
Here, the energy-$E$ represents the typical energy scale
of the hard scattering processes under consideration.
The leading $B$-term for each high energy
process is also listed and compared with the corresponding
$V_L$-amplitude. If the polarizations of the 
initial/final state gauge bosons are not distinguished 
but simply summed up, the largest $B$ 
in each process (including all possible polarization states)
should be considered for comparison.  
[If the leading $B_0$ (with just one $v_\mu$-factor, cf. eq.~(5.17b))
 happens to be zero, then the largest next-to-leading
 term, either the part of $B_0$ term that contains
$2$ (or $3$) $v_\mu$-factors or the $B_1$ term, should be considered.
Examples are the $~ZZ\rightarrow ZZ~$ and
$~f\bar{f}\rightarrow ZZZ~$ processes.]
By comparing $T_1$ with $B$ in Table~II and applying
our criterion for classifying the sensitivities, we find that
for the typical energy scale ($E$) of the relevant processes at each
collider, the leading contributions (~marked by $\surd~$) 
can be sensitively probed, while the sub-leading contributions
(~marked by $\triangle~$) can only be marginally sensitively 
probed.\footnote{  ~~The exceptions are 
$~f\bar{f}^{(\prime )}\rightarrow W^+W^-/(LT),W^\pm Z/(LT)~$ 
for which $~T_1 \leq B_0$~. Thus the probe of them is insensitive.
($L/T$ denotes the longitudinal/transverse polarizations of
$~W^\pm ,~Z^0~$ bosons.) }
~~(To save space, 
Table~II does not list those processes to which the NLO operators 
{\it only} contribute sub-leading amplitudes. These processes
are $~WW\rightarrow W\gamma ,Z\gamma +{\rm perm.}~$ 
and $~f\bar{f}^{(\prime )}\rightarrow W\gamma ,WW\gamma , WZ\gamma 
~$, which all have one external transverse $\gamma$-line and 
are at most marginally sensitive.)

From Table~VII, some of our conclusions can be drawn as follows. \\
~~~{\bf (1).}
At LC(0.5), which is a LC with $\sqrt{S}=0.5$\,TeV, $\ell_{2,3,9}$
can be sensitively probed via $e^-e^+ \rightarrow W^-_L W^+_L$. \\ 
~~~{\bf (2).}
For pure $V_L V_L \rightarrow V_L V_L$ scattering amplitudes, 
the model-dependent operators ${\cal L}_{4,5}$
and ${\cal L}_{6,7}$ can be probed 
most sensitively. 
 ${\ell}_{10}$ can only be sensitively probed 
via the scattering process $Z_LZ_L \rightarrow Z_LZ_L$ which 
is easier to detect at the LC(1.5) [a $e^-e^+$ or $e^-e^-$ collider 
with $\sqrt{S}=1.5$\,TeV] than at the LHC(14) [a pp collider with
$\sqrt{S}=14$\,TeV]. \\ 
~~~{\bf (3).}
The contributions from ${\cal L}^{(2)\prime}$~ and 
${\cal L}_{2,3,9}$ to the pure $4V_L$-scattering processes
 lose the $E$-power dependence by a 
factor of $2$ (see, e.g., Table~Ib). Hence, the pure $4V_L$-channel is 
less sensitive to these operators. 
[Note that ${\cal L}_{2,3,9}$  can be sensitively 
probed via $f {\bar f} \rightarrow W_L^-W_L^+$ process at LC(0.5) and LHC(14).]
The pure $4V_L$-channel cannot probe ${\cal L}_{1,8,11\sim 14}$ which
can only be probed via processes with $V_T$('s). 
Among ${\cal L}_{1,8,11\sim 14}$,
the contributions from $~{\cal L}_{11,12}~$ to processes 
with $V_T$('s) are most important, although their contributions 
are relatively suppressed by a factor $gf_\pi /E$  as compared to
the leading contributions from 
${\cal L}_{4,5}$ to pure $4V_L$-scatterings.
${\cal L}_{1,8,13,14}$ are generally suppressed by higher powers of
$gf_\pi /E$ and are thus the least sensitive.
The above conclusions hold for both LHC(14) 
and LC(1.5). \\
~~~{\bf (4).} 
At LHC(14), ${\ell}_{11,12}$ 
can be sensitively probed via $q \bar q' \rightarrow W^\pm Z$
whose final state is not electrically neutral. Thus, 
this final state is not accessible at LC. 
Hence, LC(0.5) will not be sensitive to these operators.
To sensitively probe ${\ell}_{11,12}$ at LC(1.5), one has to measure
$e^-e^+ \rightarrow W^-_L W^+_L Z_L$. 
\\ 
~~~{\bf (5).}
To sensitively probe ${\ell}_{13,14}$,
a high energy $~e^-\gamma~$ linear collider 
is needed for studying the processes 
$~e^-\gamma \rightarrow \nu_e W^-_LZ_L,~e^-W^-_LW^+_L~$, 
in which the backgrounds \cite{eAback}~ are much
less severe than processes like $\gamma \gamma \rightarrow W^+_L W^-_L$ at
a $\gamma\gamma$ collider~\cite{DPF,lcws93}. The amplitude of
$\gamma \gamma \rightarrow W^+_L W^-_L$ has the order of 
$~e^2\frac{E^2}{\Lambda^2}~$, to which the ${\cal L}_{13,14}$
(and also ${\cal L}_{1,2,3,8,9}$) can contribute. Thus, this process
would be useful for probing ${\ell}_{13,14}$ 
at a $\gamma\gamma$ collider
if the backgrounds could be efficiently suppressed.

We also note that to measure the individual coefficient of the 
NLO operator, one has to be able to separate , for example, the
$W^+W^- \ra Z^0 Z^0$ and the $Z^0Z^0 \ra Z^0 Z^0$ production processes.
Although this task can be easily done at the LC by detecting a forward
tagged lepton, it shall be a great challenge at the LHC because 
both the up- and down-type quarks from the initial state contribute to the 
scattering processes.
From the above conclusion,  we speculate\footnote{
  ~~To further reach a detailed quantitative conclusion, 
an elaborate and precise numerical study on all signal/background rates 
will be performed under this global guideline 
(cf. Ref.~\cite{desy}).} ~~that 
if there is no new resonance below the TeV scale and 
the coefficients of the NLO 
operators are not much larger than that suggested by 
the naive dimensional analysis~\cite{nda}, the LHC alone may not be 
able to sensitively measure all these operators,
and linear colliders are needed to {\it complementarily}
cover the rest of the NLO operators.
In fact, the different phases of 500 GeV and 1.5 TeV energies at the LC
are necessary because they will be sensitive to different 
NLO operators.  An electron-photon (or a photon-photon) 
collider is also useful for measuring
some of the NLO operators that distinguish models
of strongly coupled EWSB sector.
 
\vspace{0.7cm}
\noindent
{\bf 7. Global Analysis at the Event Rate Level}
\vspace{0.4cm}

In this section, we estimate the production rates for both
weak boson fusions and quark-anti-quark annihilations at the 
LHC (a ${\rm p}{\rm p}$ collider with $\sqrt{S}=14$\,TeV 
and an integrated luminosity of $~100\,{\rm fb}^{-1}~$) and
the LC (a $e^-e^+$ or $e^-e^-$ collider with $\sqrt{S}=1.5$\,TeV 
and an integrated luminosity of $~200\,{\rm fb}^{-1}~$).
The gauge-boson fusion process $W^+W^+ \rightarrow W^+W^+$  and the
quark-anti-quark annihilation process $~q\bar{q}^{\prime}\rightarrow W^+Z~$ 
will be separately studied at the LHC.

To calculate the event rates for gauge-boson fusions, we multiply
the luminosity of the incoming weak-boson pair $VV$
(by the effective-$W$ approximation (EWA)~\cite{effective-W}) 
and the constituent cross section of the weak-boson scattering
(from the amplitude estimated by the power counting analysis in 
the last section). Note that the validity of the EWA
requires the $VV$ invariant mass $~~M_{VV}\gg 2M_W~~$ \cite{effective-W},
which coincides with the condition in (2.3a) for the validity of
the ET.  (Here, we 
reasonably take the typical energy scale $E$ of the $VV$ scattering  
to be $M_{VV}$ for estimating the event rates.)
Thus, the EWA and the ET 
have similar precisions in computing the
event rate from $V_LV_L$ fusion process in hadron collisions.
As $M_{VV}$ increases, they become more accurate. 
It is known that the EWA is less accurate for
sub-processes involving initial transverse gauge 
boson(s)~\cite{gunion,mike-EWA}.  Nevertheless, the EWA has
been widely used in the literature for computing event rates from 
gauge-boson (either transversely or longitudinally
polarized) fusion processes because it is easy to implement and can be
used to reasonably estimate event rates before any exact calculation is 
available. As to be shown shortly, our power counting results
agree well to the existing detailed calculations  
within about a factor of $~2$. Hence,
it is appropriate to apply the power counting analysis
together with the EWA for estimating the event rates from weak-boson 
fusions at the LHC.  
The coincidence for the case of the $q\bar{q}^{(\prime )}$-annihilation 
is even better, where the EWA is not needed.

\vspace{0.5cm}
\noindent
{\bf 7.1.} Effective-$W$ Approximation
\vspace{0.25cm}

The weak gauge boson ($V^a=W^\pm ,Z^0$) radiated from the incoming fermions
(quarks or leptons)\footnote{ ~~The similar consideration can be applied
to high energy incoming photons~\cite{EPA}.} 
~~can involve into a hard scattering (like $VV$-fusion) which can be
enhanced at high energies.
Consider the most interesting $VV$-fusion mechanism. 
The cross section of the whole scattering process $~f_1+f_2\rightarrow 
f_1' + f_2' +X~$ at the c.m. energy 
$~\sqrt{s}~$ can be factorized into the product of two parts: 
(i). the probability function $~f_{f/V_{\lambda_j}} (x)~$ for 
finding a vector boson $~V_{\lambda_j}~$ 
(with helicity $\lambda_j~(j=1,2)$) inside the incoming fermion $f_j$~; 
(ii). the hard vector boson scattering cross section 
$~\hat{\sigma}(V^a_{\lambda_1}+V^b_{\lambda_2}\rightarrow X|\hat{s})~$
at the reduced c.m. energy $~\hat{s}\equiv x\cdot s~$. So, we have
$$ 
d\sigma (f_1+f_2 \rightarrow f_1'+f_2' + X|s)
=\displaystyle\int^1_{x_{\min}}dx \sum_{\lambda_j}
{\cal P}_{f_1f_2/V^a_{\lambda_1}V^b_{\lambda_2}}(x)
d\hat{\sigma}(V^a_{\lambda_1}+V^b_{\lambda_2}\rightarrow X|\hat{s}) \, ,
\eqno(7.1)                         
$$
where  $~~{\cal P}_{f_1f_2/V^a_{\lambda_1}V^b_{\lambda_2}}(x)~~$ is
the luminosity of $~VV$-pair from the two fermion system, which is 
usually factorized/approximated as the convolution of two single 
gauge-boson-distributions:\footnote{ 
~~This factorization is very convenient, 
but is not essentially necessary. In principle, 
it can be avoided though this largely
complicates the formalism and the practical applications.}
$$
\displaystyle {\cal P}_{f_1f_2/V^a_{\lambda_1}V^b_{\lambda_2}}(x)
= \displaystyle\int^1_{x}\frac{dz}{z}f_{f_1/V^a_{\lambda_1}}(z)
f_{f_1/V^b_{\lambda_2}}(x/z )
\eqno(7.2)                         
$$  

\noindent
The above is usually called the {\it Effective-$W$ Approximation} (EWA)
\cite{effective-W}.
Comparing (5.49) and (7.1), we establish the following relation for the
two differential $VV$-luminosities:
$$
{\cal P}_{f_1f_2/V^a_{\lambda_1}V^b_{\lambda_2}}(x)
~=~s\displaystyle \frac{d{\cal L}}{dM_{VV}^2} ~~,
\eqno(7.3)
$$
where we ignored the subscript ``partons'' in (5.49) for simplicity and used
the notation $~M_{VV}^2(\equiv \hat{s})~$. 

The basic assumptions of the EWA method are: 
\begin{description}

\item[(a).] The initial {\it off-shell} gauge bosons 
$~V^a_{\lambda_1}~$ and $~V^b_{\lambda_2}~$ in the 
sub-process $~V^a_{\lambda_1}+V^b_{\lambda_2}\rightarrow X~$
 are approximated as the {\it on-shell} ones.

\item[(b).] One approximates the moving direction of the radiated
 gauge boson $~V^a_{\lambda_j}~$ in the initial state 
to be along the direction of the corresponding incoming fermion.
It is known as the {\it small angle approximation}. 

\item[(c).] One approximates the  two-gauge-boson-luminosity function
as the convolution of the two single-gauge-boson-luminosity functions
as given in the above equation.
 In the currently most popular/simple formulation, the
single-gauge-boson-luminosity functions are further simplified by the
{\it Leading Logarithm Approximation} (LLA).

\item[(d).] 
One assumes that, under proper kinematic cuts, the dominant contributions
for producing the final state $~f_1'+f_2' + X~$ come from the
$VV$-fusion sub-process process. 

\end{description}

In this work, we shall use the simplest form of the EWA method, i.e.
the LLA EWA, to estimate the production rates of signal and background
from the $VV$ fusion processes.
In this case, the probability function to be used in the above 
equation is\footnote{ 
~~For example, $g_L^a=g/\sqrt{2}$ and $g_R^a=0$ for $V^a=W^\pm~$.
}
 $$                                                                    
\begin{array}{l}
f_{f/V^a_L}=\displaystyle\frac{(g_L^a)^2+(g_R^a)^2}{8\pi^2}\frac{1-x}{x}
\\[0.4cm]
f_{f/V^a_{T}}=\displaystyle\frac{(g_L^a)^2+(g_R^a)^2}{16\pi^2}
\frac{x^2+2(1-x)}{x}\ln\frac{\hat s}{M_V^2}
\end{array}
\eqno(7.4a,b)    
$$
Here, we have averaged over the left/right-handed 
$V_T^a$ polarizations.  
We note that the longitudinal luminosity function $f_{f/V^a_L}$
does not have any logarithmic
dependence in the leading-Log limit, while the transverse luminosity
has. Hence,  the transverse luminosity function $f_{f/V^a_{T}}$ is much
less accurate than the longitudinal one
because the scale choice inside the logarithmic argument
is very uncertain. For instance, adding a non-leading term like 
$$
-\frac{x^2+2(1-x)}{x}\ln\frac{1}{x}
\eqno(7.5)   
$$
to (7.4b), we find the log-factor $~~\ln\frac{\hat s}{M_V^2}~~$ changing
into a larger one $~~\ln\frac{s}{M_V^2}~~$.
In the literature, different choices for the logarithmic argument
were suggested.\footnote{ ~~See: J. Lindfors, in Ref.~\cite{effective-W}.}
~~Also one may use different definitions for the scaling variable $~x~$.
If one defines $~x~$ as the ratio of the vector-boson energy ($E_V$) over
that of the corresponding incoming fermion ($E_f$),\footnote{ ~~See: 
S. Dawson (1985), and G.L. Kane et al (1984), in Ref.~\cite{effective-W}.}
 ~~then the relation
$~s = \hat{s}/x~$ between the $f$-$f$ system and the sub-system $V$-$V$
exactly holds only when the gauge boson is collinearly radiated (i.e., 
when the small angle approximation holds).

Substituting (7.4a,b) into (7.2)-(7.3), we get
$$
\begin{array}{ll}
\dis \frac{d{\cal L}_{LL}}{dM_{VV}^2} & =~\dis
\frac{\left[(g_{L1}^a)^2+(g_{R1}^a)^2\right]
     \left[(g_{L2}^a)^2+(g_{R2}^a)^2\right]}{(8\pi^2)^2(s\cdot x)}
\left[ (1+x)\ln\frac{1}{x}-2(1-x)\right] ~,
    \\[0.5cm]
\dis \frac{d{\cal L}_{LT}}{dM_{VV}^2} & =~\dis
\frac{\left[(g_{L1}^a)^2+(g_{R1}^a)^2\right]
     \left[(g_{L2}^a)^2+(g_{R2}^a)^2\right]}{(8\pi^2)^2(s\cdot x)}
\left[(1+x)\ln\frac{1}{x}-\frac{1}{4}(1-x)(7+x)\right]
\left[\ln\frac{s}{M_V^2}\right] ~,
     \\[0.5cm]
\dis \frac{d{\cal L}_{TT}}{dM_{VV}^2} & =~\dis    
\frac{\left[(g_{L1}^a)^2+(g_{R1}^a)^2\right]
     \left[(g_{L2}^a)^2+(g_{R2}^a)^2\right]}{(8\pi^2)^2(s\cdot x)}
\left[\left(1+\frac{x}{2}\right)^2\ln\frac{1}{x}-
     \frac{1}{2}(1-x)(3+x)\right]\left[\ln\frac{s}{M_V^2}\right]^2 ~.
\\[0.5cm]
\end{array}
\eqno(7.6)
$$
From (7.6), we plot in Fig.~3(a)-(f) 
the luminosity distribution
($~d{\cal L}/dM_{VV}^2~$)
of the polarized $~VV$-pair from the two fermion system,
as a function of the invariant mass $M_{VV}$ for 
the LHC and the LC. 

\vspace{0.5cm}
\noindent
{\bf 7.2.} At the LHC
\vspace{0.3cm}

\noindent
{\it 7.2.1. Preliminaries}
\vspace{0.2cm}

For the purpose of systematically analyzing the 
sensitivity to each bosonic operator in (2.2), 
in what follows, we separately compare the rates contributed by each 
individual operator with that by the $B$-term.
The actual experiments contain the contributions from {\it all} 
possible operators and are thus more complicated. For simplicity and
clearness, we follow the well-known naturalness 
assumption (i.e., contributions from different operators do not 
accidently cancel each other) and estimate the 
contributions from each operator separately. 

Let us denote the production rate for the scattering 
process $~W_{\alpha}^+W_{\beta}^+\rightarrow W_{\gamma}^+W_{\delta}^+~$
as $~R_{\alpha\beta\gamma\delta (\ell)}~$,
where $\alpha ,\beta ,\gamma ,\delta =L,T$ label the polarizations
of the $W$-bosons 
and $~\ell=0,1$ indicates contributions from
leading order and next-to-leading order, respectively. 
Up to the one-loop level, we define
$$
\begin{array}{lll}
R_{\alpha\beta\gamma\delta}
& = & R_{\alpha\beta\gamma\delta (0)} +R_{\alpha\beta\gamma\delta (1)} ~~,
\\[0.2cm]
 R_{\alpha\beta\gamma\delta (\pm )} 
& = & R_{\alpha\beta\gamma\delta (0)} \pm 
|R_{\alpha\beta\gamma\delta (1)}| ~~.
\end{array}
\eqno(7.7a,b)
$$
Also, $~R_B~$ denotes the rate contributed by the largest $B$-term.
For convenience, we use the subscript ``$_S$'' to stand for 
summing up the polarizations of the corresponding gauge boson. For the
$~q\bar{q}^{(\prime)}\rightarrow V_\alpha V_\beta~$ processes, the rates are 
denoted by $~R_{\alpha\beta(\ell )}~$, and correspondingly we define
$~R_{\alpha\beta}=R_{\alpha\beta(0)}+R_{\alpha\beta(1)}~$ and
$~R_{\alpha\beta (\pm )}=R_{\alpha\beta(0)}\pm |R_{\alpha\beta(1)}|~$.
We also note that when applying the power counting analysis,
we have ignored the angular dependence in the scattering 
amplitudes (cf. Tables~I$\sim$II)
because it does not affect the order of magnitude estimates
for the total cross sections (or the event rates). 

To check the reliability of our power counting method,
we have compared our numerical results for the 
$~W^+W^+\rightarrow W^+_LW^+_L$ fusion and 
the $q\bar{q}\rightarrow W^+_LW^-_L$ annihilation with 
those in Fig.~8 and Fig.~5 of Ref.~\cite{BDV} 
in which the above constituent amplitudes were explicitly calculated and
the polarizations of the initial weak-bosons were summed over 
for the fusion process.
As shown in Figs.~4a and 4b, both results coincide
well within about a factor of $2$ or better.
These are two typical examples showing that the correct physical picture
can be consistently and quickly grasped by our power counting analysis. 

We conclude this subsection by briefly commenting on  
a recent paper [by A.~Dobado et al, {\it Z.~Phys.}~C17~(1996)~965]. 
The sole purpose of that paper 
was to avoid using the ET, but still within the EWA, 
to increase the calculation precision and to extend the results 
to lower energy regions.
This approach is, however, inconsistent  because the validity of
the EWA also requires the {\it same high energy condition
$~E\gg M_W~$} as that of the ET.
To study  the operators ${\cal L}_{4,5}$ (dominated by pure $V_L$-modes),
the approach in that paper cannot really get 
higher precision than previous studies~\cite{BDV,wwlhc} 
(using the ET combined with the EWA) except 
making unnecessary complications and confusions in their calculation.
To study other operators like ${\cal L}_{1,2,3,8,9,11}$, 
the $V_T$-modes must be included, for which the EWA is much 
worse~\cite{gunion,mike-EWA}. 
Hence, to get the {\it consistent and precise results}
for these operators, one must go beyond the EWA for full calculations.
This was not contained in the paper of A.~Dobado et al. On the contrary,
our present global power counting analysis
provides the first complete and consistent estimate for all 
NLO operators in the EWCL (2.2).

\vspace{0.4cm}
\noindent
{\it 7.2.2. Analyzing the Model-Independent Contributions 
          to the Event Rates}
\vspace{0.2cm}

In Fig.~5a we give the power counting estimates 
for the production rates
of the $W^+_L W^+_L$ pairs at the LHC 
from the initial state $W$-bosons with different polarizations.
In this figure, setting the renormalized coefficients 
$~\ell_{0\sim 14}$ to be zero, we include only the model-independent 
contributions up to one-loop.\footnote{ ~~It is understood 
that the divergent pieces 
from one-loop calculations have been absorbed by the
coefficients of the corresponding 
NLO effective operators~\cite{app,georgi}.}~~
As clearly shown in Fig.~5a, the rate from $4W_L$ scattering
dominates over the rate from $W_T+3W_L$ scattering. The latter
is lower by about
an order of magnitude for large $M_{WW}$ in spite of the fact that
the $W_T W_L$ luminosity is larger than the $W_LW_L$
luminosity in the initial state.
Also separately shown in the same figure is the 
event rate $~|R_B|~$ contributed by the 
largest $B$-term [cf. (5.17) and (5.45)]
which is even significantly lower than that from the $W_T+3W_L$ scattering
by a factor of ~$2\sim 7$~ for $~M_{WW}>500$\,GeV.
However, the rate from $W_TW_T$ initial state is lower than 
that from the $B$-term in the $4W_L$ amplitude as $~E\geq 600$\,GeV.
This implies that if the contribution from $W_TW_T$ initial state
is to be included in calculating the total production rate
of the  $W_LW_L$ pair,
the contribution from the $B$-term in the $4W_L$ amplitude 
should also be included because
they are of the same order in magnitude. 
If, however, only the pure Goldstone boson amplitude
$~T[\pi^+ \pi^+ \rightarrow \pi^+ \pi^+]~$
is used to calculate the $4W_L$-amplitude 
(with the $B$-term ignored) 
the contribution from $~T[W^+_T W^+_T \rightarrow W_L^+ W_L^+]~$
should also be consistently ignored for computing the 
total rate of $W^+_L W^+_L$ pair production 
via the weak-boson fusion mechanism.

As shown in Ref.~\cite{wwlhc}, it is possible to statistically,
though not on the event-by-event basis,
choose event with longitudinally polarized 
$W$-bosons in the initial state 
by applying the techniques of 
forward-jet tagging and central-jet vetoing.
In this work we do not intend to study the details of the 
event kinematics, and 
we shall sum over all the initial state polarizations 
for the rest of discussions. 
Let us first compare the rates for different polarizations in the final 
state. Fig.~5b
shows that the rate of $W_LW_L$ final state dominates, while the
rate of $B$-term and the
rates of $W_LW_T$ and $W_TW_T$ final states are of the same order, 
and all of them are about an $O(10)$ to $O(10^2)$  
lower than the rate of $W_LW_L$ final state in the energy region 
$~E=M_{WW} > 500\,{\rm GeV}$. 
Therefore, if one wants to increase the precision
in calculating the total event rates by 
including the small contribution from 
the $B$-term in pure $4W^+_L$ scattering,
then the contributions from $W^+_SW^+_S\rightarrow W^+_TW^+_T$
and $W^+_SW^+_S\rightarrow W^+_LW^+_T$ scatterings should also be 
consistently included. Otherwise, they must be neglected all together.
{}From Figs.~5a and 5b, we conclude that the scattering process
$W^+_L W^+_L \rightarrow W^+_L W^+_L$ dominates the $W^+W^+$-pair
productions when the model-dependent coefficients $~\ell_{0\sim 14}$ 
in (2.2) are set to be zero.

Similarly, we apply the power counting analysis to 
estimate the model-independent contribution to the rates of 
polarized $W^+Z^0$ pair produced from 
$q {\bar{q}}'$ fusion up to one-loop order. 
 The results are plotted in Fig.~6. 
It shows that the $~W_T^+Z^0_T~$ and $~W_L^+Z^0_L~$ rates (i.e.,
$R_{TT}$ and $R_{LL}$) are of the same order. They are much larger
than the rate $~R_{LT}~$ (from the $~W^+_LZ^0_T~$ and $~W^+_TZ^0_L~$
final states) and the rate $~|R_B|~$ (from the largest $B$-term
contained in the $~q\bar{q}'\rightarrow W^+_LZ^0_T,W^+_TZ^0_L~$
amplitudes). The rate $~R_{LT}~$ is only slightly above the $~|R_B|~$ 
because the leading GB-amplitude $~T_0[q\bar{q}';\pi^+ Z_T/\pi^0 W_T^+]~$ 
is of the same order as
the term $~B^{(1)}_0~$ (cf. Table~IIIa). Here we see that for the
lowest order total signal rate, both $~R_{TT}~$ and $~R_{LL}~$ have to be
included since they are of the same order and larger than the
NLO model-dependent contributions. If one wants
to further include $~R_{LT}~$, then $~R_B~$ should also be included.

\vspace{0.4cm}
\noindent
{\it 7.2.3. Estimating Sensitivities for Probing 
          the Model-Dependent Operators}
\vspace{0.2cm}

In this section, we classify the sensitivities to all the 
NLO bosonic operators at the LHC. Without knowing the values of the 
{\it model-dependent coefficients} ($\ell_n$~'s), we shall take them to
vary from $O(1)$ to $O(10)$  except that $~\ell_0,
~\ell_1$ and $\ell_8$ are bounded to be of $O(1)$ 
by the low energy data [cf. (3.5)].

We first consider the scattering process $W^+W^+ \rightarrow W^+W^+$.
Our theoretical criterion for discriminating different sensitivity levels
({\it sensitive, marginally sensitive}, or {\it insensitive})
to probe a particular operator via the production of $W^+W^+$ pairs
is to compare its contribution to the event rate 
( $|R_{\alpha\beta\gamma\delta (1)}|$ ) with that from the
largest model-independent contribution of the LNI $B$-term ( $|R_B|$ ),
according to the Sec.5. 
In Figs.~7-10, we show the results
for varying $~|\ell_n|~$ from $~O(1)$ to $~O(10)$ (except $\ell_0,
~\ell_1$ and $\ell_8$). Here, the polarizations of the initial and the final 
states have been summed over. 
In Figs.~7a and 7b, we consider the coefficients ( $\ell_n$'s ) 
to be naturally of ~$O(1)$~ according to 
the naive dimensional analysis~\cite{georgi}.
Fig.~7a shows that the event rates/($100$\,${\rm fb}^{-1}$GeV) from
operators $~{\cal L}_{4,5}~$ are larger than that from the $B$-term 
when $~E=M_{WW} >600~$GeV, while the rates from operators
$~{\cal L}_{3,9,11;12}~$ can exceed $~|R_B|~$ only if $~E=M_{WW}>860$~GeV.
As $M_{WW}$ increases, the 
rates contributed by $~{\cal L}_{4,5}~$ remain flat,
while the rates by $~{\cal L}_{3,9,11;12}~$  and the $B$-term decrease.
The ratio of the event rates from $~{\cal L}_{4,5}~$ 
to $~|R_B|~$ is $~5.0$ at $~E=M_{WW}=1$\,TeV, and rapidly increases 
to $~19.6$ at $~E=M_{WW}=1.5~$\,TeV.
In contrast, the ratio between the rates from
$~{\cal L}_{3,9,11;12}~$  and the $B$-term 
only varies from $~1.4$ to $3.0$ for $~E=M_{WW}=1\sim 1.5$\,TeV.
Fig.~7b shows that for the coefficients of $~O(1)~$, the event rates
contributed by operators $~{\cal L}^{(2)\prime}~$ and 
$~{\cal L}_{1,2,8;13,14}~$  are all below $~|R_B|~$ for a wide 
region of energy up to about $~2$\,TeV, 
so that they cannot be sensitively probed in this case. 
Especially, the contributions from
$~{\cal L}_{1,13}~$  are about two orders of magnitude 
lower than that from the $B$-term. 
This suggests that $~{\cal L}_{1,13}~$   must be tested via
other processes~\cite{global2}.
In Figs.~8a and 8b, different event rates are compared for 
the coefficients (except $\ell_0,~ \ell_1,~\ell_8$) to be of $O(10)$. 
Fig.~8a shows that 
the rates from $~{\cal L}_{3,9,11;12}~$ could significantly dominate 
over $~|R_B|~$ by an order of magnitude for 
$E=M_{WW}\sim 1$\,TeV
if their coefficients are increased by a factor
of $~10$ relative to the natural size of $~O(1)~$.
Fig.~8b shows that the rates from 
$~{\cal L}_{13}~$ is still lower than $~|R_B|~$ by about an order of 
magnitude,
while the rate from $~{\cal L}_{2}~$ agrees with
 $~|R_B|~$  within a factor of $~2$. 
The contribution from $~{\cal L}_{14}~$ exceeds 
$~|R_B|~$ by about a factor $2\sim 3$ at $~E=M_{WW}=1$\,TeV and a
factor of $3\sim 5$ at $~E=M_{WW}=1.5$\,TeV when its coefficients is of 
$~O(10)~$.

As discussed above, the cutoff scale $\Lambda$ can be lower than
$~\Lambda_0=4 \pi f_\pi \simeq 3.1~$TeV if there is any new 
heavy resonance below $\Lambda_0$.
In that case, the signal rates $|R_1|$ will be higher than 
that reported in Figs.~7 and 8 by about a factor of 
$~\left(\frac{\Lambda_0}{\Lambda}\right)^2~$.
For comparison, we repeat the above calculations for 
$\Lambda=2~$TeV in Figs.~9 and 10.
As indicated in Fig.~9a and 10b, the sensitivities to probing 
$~\ell_{3,9,11,12}\sim O(1)~$ and $~\ell_{14,2}\sim O(10)~$ 
via $W^+W^+ \rightarrow W^+W^+$ 
process increase for a lower  ~$\Lambda$~ value.

{}From the above analyses, we conclude that studying the 
$~W^+W^+\rightarrow W^+W^+~$ process can sensitively probe
the operators $~{\cal L}_{4,5}~$, but is only marginally sensitive for
probing $~{\cal L}_{3,9,11;12}~$ and insensitive for 
$~{\cal L}^{(2)\prime}~$ and $~{\cal L}_{1,2,8;13,14}~$, if their 
coefficients are naturally of $~O(1)~$. 
In the case where these coefficients are of $~O(10)~$, the probe of 
$~{\cal L}_{14}~$ (for lower $\Lambda$) and $~{\cal L}_{3,9,11;12}~$
could become sensitive and that of $~{\cal L}_{2}~$ (for lower $\Lambda$)
could become marginally sensitive, while 
 $~{\cal L}_{13}~$ still cannot be sensitively or marginally sensitively
measured.

Moreover, we note that the operators $~{\cal L}_{6,7,10}~$, 
which violate the custodial $SU(2)_C$ symmetry,
do not contribute to the $W^+W^+$ pair productions up to
$~O(1/\Lambda^2)~$.
They can however contribute to the other scattering channels such as
$~WZ\rightarrow WZ~$, $~WW\rightarrow ZZ~$,
$~ZZ\rightarrow WW~$ and $~ZZ\rightarrow ZZ~$,
cf. Table~III. (Here, $~\,{\cal L}_{10}$ only contributes 
to $~ZZ\rightarrow ZZ~$ channel.)
By our order of magnitude estimates,
we conclude that they will give the similar kind of contributions
to the $WZ$ or $ZZ$ channel as
$~{\cal L}_{4,5}~$ give to the $W^+W^+$ channel.
This is because all these operators
contain four covariant derivatives [cf. (3.1.2)]
and thus become dominant in the high energy $VV$-fusion processes. 

Let us consider the $W^-W^- \rightarrow W^-W^-$ production process.
At the LHC, in the TeV region,
the luminosity of $W^-W^-$ is 
typically smaller than that of $W^+W^+$ by a factor of $3 \sim 5$.
This is because in the TeV region, where the fraction
of momentum ($x$) of proton carried by the quark (which 
emitting the initial state $W$-boson) is large 
(for $x=\frac{E}{\sqrt{S}}\sim 0.1$), the parton luminosity is
dominated by the valence quark contributions.
Since in the large-$x$ region, the probability of finding a down-type 
valence quark in the proton is smaller than finding an up-type 
valence quark, 
the luminosity of $W^-W^-$ is smaller than that of $W^+W^+$.
However, as long as there are enough $W^-W^-$ pairs detected,
which requires a large integrated luminosity of the machine and a 
high detection efficiency of the detector, conclusion similar to probing 
the effective operators for the $W^+W^+$ 
channel can also be drawn for this channel.
For ~$M_{WW} > 1.5 $~TeV, the $W^-W^-$ production rate becomes 
about an order of magnitude smaller than the $W^+W^+$ rate for any given 
operator. Thus, this process will not be sensitive to probing
the NLO operators when ~$M_{WW} > 1.5 $~TeV.

Next, we examine the $W^+Z^0$ production
 rates in the $~q\bar{q}^{\prime}\rightarrow W^+ Z^0~$
channel. As shown in Figs.~11, for
~$\Lambda=3.1~$TeV, when the coefficients are of
~$O(1)~$, the probe of ~${\cal L}_{3,11,12}~$ is sensitive 
when $~E>750~$GeV, while that of ~${\cal L}_{8,9,14}~$ is 
marginally sensitive if ~$E>950$~GeV. 
The probe of ~${\cal L}^{(2)\prime}$~ 
becomes marginally sensitive if ~$E>1.4$~TeV, and
that of ${\cal L}_{1,2,13}$ is insensitive for ~$E<1.9~$TeV. When the
coefficients other than $~\ell_0,~\ell_1,~\ell_8~$ are of ~$O(10)$~, 
~${\cal L}_{3,11,12,9,14}$~ 
could all be sensitively probed when $~E>500~$GeV, 
and the probe of ~${\cal L}_{2,13}$~ could be sensitive when ~$E>1.2~$TeV. 
For ~$\Lambda=2~$TeV, the sensitivities are increased overall by a
factor $~\left(\frac{\Lambda_0}{\Lambda}\right)^2\simeq 2.4~$,
as shown in Fig.~12.   The event rate for
$~q\bar{q}^{\prime}\rightarrow W^- Z^0~$ is slightly lower than that of
$~q\bar{q}^{\prime}\rightarrow W^+ Z^0~$ by 
only about a factor of $1.5$ due
to the lower luminosity for producing $W^-$ bosons in 
${\rm p}{\rm p}$ collisions. Hence, the 
above conclusion also holds for the 
$~q\bar{q}^{\prime}\rightarrow W^- Z^0~$ process.

We note that the $~q\bar{q}^{\prime}\rightarrow W^\pm Z^0~$ annihilation
provides {\it complementary} information on probing 
the EWSB sector, in comparison with 
the ~$W^\pm W^\pm\rightarrow W^\pm W^\pm$~ fusion. 
In the former, $~{\cal L}_{3,11,12}$~ can be 
unambiguously probed, while in the latter, 
~${\cal L}_{4,5}~$ can be sensitively probed.
Furthermore, as shown in Table~V and VI, the operators
$~{\cal L}_{6,7}~$ can be probed from either 
~$T_1[2 W_L, 2 Z_L]~$ or $~T_1[4 Z_L]$~, and ~${\cal L}_{10}$~ can 
only be probed from ~$T_1[4 Z_L]$~, while ~${\cal L}_{2,9}$~
can be tested from $~T_1[q\bar{q};W^+_LW^-_L]~$.
It is therefore necessary and useful to measure all the 
the gauge boson fusion and quark-anti-quark annihilation
processes for completely exploring the EWSB sector.    \\

\vspace{0.4cm}
\noindent
{\bf 7.3.} At the LCs
\vspace{0.2cm}

At the LCs, by our global power counting
analysis, we only need to take into account the corresponding collider
energy and its integrated luminosity. All analytic power counting formula for
the scattering amplitudes of the $V$-$V$ fusion and the
$f$-$\bar{f}^{(\prime )}$
annihilation processes (given in Sec.~6) are still the same.
For the 14\,TeV LHC  and the 1.5~TeV $e^-e^+$ LC, 
the c.m. energy of the $V$-$V$ fusion and the $f$-$\bar{f}^{(\prime )}$ 
annihilation processes is typically of $O(1)$~TeV. 
Hence, the different sensitivities of the LHC and  the LCs to probing
the complete set of the NLO operators can be easily obtained by comparing
their $VV$-pair luminosities at the TeV scale, which
have been shown in Fig.~3. 

Needless to say, the background processes to the $VV$ fusion signal event
at the LHC and the LCs can be very different. Hence, it may require 
different techniques to suppress the backgrounds and enhance
the ratio of signal to background to detect the strongly
coupled EWSB sector and measure all these model-dependent
NLO operators.
Although we do not intend to perform such a detailed Monte Carlo study 
in this work to illustrate how to extract out the signal rate, 
we shall briefly discuss various techniques useful for studying
the $VV \ra VV$ processes. 
This is the subject of the next section. (For more elaborate numerical
analyses for LCs, see Refs.~\cite{lcllll,desy}.)

\vspace{0.7cm}
\noindent
{\bf 8. How to Study TeV $\WWWW$ Interactions Experimentally}
\vspace{0.3cm}

In this section, we discuss how to experimentally 
study the EWSB sector by observing $\WWWW$ interactions in the TeV region. 
(Here, $W$ symbolizes either $W^\pm$ or $Z^0$ boson unless specified
otherwise.) 
We analyze some general features of the event structure in the 
signals and backgrounds. Various techniques 
to enhance the signal-to-background ratio are also presented. 

For completeness, we shall also discuss in this section  
the experimental techniques for detecting heavy resonance(s) 
(with mass of the order of TeV) predicted 
in some models of strongly coupled EWSB sector, although 
in the rest part of this article we have discarded such models.
Since for models with light resonance(s) in the symmetry breaking sector,
the interactions among the Goldstone bosons in the TeV region 
cannot become strong,\footnote{
 ~~The potentially bad high energy behavior of 
the scattering matrix element (if without any light resonance)
is cut off by the tail of the light resonance.
}  ~~we shall not consider that class of models here.

\vspace{0.5cm}
\noindent
{\bf 8.1.} Signal 
\vspace{0.3cm}

The event signature of the signal is mainly 
a longitudinal $W$-pair produced in the final state. 
Assuming no light resonance(s) \cite{light}, 
the electroweak symmetry breaking sector in the TeV region
may either contain a scalar- or vector-resonance, \etc, 
or no resonance at all.
For a model with a TeV scalar (spin-0,isospin-0) resonance, 
the most useful detection
modes  are the $W^+W^-$ and $Z^0Z^0$ modes
which contain large isospin-0 channel contributions.
For a model with a TeV vector (spin-1,isospin-1) resonance, the most
useful mode is the $W^\pm Z^0$ mode because it contains 
a large isospin-1 channel contribution.
If there is no resonance present in the symmetry breaking sector, all the 
$WW$ modes are equally important, so the $W^\pm W^\pm$ mode is also 
useful. Actually, because of the small SM backgrounds for the 
$W^\pm W^\pm$ mode, this can become the most important 
detection mode if no TeV resonance is present in the EWSB sector. 

Before we discuss the backgrounds, we have to specify the decay mode of 
the $W$-bosons in the final state. 
Let's first concentrate on the cleanest final state, i.e. the pure
leptonic decay mode. 
The branching ratios of $W^+ \ra \ell^+ \nu$ and 
$Z^0 \ra \ell^+\ell^-$ are 2/9 and 0.06, respectively, for
$\ell^+=e^+$ or $\mu^+$.
If the $Z^0Z^0$ pair signal is large enough, the 
$Z^0(\ra \ell^+\ell^-)Z^0(\ra \ell^+\ell^-)$
and $Z^0(\ra \ell^+\ell^-)Z^0(\ra \nu \bar \nu)$ modes would be most
useful at hadron colliders \cite{ZZllnn}.
Otherwise, it is also necessary to include the 
$W^\pm W^\mp$, $W^\pm Z^0$ and $W^\pm W^\pm$ modes \cite{llll}.
Although the pure leptonic mode gives the cleanest signal signature,
its event rate is small 
because of the relatively small branching ratio.
To improve the signal event rate for discriminating models of EWSB, 
one should also study the other decay modes, such as the lepton plus jet 
modes at the LHC and the LC, or the pure jet mode at the LC \cite{lcllll}.
(Because of the large QCD background rate, the pure jet mode will be 
extremely difficult to utilize at hadron colliders.)

\vspace{0.5cm}
\noindent
{\bf 8.2.}  Backgrounds 
\vspace{0.3cm}

For each decay mode of the $WW$ pair, the relevant backgrounds vary. But, 
in general, one of the dominant background processes is the {\it intrinsic}
electroweak background, which contains the same final state 
as the signal event.
This background rate in the TeV region can be generated by calculating
the Standard Model production rate of $f {\bar f} \ra f {\bar f} WW$ with a
light (e.g., 100\,GeV) SM Higgs boson \cite{llll}.
For example, the $W_LW_L$ signal rate in the TeV region from a 
1 TeV Higgs boson is equal to the difference between the event rates 
calculated using a 1 TeV Higgs boson and a 100 GeV Higgs boson \cite{llll}.

The other important backgrounds are:
the electroweak-QCD process $W+\,{\rm jets}$ 
(which contains a ``fake $W$'' mimicked by two QCD jets),  
and the $t \bar t$ pair (which subsequently decays to a $WW$ pair),
{\it etc}.
We now discuss these backgrounds in various $WW$ decay modes.
Without the loss of generality, in the rest of this article
we shall only consider the major background processes at the hadron 
collider LHC.
The backgrounds at the Linear Collider (LC) 
are usually easier to deal with 
because the initial state does not involve 
strong QCD interactions.

\vspace{0.4cm}
\noindent
{\it 8.2.1. $Z^0(\ra \ell^+\ell^-)Z^0(\ra \ell^+\ell^-)$
and $Z^0(\ra \ell^+\ell^-)Z^0(\ra \nu \bar \nu)$ modes }
\vspace{0.2cm}

The signature for the signal in this mode is either an event with 
four isolated leptons with high transverse momenta, 
or two isolated leptons associated with large missing transverse momentum
in the event.
The dominant background processes for this mode are
$q \bar q \ra Z^0 Z^0 X$, $g g \ra Z^0 Z^0 X$ \cite{llll},
where $X$ can be additional QCD jet(s).
The final state $Z^0$ pairs produced from the above processes 
tend to be transversely polarized.
Similarly, the $Z^0$ pairs produced from the 
{\it intrinsic} electroweak background process are also mostly 
transversely polarized.
This is because the coupling of a transverse $W$ boson 
to a light fermion (either quarks or leptons) is stronger than that of 
a longitudinal $W$ in the high energy region.
Hence, to discriminate the above backgrounds
from the signals, we have to study the polarization 
of the final state $W$ boson.
For the same reason, the gauge boson emitted from the initial state fermions
are likely to be transversely polarized.
To improve the ratio of signal to background rates, 
some kinematic cuts (to be discussed in the next section)
can be used to enhance the event sample in which the $W$ bosons
emitted from the incoming fermions are mostly longitudinally
polarized, and therefore can enhance the $\WWWW$ signal event.

Since it is easy to detect $Z^0 \ra \ell^+\ell^-$ with a good accuracy,
we do not expect backgrounds other than those discussed above to be large. 
Similarly, because of the  large
missing transverse momentum of the signal event,
the $Z^0 \ra \nu \bar \nu$ signature is difficult
to mimic by the other SM background processes.
 
\vspace{0.4cm}
\noindent
{\it 8.2.2. $W^+(\ra \ell^+\nu) W^-(\ra \ell^- \bar \nu)$ mode }
\vspace{0.2cm}

For this mode, in addition to the background processes
$q \bar q \ra q {\bar q} W^+W^-$, $q \bar q \ra W^+W^- X$ and
$g g \ra W^+W^- X$, the $t \bar t + \, {\rm jet}$ process
can also mimic the signal event because the 
final state top quark pair can decay into 
a $W^+W^-$ pair for the heavy 
top quark \cite{llll}.

\vspace{0.4cm}
\noindent
{\it 8.2.3. $W^\pm (\ra \ell^\pm \nu) Z^0(\ra \ell^+ \ell^-)$   
and $W^\pm(\ra \ell^\pm \nu) W^\pm (\ra \ell^\pm \nu)$ modes}
\vspace{0.2cm}

Besides the background processes similar to those discussed
above, the $Z^0 t \bar t$
event can also mimic the $W^\pm Z^0$ signal.

For the purely leptonic decay 
mode of $W^\pm W^\pm$ \cite{wpwpll}, the signature is two like-sign 
isolated leptons with high $P_T$ and large $\ETslash$.
There are no low-order backgrounds from quark-antiquark or gluon-gluon
fusion processes. However, other backgrounds can be important, such as
the production of the transversely polarized 
$W$-pairs from the {\it intrinsic} electroweak background 
process \cite{wpwpew} or from the QCD-gluon exchange 
process \cite{gluonex}, 
and the $W^\pm t \bar t$ production from the electroweak-QCD process
\cite{llll}.

\vspace{0.4cm}
\noindent
{\it 8.2.4. $\WPWM$ mode }
\vspace{0.2cm}

The dominant background processes for this mode are
$q \bar q \ra q {\bar q} W^+W^-$, $q \bar q \ra W^+W^- X$, 
and $g g \ra W^+W^- X$ \cite{qqww,ggww,wpwmew,wwjet}.
The signature for the signal in this mode is an 
isolated lepton with high transverse momentum $P_T$, 
a large missing transverse energy $\ETslash$, 
and two jets whose invariant mass is about the mass of the $W$-boson. 
The electroweak-QCD process $W^++ \, {\rm jets}$ 
 can mimic the signal when the invariant mass of the two 
QCD jets is around $M_W$ \cite{wtwoj,wthreej}.
Other potential background processes for this mode are the QCD processes
$ q \bar q, \, gg \ra t \bar t X $, $W t \bar b$ and 
$t {\bar t} + \, {\rm jet(s)}$ 
\cite{ttbaro,ttbarqcd,ttbar,wt,ttbarjet}, in which
a $W$ boson can come from the decay of $t$ or $\bar t$.

\vspace{0.4cm}
\noindent
{\it 8.2.5. $\WPZ$ mode }
\vspace{0.2cm}

The signature of the signal in this mode 
is an isolated lepton with high $P_T$, 
a large missing transverse energy $\ETslash$, 
and a two jet invariant mass around $M_Z$. 
The dominant background processes for this mode are similar to those for the 
$\WPWM$ mode discussed above. They are $ q_1 \bar q_2 \ra W^+ Z^0$,
$W^+Z^0 + \, {\rm jet(s)}$, $W^+ + \, {\rm jets}$ 
and $Z t \bar t$ production processes \cite{wwjet,wtwoj,wthreej,qqwz,zttbar}.

To separate this signal from $\WPWM$ a good jet energy resolution is needed
to distinguish the invariant mass of the two jets between $M_Z$ and $M_W$,
which differ by about 10 GeV.  Another technique to distinguish these
two kinds of events is to measure the average electric charge of the jets,
which has been applied successfully at LEP experiments \cite{aleph}.

As noted above, because of the large branching ratio,
the pure jet mode from the $W$ boson decay can also be useful
at the future lepton colliders \cite{lcllll}, where the dominant background 
for the detection of the $\WWWW$ signal event
is again the {\it intrinsic} electroweak process.

In general, without imposing any kinematic cuts,
the raw event rate of the signal is significantly
smaller than that of the backgrounds. 
However, the signature of the signal can actually be distinguished from that 
of the backgrounds so that some kinematic cuts can be applied
to suppress the backgrounds and enhance the signal-to-background ratio.
We shall examine the characteristic differences between 
the event structures of the signal and the backgrounds in the next section.

\vspace{0.5cm}
\noindent
{\bf 8.3.} How to Distinguish Signal from Background 
\vspace{0.3cm}

The signature of the signal event can be 
distinguished from that of the background events in 
many ways. We first discuss differences in the global features of the 
signal and the background events, then point out 
some distinct kinematics of the signal events.
To simplify our discussion, we shall only concentrate on the $\WPWM$ mode
in this section.

\vspace{0.4cm}
\noindent
{\it 8.3.1. Global Features }
\vspace{0.2cm}

The signal of interest is the $WW$ pair produced from the $W$-fusion
process. The spectator quark jet that emitted the
$W$-boson in the $W$-fusion 
process tends to go into the 
high rapidity region. This jet typically has a high 
energy, about 1 TeV, for $\MWW \sim 1$ TeV  ($\MWW$ is the invariant mass 
of the $WW$ pair.) Therefore, one can tag this forward jet to suppress 
backgrounds \cite{wwpt,tag,llll}.
As noted in the previous section, the $W$ boson pairs produced from the 
{\it intrinsic} electroweak process 
$ q \bar q \ra q \bar q W^+ W^-$ tend to be transversely polarized,
and the initial state gauge bosons are likely to be transversely 
polarized as well. 
To see how the forward jet can be used to discriminate the signal from 
the background events, we consider
the $W^+W^-$ fusion process as an example.
Since the coupling of the $W^\pm$ boson and the incoming quark is
purely left-handed, the out-going quark tends to go along with
the incoming quark direction when emitting a longitudinal 
$W$ boson, and opposite direction when emitting a transverse (left-handed) 
$W$. This can be easily understood from the helicity conservation. 
Hence, in the {\it intrinsic} background event, the out-going quark jet
is less forward (and less energetic) than that in the
signal event.

Furthermore, because the production mechanism of 
the signal event is purely electroweak, 
the charged particle multiplicity of the signal event is smaller than 
that of a typical electroweak-QCD process such as 
$q \bar q \ra g W^+W^-(\ra q_1 \bar q_2)$ 
or $qg \ra q W^+ q_1 \bar q_2$.
Because of the small hadronic activity in the signal event, in the central 
rapidity region there will be fewer hard QCD jets produced. 
At the parton level, they are the two quark jets produced from the $W$-boson 
decay plus soft gluon radiation.
 However, for the background process, such as $t \bar t$ production, 
there will be more than two hard jets 
in the central rapidity region both because 
of the additional jets from the decay of $t$ and $\bar t$ 
and because of the stronger hadronic 
activity from the effect of QCD color structure 
of the event.
Therefore, one can reject events with more than two 
hard jets produced in the central rapidity
region to suppress the backgrounds. This was first suggested in 
Ref.~\cite{ttbar} using a hadron level analysis to show how the $t \bar t$ 
background can be suppressed.

A similar trick of vetoing extra jets in the central rapidity region
was also applied at the parton level \cite{veto,llll}
for studying the pure leptonic decay mode of $W$'s.
An equivalent way of making use of the global 
difference in the hadronic activity of the events is to apply cuts on the 
number of charged particles. This was first pointed out in 
Refs.~\cite{multi} and \cite{kane}.
The same idea was later packaged in the context of 
selecting events with ``rapidity gap'' to enhance
the signal-to-background ratio \cite{gap}.

In the $W$-fusion process, the typical transverse momentum
of the final state $W$-pair is about $M_W/2$ \cite{wwpt}.
However, in the TeV region, the $P_T$ of the $W$-pair produced from the 
background process, such as $q \bar q \ra gWW$, can be 
of a few hundred GeV.
Therefore, the two $W$'s (either both real or one real and one fake)
produced in the background event are less back-to-back
in the transverse plane than those in the signal event.

\vspace{0.4cm}
\noindent
{\it 8.3.2. Isolated Lepton in $W^+ \ra \ell^+ \nu$}
\vspace{0.2cm}

Because the background event typically has more hadronic activity in the 
central rapidity region, 
the lepton produced from the $W$-boson decay is 
usually less isolated than that in the signal event.
 Therefore, requiring an isolated 
lepton with high $P_T$ is a useful method to suppress
the backgrounds. This requirement together with 
large missing transverse energy in the event 
insures the presence of a $W$-boson.
Finally, it is also important to be able to measure 
the sign of the lepton charge to reduce the backgrounds for the
detection of the $\WPWP$ mode \cite{llll}

\vspace{0.4cm}
\noindent
{\it 8.3.3. $W \ra q_1 \bar q_2$ decay mode}
\vspace{0.2cm}

To identify the signal, we have to reconstruct the
two highest $P_T$ jets in the 
central rapidity region to obtain the invariant mass of the
$W$-boson. It has 
been shown \cite{kane}
 that an efficient way of finding these two jets is to first 
find a big cone jet with invariant mass around $M_W$, then demand that there 
are two jets with smaller cone size inside this big cone jet. 
Because we must measure any new activity in $\WWWW$, and
because the 
$W$-boson in the background event is mainly 
transversely polarized \cite{kane}, \footnote{
 ~~The same conclusion also holds for the QCD background event with 
``fake $W$'', which usually consists of one hard jet and one soft jet.
Hence, after boosting these two jets back into the rest frame of the
``fake $W$'', its angular distribution resembles that from a 
transverse $W$ boson.}
~~one must measure the fraction of longitudinally 
polarized $W$-bosons in the $WW$ pair data sample and compare with that 
predicted by various models of EWSB sector.
                    
It was shown in Ref. \cite{toppol} that a large fraction
($\sim 65\%$ for a 175 GeV top quark)
of the $W$ bosons from the top quark decays is longitudinally 
polarized.\footnote{
 ~~The ratio of the longitudinal versus the transverse $W$'s from top
quark decays is about $m_t^2/(2 M_W^2)$.}
~~This can in principle complicate the above method of using the 
fraction of longitudinally $W$ bosons to detect the signal.
Fortunately, after imposing the global cuts such as vetoing the central 
jet and tagging the forward jet, the $t \bar t$ backgrounds are small.
(If necessary, it can be further suppressed by vetoing event with $b$ jet,
because for every background event with top quark there is always a $b$ quark
from the SM top decay.)
To suppress the $W^+ t \bar b$ and $W^- {\bar t} b$ backgrounds 
\cite{wt}, which have smaller raw production rate than the $t \bar t$ event, 
the same tricks can be used. Furthermore, the top quark produced
in the $W^+ t \bar b$ event is mostly left-handed polarized
because the coupling of $t$-$b$-$W$ is purely left-handed in the 
SM. The kinematic cut of vetoing events with additional 
jets in the central rapidity region will reduce the fraction
of the events in which the $W$ boson from the top 
quark (with energy of the order 1 TeV) decay is longitudinally polarized.
This is because in the rest frame of
a left-handedly polarized top quark, 
which decays into a longitudinal $W$-boson, the decay
$b$-quark prefers to move along the moving direction of the top quark 
in the center-of-mass frame of the $W^+ t$ pair. Hence, such a
background event will produce an additional hard jet in the central 
rapidity region \cite{toppol}.

In the next section, we show how to observe the signals predicted by various 
models of the
symmetry breaking sector. Some of them were studied at the hadron 
level, some at the parton level. We shall  not reproduce those 
analyses but only sketch the ideas of various techniques used in detecting  
$\WWWW$ interactions. The procedures discussed here are not necessarily
the ones used in the analyses previously performed in the literature.
If the signal event rates are large enough to observe the purely
leptonic mode, then studying the symmetry breaking sector at the LHC 
shall be possible. However, a parton level study in Ref.~\cite{llll}
shows that the event rates are generally small after imposing the necessary
kinematic cuts to suppress the backgrounds. To clearly identify the
EWSB mechanism from the $\WWWW$ interactions, 
the $\ell^\pm + \, {\rm jet}$ mode of the $WW$ pair should also be studied
because of its larger branching ratio than the pure leptonic 
mode.\footnote{
 ~~At the LC, because its initial state is colorless, the pure jet decay
mode (with the largest branching ratio) of the $W$ boson can also
be useful. }
~~That is the decay mode we shall concentrate on in the following section
for discussing the detection of various models of EWSB sector.

\vspace{0.5cm}
\noindent
{\bf 8.4.} Various Models 
\indent\indent

\vspace{0.3cm}
\noindent
{\it 8.4.1. A TeV Scalar Resonance }
\vspace{0.2cm}

Based on the triviality argument \cite{trivial},
the mass of the SM Higgs boson cannot be much larger than 
$\sim 650$ GeV, otherwise the theory would be inconsistent. 
(If the SM is an effective theory valid up to the energy scale much 
higher than 1 TeV, then this number is even lower.)
However, one may consider an effective theory, such as an 
 electroweak chiral lagrangian, in which 
a TeV scalar (spin-0,isospin-0) resonance couples
to the would-be Goldstone bosons in the same way as the Higgs boson 
in the Standard Model \cite{chiralone,chiral,llll}.
(The mass and the width of the scalar resonance are the two free parameters 
in this model.)
Then one can ask how to detect such a TeV 
scalar resonance. This study was
already done at the hadron level in Ref.~\cite{kane}. 

The tricks of enhancing the ratio of signal 
to background are as follows.
First of all, we trigger on a
high $P_T$ lepton. The lepton is said to be isolated if there is no more than
a certain amount of hadronic energy inside a cone 
of size $\Delta R$ surrounding 
the lepton. ($\Delta R=\sqrt{(\Delta \phi)^2+(\Delta \eta)^2}$,
$\phi$ is the azimuthal angle and $\eta$ is the pseudo-rapidity.)
A TeV resonance produces a $W$-boson with typical $P_T$ at the order of 
$\sim 1/2$ TeV, therefore, the $P_T$ of the
lepton from the $W$-decay is at the order
of a few hundred GeV. 
The kinematic cut on the $P_T$ of an isolated lepton alone can
suppress a large fraction of $t \bar t$ background events because the lepton
produced from the decay of the $W$-boson typically has $P_T \sim m_t/3$, where
$m_t$ is the mass of the top quark. Furthermore, 
the lepton is also less isolated in
the $t \bar t$ event than that in the signal event.
After selecting the events with an
isolated lepton  with high $P_T$, we can make use of
the fact that the background event contains more hadronic activity than the
signal event to further suppress the background. One can make a cut on the
charged particle multiplicity of the event to enhance the 
signal-to-background ratio. 
The alternative way of making use of this fact is to
demand that there is only one big cone jet in the central rapidity region
of the detector \cite{kane}.
 The background process typically produces more hard jets
than the signal, hence vetoing the events with more than one big cone jet
in the central rapidity region is also a useful technique. 
The $W^++\,{\rm jets}$ and 
$t \bar t$ background
processes can further be suppressed by demanding that
the large cone jet has
invariant mass $\sim M_W$ and high $P_T$.
Inside this big cone jet, one requires two small cone 
jets corresponding to the two decay quark jets of the $W$-boson. 

As discussed above, measuring the polarization of the $W$ bosons in the 
final state can be a very useful tool for detecting and discriminating 
mechanisms of EWSB. Therefore, the best strategy for analyzing the
the experimental data is not to bias the information on the
polarization of the $W$ boson.
Some of the methods that can preserve the information on the
polarization of the $W$ boson were presented in Ref. \cite{kane}.
It was shown that it is possible to
measure the fraction of longitudinal $W$'s in the candidate $W$ samples to
distinguish various models of EWSB sector.
One of the techniques which would not bias the polarization of the $W$-boson
is to count the charged particle multiplicity inside the big cone jet. A real
$W$-boson decays into a color singlet state of $q \bar q$
with the same multiplicity regardless of its energy, hence the
charged particle multiplicity of these two jets is less than that of a pair
of non-singlet QCD jets (which form the
``fake $W$''), either quark or gluon jets. 
Furthermore, the QCD background events usually have more complicated 
color structure at the parton level, so that the hadron multiplicity of
the background event is generally larger than that of the signal event
in which the $WW$ system is a color singlet state.
Since the above methods only rely on the global features of the events,
they will not bias the information on the $W$ boson polarization.

Up to this point, we have only discussed the event structure in the central
rapidity region. As discussed in the previous section, in the large rapidity
region the signal event tends to have an energetic forward jet. It has been
shown that tagging one such forward jet can further suppress the background
at very little cost to the signal event rate \cite{llll}.

Furthermore, with rapidity coverage down to 5, one can have a good
measurement on the missing energy
($\ETslash$). Because 
the typical $\ETslash$ due to the neutrino
from the $W$-boson, with energy $\sim 1$\,TeV,
decay is of the order of a few hundred GeV, the
mis-measurement of neutrino transverse momentum 
due to the underlying hadronic activity is negligible.
 Knowing $\ETslash$ and the momentum of the lepton, one
can determine the longitudinal momentum of the
neutrino up to a two-fold solution
by constraining the invariant mass of the lepton and neutrino to
be $M_W$ \cite{kane}.
{}From the invariant mass of
 $\ell,\,\nu,\,q_1$, and $\bar q_2$, one can reconstruct 
$M_{WW}$ to discriminate background from signal events. 
If the width of the heavy resonance is small,\footnote{
 ~~For a SM Higgs boson with mass $m_H$ in unit of TeV, its 
decay width would be about equal to $m_H^3/2$ (in TeV), which is not small.
}   ~~then one can detect a ``bump'' in the $M_{WW}$ distributions. 
However, if its width is too large, then the best way to detect 
this new physics effect is to measure the fraction ($f_L$) of longitudinal 
$W$'s in the event sample. 

\vspace{0.4cm}
\noindent
{\it 8.4.2. A TeV Vector Resonance }
\vspace{0.2cm}

An example of this type of resonance is a techni-rho in the techni-color
model \cite{techni}.
 What we have in mind here
is a vector (spin-1,isospin-1) resonance in the electroweak chiral
lagrangian. The mass and the width of the vector resonance 
are the two free parameters
in this model. Because this resonance gives a large contribution in the 
isospin-1 channel, 
the most useful mode to look for such a resonance is the $W^\pm Z^0$ mode. 
If the signal event rate is large enough, the resonance can be observed by
the pure leptonic decay mode
$W^+(\ra \ell^+\nu)Z^0(\ra \ell^+\ell^-)$ in which all 
the leptons have
$P_T\sim$ few hundred GeV and are well isolated. If the $\WPZ$ mode is
necessary for the signal to be observed, the strategies discussed in the
previous subsection for the $\WPWM$ mode can be applied in this case as well.
Needless to say, in this case,
the invariant mass of the two jets peaks 
around $M_Z$ not $M_W$. It could be very valuable to improve the
techniques that separate $W( \ra jj)$ from $Z(\ra jj)$ by identifying
the average electric charge of each of the two decay jets.\footnote{
 ~~For the techniques used in identifying the average electric charge of 
a QCD jet, see, for example, Ref. \cite{aleph}.  }
~~ Obviously, the two jets from the $Z^0$ boson decay should have the same 
electric charges.

\vspace{0.4cm}
\noindent
{\it 8.4.3. No Resonance }
\vspace{0.2cm}

If there is no resonance at all,
the interactions among the longitudinal $W$'s become strong in the TeV region. 
Although the non-resonance scenario is 
among the most difficult cases to be probed, 
this does {\it not} imply that 
it is less likely than the others to describe the underlying dynamics
of the electroweak symmetry breaking. 
For example, it was argued in Ref.~\cite{velt} that 
the non-resonance scenario may likely be realized.
Within this non-resonance scenario, the electroweak chiral lagrangian (EWCL)
provides the most economic way to parameterize models of strongly coupled
EWSB sector.
The model with only the lowest order term 
(containing two derivatives) in the EWCL
is known as the low energy theorem model.
The signal of this model can be detected from studying the $\WPWM$ mode 
in the TeV region Ref.~\cite{kane}.
The techniques of observing this signal are identical to 
those discussed above.

In Ref. \cite{llll}, it was shown that it is possible to  
study the pure leptonic mode $\WPWP$ in
the multi-TeV region  to test 
the low energy theorem model as long as the integrated luminosity
 (or, the event rate) is large enough. 
The dominant backgrounds for this mode are 
the {\it intrinsic} background, $W^+ t \bar t$, and QCD-gluon
exchange processes.
The signal event can be triggered by two like-sign
charged leptons with high $P_T$ ($\sim $ few hundred GeV). One can 
further require
these leptons to be isolated and veto events with additional high $P_T$ jets in
the central rapidity region.
There are two missing neutrinos in the event 
so that it is difficult to reconstruct the $W$-boson and
measure its polarization. Hence,  in the
absence of a ``bump'' structure in any
distribution, one has to know the background event rate
well to study the EWSB sector,
unless the signal rate is very large.
Similarly, measuring the charged or total
particle multiplicity of the event and tagging
a forward jet can further improve the signal-to-background ratio.

Particularly for the case of no resonance, when the signal rate is not 
large, it is important to avoid imposing kinematic cuts which greatly
reduce the signal or  bias the polarization information 
of the $W$ bosons in the data sample.
The specific technique for measuring $f_L$, as proposed 
in Ref.~\cite{kane},  will probably have to be used to study the 
non-resonance case, and to probe the EWSB sector.
This technique takes advantage of the fact that the SM
is well tested, and will be much better tested in the 
TeV region by the time the study of $W_LW_L$ 
interactions is under way.
Every event of a real or fake $W_LW_L$ interaction will be
clearly identified as originating either from SM or new physics.
The real SM events 
(from $q \bar q , \, gg \ra WW$, $Wjj$, $t \bar t$, \etc)
can all be calculated and independently measured. 
Thus, one can first make global cuts such as requiring a high energy 
spectator jet and low total particle multiplicity,
and then examine all remaining candidate events 
to see if they are consistent with SM processes
or if they suggest new physics, in particular new sources of longitudinal
$W$'s. In principle, only one new quantity needs to be measured: the fraction
of $W_LW_L$ events compared to the total number of all $WW$ events 
including real and fake $W$'s. This can be done by the
usual approach of a maximum likelihood analysis, or
probably even better by the emerging neural network 
techniques~\cite{neural},
for which this analysis appears to be ideally suited.

Ultimately, recognizing that {\it in the TeV region}
every event must originate from either the
well understood Standard Model physics or beyond
 will be the most powerful approach to discovering
any deviations from the perturbative Standard Model predictions. 
Most of the proposals discussed here have been examined at the parton level
but not in detector simulations~\cite{sdcgem}.
 They have been demonstrated to be 
promising techniques, but we cannot be sure they will work 
until the detector simulations are carried out by experimentalists.
Fortunately, there will be plenty of time to do those studies before the
data is available.

\vspace{0.7cm}
\noindent
{\bf 9. Conclusions}
\vspace{0.3cm}

In the absence of light resonance in the EWSB sector, the 
longitudinal-$W$ boson scatterings must become strong in the TeV region.
In this review, after introducing the EWCL formalism to economically
describe the EWSB sector in Sec.~2, we first discuss the current low
energy bounds on these NLO EWSB parameters in Sec.~3.
Then, in Sec.~4, we systematically develop a precise electroweak power 
counting rule (4.5), from a natural generalization of Weinberg's counting 
method for the ungauged non-linear sigma model. For completeness
and for other possible applications, we also generalize our power 
counting rule for a linearly realized effective Lagrangian~\cite{linear}
which is often studied in the literature and called the decoupling scenario. 
The renormalizable SM with a light Higgs boson is
included in the linear effective Lagrangian formalism at the lowest order.
The corresponding power counting analysis for this decoupling scenario is
given in Sec.~4.2.2. 
Furthermore, based upon our recent study 
on the intrinsic connection between
the longitudinal weak-boson scatterings and probing the EWSB sector, we
formulate, in Sec.~5, the physical content of the 
electroweak equivalence theorem (ET) as a criterion for
discriminating processes which are sensitive/insensitive to probing
the EWSB mechanism. Our recent works on 
the other aspects of the precise formulation and application 
of the ET, including its deep connection with the underlying Higgs mechanism
(with or without elementary Higgs boson), are comprehensively reviewed 
in the rest parts of Sec.~5 for the first time.

Armed with the powerful precise counting method 
and using the ET as a necessary
criterion for probing the EWSB sector, we then systematically classify
the sensitivities of various scattering processes to 
the complete set of bosonic operators at the level of $S$-matrix
elements (cf. Tables~I-VII and Sec.~6). 
The power counting hierarchy in (6.1) governs the 
order of magnitude of all relevant scattering amplitudes.

Finally, based on the above power counting analysis combined with the EWA, 
we study the phenomenology for probing the EWSB sector
at the LHC and the LC via the $~V^aV^b\rightarrow V^cV^d~$ fusion 
and the $~f\bar{f}'\rightarrow V^aV^b~$  annihilation processes.
In this simple power counting analysis,
our numerical results for the production rates agree, 
within about a factor of $2$ (cf. Fig.~4.),
with the explicit calculations performed in the literature
in which only a small subset of the NLO operators were studied.
This indicates that our power counting analysis conveniently and reasonably
grasps the overall physical picture.
With this powerful tool, we perform the first complete survey on 
estimating the sensitivities
of all fifteen next-to-leading order 
$CP$-conserving and $CP$-violating effective operators at the LHC 
via $W^\pm W^\pm$-fusions and 
$q\bar{q}'\rightarrow W^\pm Z^0~$ 
annihilations. The results are shown in Figs.~7-10 and Fig.~11-12, 
respectively.
We find that, for $W^+W^+$-channel, when the  coefficients $\ell_n$'s 
are naturally of $O(1)$,
$~{\cal L}_{4,5}~$ are most sensitive, 
$~{\cal L}_{3,9,11;12}~$ are marginally sensitive, and
$~{\cal L}^{(2)\prime}~$ and $~{\cal L}_{1,2,8;13,14}~$ are insensitive.
For the case where
the coefficients other than ~$\ell_0,~\ell_1,~\ell_8$~ are of $~O(10)~$, 
the probe of $~{\cal L}_{14}~$ (for lower $\Lambda$) and 
$~{\cal L}_{3,9,11;12}~$ could become sensitive and that of 
$~{\cal L}_{2}~$ (for lower $\Lambda$) could become 
marginally sensitive. However,
$~{\cal L}_{13}~$ cannot be sensitively
probed via this process so that it must be 
measured via other processes.\footnote{  ~~We note that 
$~{\cal L}_{13}~$ (and $~{\cal L}_{14}~$) can be sensitively probed via
$~e^- \gamma \rightarrow \nu_e W^-_L Z^0_L \, ~{\rm or}~ \,
e^- W^-_L W^+_L~$ processes at the future TeV 
linear collider~\cite{global2}.}~~ A similar conclusion holds for the 
$W^-W^-$ channel except that the event rate is lower by about a factor 
of $3 \sim 5$ in the TeV region because the quark 
luminosity for producing a $W^-W^-$ pair is smaller 
than that for a $W^+W^+$ pair in pp collisions. 
 Up to the next-to-leading order,
the $SU(2)_C$-violating operators $~{\cal L}_{6,7,10}~$
do not contribute to the $W^\pm W^\pm$ channel. They, 
however, can be probed 
via the $WZ \rightarrow WZ$,~ $WW \rightarrow ZZ$,~
$ZZ \rightarrow WW$,  and  $ZZ \rightarrow ZZ$ processes~\cite{global2}.

For the ~$q\bar{q}^{\prime}\rightarrow W^\pm Z^0$~ process,
the conclusion is quite different. 
The operators $~{\cal L}_{4,5,6,7,10}~$ do not contribute at the tree level. 
Using this process, $~{\cal L}_{3,11,12}~$ can be 
sensitively probed in the high energy range ($E>750~$GeV), 
and the probe of $~{\cal L}_{8,9,14}~$ can be marginally sensitive 
for $~E>950$~GeV if their coefficients are of $~O(1)~$ and that of
~${\cal L}_{9,14}$~ can be sensitive if their coefficients are of ~$O(10)$~. 
The results are plotted in Figs.~11-12.
We conclude that the $VV$-fusion and the 
$q {\bar q}^{(\prime )}$-annihilation processes are 
{\it complementary} to each other 
for probing the complete set of the 
NLO effective operators in the electroweak chiral Lagrangian (2.2).
A comparative analysis for the LHC and LCs is further performed and 
reviewed in Secs.~6.3 and 7.3.

From the above global analysis,  we speculate\footnote{
 ~~To further reach a detailed quantitative conclusion, 
an elaborate and precise numerical study on all signal/background rates 
should be performed under the guideline of this global analysis
(cf. Ref.~\cite{desy}).} ~~that 
before having a large number of signal events at the LHC (i.e. with 
large integrated luminosity), the LHC alone will not be 
able to sensitively measure all these operators,
the linear collider is needed to {\it complementarily}
cover the rest of the NLO operators.
In fact, the different phases of 500 GeV and 1.5 TeV energies at the LC
are necessary because they will be sensitive to different 
NLO operators in the EWCL.  An electron-photon (or a photon-photon) 
collider is also very useful for measuring
all the NLO operators which distinguish different models
of the EWSB in the strongly interacting scenario. 

Finally, we have discussed, in section~8, 
how to experimentally test the strong  
$\WWWW$ interactions in the TeV region, with
emphasis on the general features of the event structure for the 
signals and backgrounds. 
Various techniques for enhancing the ratio of signal to background  
have also been presented. 
We show that it is possible to probe the electroweak
symmetry breaking (EWSB) sector in the TeV regime
even when the $W_LW_L$ scattering is non-resonant, as maybe the most 
likely outcome of a strong EWSB dynamics~\cite{velt}.

\vspace{1.0cm}
\noindent
{\bf Acknowledgements}~~~
C.P.Y. would like to thank Professors Yu-Ping Kuang and
Qing Wang, and all those involved in organizing this Workshop for 
their warm hospitality.  We also thank many colleagues, especially
Michael Chanowitz, John Donoghue, Tao Han and Peter Zerwas, for useful
discussions on this subject.
H.J.H. is supported by the AvH of Germany.
Y.P.K. is supported by the NSF of China 
and the FRF of Tsinghua University;
C.P.Y. is supported in part by the NSF grant No. PHY-9507683.


\newpage
\vspace{1.2cm}

\newpage
\noindent
{\bf Table Captions}
\vspace{0.2cm}

\noindent
{\bf Table I.}  Estimates of amplitudes 
for $~W^\pm W^\pm \rightarrow W^\pm W^\pm ~$
scattering.\\[0.35cm]
{\bf Table Ia.} Model-independent contributions from 
$~{\cal L}_G +{\cal L}_F +{\cal L}^{(2)}~$.\\[0.20cm]
{\bf Table Ib.} Model-dependent contributions from the 
next-to-leading order operators.

\noindent
{\bf Table II.} Order estimates of $B$-terms for 
$~W^\pm W^\pm \rightarrow W^\pm W^\pm ~$ scattering.\\[0.35cm]
{\bf Table IIa.}  Model-independent contributions.\\[0.20cm]
{\bf Table IIb.}  Relevant operators for model-dependent 
                 contributions.$^{(a)}$

\noindent
{\bf Table III.}~ Estimates of amplitudes for $~q\bar{q}^{(\prime )}   
      \rightarrow V^aV^b~$: Model-independent contributions.\\[0.35cm]
{\bf Table IIIa.} ~For $~q\bar{q}^{(\prime )}\rightarrow W^{+}W^{-}~,
                ~W^{\pm}Z ~$.\\[0.20cm] 
{\bf Table IIIb.} ~For $~q\bar{q}\rightarrow ZZ~$.

\noindent
{\bf Table IV.}~ 
 Estimates of amplitudes for $~q\bar{q}^{(\prime )}\rightarrow V^aV^b~$: 
 Model-dependent contributions.$^{(a)}$\\[0.35cm] 
{\bf Table IVa.} ~For $~q\bar{q}\rightarrow W^{+}W^{-}~ $.\\[0.20cm]
{\bf Table IVb.} ~For $~q\bar{q}^{\prime}\rightarrow W^{\pm}Z~ $.

\noindent
{\bf Table V.} Global classification of sensitivities 
to probing direct and indirect EWSB information from effective 
operators at the level of $S$-matrix elements (A).$^{(a)}$\\
Notes:\\ {\footnotesize
$^{(a)}$ The contributions from
${\cal L}_{1,2,13}$ are {\it always} associated 
with a factor of $\sin^2\theta_W$, unless specified otherwise. 
Also, for contributions to the $B$-term in 
a given $V_L$-amplitude, we list them
separately with the $B$-term specified.\\ 
$^{(b)}$ MI $=$ model-independent, MD $=$ model-dependent.\\
$^{(c)}$ There is no contribution when all the external lines are 
electrically neutral.\\
$^{(d)}$ $B_0^{(1)}\simeq T_0[2\pi ,v,V_T]~(\neq T_0[2\pi^0 ,v^0,Z_T])$,~
$B_0^{(3)}\simeq T_0[v,3V_T]~(\neq T_0[v^0,3Z_T])$.\\
$^{(e)}$  $T_1[2V_L,2V_T]=T_1[2Z_L,2W_T],~T_1[2W_L,2Z_T]$, ~or
$~T_1[Z_L,W_L,Z_T,W_T]$.\\
$^{(f)}$ ${\cal L}_2$ only contributes to $T_1[2\pi^\pm ,\pi^0,v^0]$ and
$T_1[2\pi^0,\pi^\pm ,v^\pm ]$ at this order; 
${\cal L}_{6,7}$ do not contribute
to $T_1[3\pi^\pm ,v^\pm ]$.\\
$^{(g)}$  ${\cal L}_{10}$ contributes only 
to $T_1[\cdots ]$ with all the external
lines being electrically neutral.\\
$^{(h)}$ $B_0^{(2)}$ is dominated by $~T_0[2V_T,2v]~$ since 
$~T_0[\pi,2V_T,v]~$ contains a suppressing factor $\sin^2\theta_W$ 
as can be deduced from $~T_0[\pi,3V_T]~$ (cf. Table~1a) 
times the factor $~v^\mu =O\left(\frac{M_W}{E}\right)~$.\\
$^{(i)}$ Here, $T_1[2W_L,2W_T]$ contains a coupling 
$e^4=g^4\sin^4\theta_W$.\\
$^{(j)}$ ${\cal L}_2$ only contributes to $T_1[3\pi^\pm ,v^\pm ]$.\\
$^{(k)}$ ${\cal L}_{1,13}$ do not contribute to 
$T_1[2\pi^\pm ,2v^\pm ]$.  }

\noindent
{\bf Table VI.} 
Global classification of sensitivities to probing direct and indirect EWSB\\ 
information from effective operators at the level of $S$-matrix elements (B). 
$^{(a)}$

\noindent
{\bf Table VII.} 
Probing the EWSB sector at high energy colliders:
a global classification for the NLO bosonic operators.

\newpage
\vspace{1.8cm}
\noindent
{\bf Figure Captions}
\vspace{0.2cm}

\noindent
{\bf Fig.~1.} Examination on the kinematic dependence and the 
validity of the ET
for the $~W^+_LW^-_L \rightarrow W^+_LW^-_L ~$ scattering 
process .   \\[0.20cm]
{\bf (1a).}  The ratio $~|B/g^2|~$ for $~\theta 
= 2^{\circ},~10^{\circ},~45^{\circ},~
90^{\circ},~100^{\circ},~120^{\circ},~ 135^{\circ},~ 150^{\circ},~ 
180^{\circ}~$. \\
{\bf (1b).}  Same as (10a), but for the ratio $~|B/T[W_L]|~$.\\
{\bf (1c).}  Comparison of the $W_L$-amplitude (solid lines) and 
the corresponding GB-amplitude (dotted lines) for 
$~\theta = 10^{\circ},~45^{\circ},~100^{\circ},~150^{\circ}~$. 
Here, $~B\,[150^{\circ}]~$ denotes
the $B$-term at $~\theta =150^{\circ}~$.

\noindent
{\bf Fig.~2.} Comparison of the LHC production rates 
contributed by the exact $W_L$-amplitude (solid line) and the
GB-amplitude (dotted line) from eq.~(5.43).\\
{\bf (2a).} For the $WW$ invariant mass distribution.\\
{\bf (2b).} For the angular distribution.

\noindent
{\bf Fig.~3.}
The differential luminosity distribution 
($~d{\cal L}/dM_{V V}^2~$)
of the polarized $~VV$-pair from the two fermion system,
as a function of the invariant mass $M_{VV}$, for 
the LHC (a 14\,TeV pp collider) and 
the LC (a 1.5\,TeV $e^-e^+$ or $e^-e^-$ collider). 
In each plot, three curves are presented for different combinations,
i.e. $LL$, $LT$ and $TT$ (from the bottom to top),
of the polarizations of the $VV$ pair at LHC (solid lines) and LC 
(dashed lines).
\\[0.20cm]
{\bf (3a).} For the $W^-W^-$ mode. \\
{\bf (3b).} For the $W^+W^+$ mode (LHC only). \\
{\bf (3c).} For the $W^+W^-$ mode. \\
{\bf (3d).} For the $Z^0Z^0$ mode.\\
{\bf (3e).} For the $W^+Z^0$ mode. \\
{\bf (3f).} For the $W^-Z^0$ mode. 

\noindent
{\bf Fig.~4.}  Comparison of the power counting predictions with the
corresponding ones in Fig.~8 of Ref.~\cite{BDV} up to one-loop for 
a pp collider with
$\sqrt{S}= 40$\,TeV. The solid lines are given by our power counting 
analysis; the dashed lines are from Ref.~\cite{BDV}. [The meanings of the 
rates $R_{\alpha\beta\gamma\delta}$'s and $R_{\alpha\beta}$'s
 are defined in eq.~(7.7a,b) and below.]
\\[0.20cm]
{\bf (4a).}  $W^+W^+\rightarrow W^+_L W^+_L$.\\
{\bf (4b).}  $q\bar{q}^{\prime}\rightarrow W^+_L W^-_L$.

\noindent
{\bf Fig.~5.}\\ 
{\bf (5a).} Comparison of the $W^+_LW^+_L$ production rates up to 
one-loop (for $\ell_{0-14}=0$)
with $W^+_LW^+_L$, $W^+_LW^+_T$ and $W^+_TW^+_T$ initial states,
at the  $14$\,TeV LHC.\\
{\bf (5b).} Comparison of the production rates for
different final-state polarizations up to one-loop (for $\ell_{0-14}=0$)
after summing over the polarizations of the initial states, at the 
$14$\,TeV LHC.

\noindent
{\bf Fig.~6.} Comparison of the production rates for different
final-state polarizations up to one-loop (for $\ell_{0-14}=0$)
via $q\bar{q}^{\prime}\rightarrow W^+ Z^0$ at the  $14$\,TeV LHC.

\noindent
{\bf Fig.~7.} Sensitivities of the operators ${\cal L}^{(2)\prime}$ and 
${\cal L}_{1\sim 14}$ at the $14$ TeV LHC with $\Lambda=3.1~$TeV. 
The coefficients $\ell_n$'s are
taken to be of $O(1)$,\\[0.20cm]
{\bf (7a).} For operators ${\cal L}_{3,4,5,9,11,12}~$.\\
{\bf (7b).} For operators 
${\cal L}^{(2)\prime}$ and ${\cal L}_{1,2,8,13,14}~$.

\noindent
{\bf Fig.~8.}  Same as Fig.~7, but the coefficients $\ell_n$'s are 
taken to be of $O(10)$ except $\ell_{0,1,8}$ which are already 
constrained by low energy data to be of $O(1)$.\\[0.20cm]
{\bf (8a).} For operators ${\cal L}_{3,4,5,9,11,12}~$.\\
{\bf (8b).} For operators ${\cal L}_{2,13,14}~$.

\noindent
{\bf Fig.~9} Same as Fig.~7, but with $\Lambda=2.0~$TeV~.

\noindent
{\bf Fig.~10.} Same as Fig.~8, but with $\Lambda=2.0~$TeV~.

\noindent
{\bf Fig.~11.} Sensitivities of the operators ${\cal L}^{(2)\prime}$ and
${\cal L}_{1-14}$ in $q\bar{q}^{\prime}\rightarrow W^+ Z^0$ at the 
$14~$TeV LHC with $\Lambda=3.1~$TeV~.
\\[0.20cm]
{\bf (11a).}  The coefficients $\ell_n$~'s are taken to be of $O(1)$.\\
{\bf (11b).}  The coefficients $\ell_n$~'s are taken to be of $O(10)$
except $\ell_{0,1,8}$ which are already 
constrained by low energy data to be of $O(1)$.

\noindent
{\bf Fig.~12.} Same as Fig.~11, but with $\Lambda=2.0~$TeV~.


\addtolength{\textheight}{2cm}
\addtolength{\topmargin}{-1.1cm}

\addtolength{\textwidth}{2.0cm}
\addtolength{\oddsidemargin}{1.1cm}
\addtolength{\oddsidemargin}{-0.9cm}
\evensidemargin=\oddsidemargin
\addtolength{\textheight}{-0.5cm}
\addtolength{\topmargin}{1.0cm}

\newpage

\renewcommand{\baselinestretch}{1}


\begin{table}[t]  
\begin{center}

\tcaption{
Estimates of
amplitudes for $~W^{\pm}W^{\pm}\rightarrow W^{\pm}W^{\pm}~$ scattering. }
\vspace{0.8cm}

{\small Table~Ia.
~Model-independent contributions from 
$~{\cal L}_G+{\cal L}_F +{\cal L}^{(2)}~$. }
\vspace{0.5cm}

\small

\begin{tabular}{||c||c|c|c|c|c||} 
\hline\hline
& & & & &  \\
${\cal L}_G +{\cal L}_F + {\cal L}^{(2)}$  
&  $~~~T_{\ell}[4\pi]~~~     $  
&  $~T_{\ell}[3\pi,W_T]~ $  
&  $~T_{\ell}[2\pi,2W_T]~$  
&  $~T_{\ell}[\pi,3W_T]~ $  
&  $~T_{\ell}[4W_T]~     $  \\ 
& & & & &  \\
\hline\hline
& & & & &\\
Tree-Level
&  $ \frac{E^2}{f_{\pi}^2} $
&  $ g\frac{E}{f_\pi} $
&  $ g^2 $
&  $ e^2g\frac{f_\pi}{E} $ 
&  $g^2$ \\
( ${\ell}=0$ ) & & & & &\\
& & & & &\\
\hline
& & & & &\\
One-Loop 
& $\frac{E^2}{f_{\pi}^2}\frac{E^2}{\Lambda_0^2}$ 
& $g\frac{E}{f_{\pi}}\frac{E^2}{\Lambda_0^2}$ 
& $g^2\frac{E^2}{\Lambda_0^2}$  
& $g^3\frac{f_{\pi}E}{\Lambda_0^2}$
& $g^4\frac{f_{\pi}^2}{\Lambda_0^2}$   \\
( ${\ell}=1$ ) & & & & &\\
& & & & & \\
\hline\hline 
\end{tabular}
\end{center}
\end{table}

\vspace{0.8cm}


\newpage
\begin{table}[t]  
\begin{center}

{\small Table~Ib.
Model-dependent contributions from the 
next-to-leading order  operators. }
\vspace{0.8cm}

\renewcommand{\baselinestretch}{1.0}
\small
\begin{tabular}{||c||c|c|c|c|c||} 
\hline\hline
& & & & &  \\
Operators 
& $ T_1[4\pi] $
& $ T_1[3\pi,W_T] $
& $ T_1[2\pi,2W_T] $
& $ T_1[\pi,3W_T] $
& $ T_1[4W_T] $ \\
& & & & &  \\
\hline\hline
& & & & &  \\
$ {\cal L}^{(2)\prime} $
& $ \ell_0 ~\frac{E^2}{\Lambda^2} $
& $ \ell_0~g\frac{f_\pi E}{\Lambda^2} $ 
& $ \ell_0~g^2\frac{f_\pi^2}{\Lambda^2} $
& $ \ell_0~g^3\frac{f_\pi^3}{E \Lambda^2} $
& /  \\
& & & & &  \\
\hline
& & & & &  \\
$ {\cal L}_{1,13} $
&  /
& $ \ell_{1,13}~e^2g\frac{f_\pi E}{\Lambda^2} $
& $ \ell_{1,13}~e^4\frac{f_\pi^2}{\Lambda^2} $
& $ \ell_{1,13}~e^2g\frac{f_\pi E}{\Lambda^2} $
& $ \ell_{1,13}~e^2g^2\frac{f_\pi^2}{\Lambda^2} $  \\   
& & & & &  \\
\hline
& & & & &  \\
$ {\cal L}_2 $
& $ \ell_2~e^2\frac{E^2}{\Lambda^2} $ 
& $ \ell_2~e^2g\frac{f_\pi E}{\Lambda^2} $
& $ \ell_2~e^2\frac{E^2}{\Lambda^2} $
& $ \ell_2~e^2g\frac{f_\pi E}{\Lambda^2} $
& $ \ell_2~e^2g^2\frac{f_\pi^2}{\Lambda^2} $ \\
& & & & &  \\
\hline
& & & & &  \\
$ {\cal L}_3 $
& $ \ell_3~g^2\frac{E^2}{\Lambda^2} $
& $ \ell_3~g\frac{E}{f_\pi}\frac{E^2}{\Lambda^2} $
& $ \ell_3~g^2\frac{E^2}{\Lambda^2} $
& $ \ell_3~g^3\frac{f_\pi E}{\Lambda^2} $
& $ \ell_3~g^4\frac{f_\pi^2}{\Lambda^2} $ \\
& & & & &  \\
\hline
& & & & &  \\
$ {\cal L}_{4,5} $
& $ \ell_{4,5}~\frac{E^2}{f_\pi^2}\frac{E^2}{\Lambda^2} $
& $ \ell_{4,5}~g\frac{E}{f_\pi}\frac{E^2}{\Lambda^2} $
& $ \ell_{4,5}~g^2\frac{E^2}{\Lambda^2} $
& $ \ell_{4,5}~g^3\frac{f_\pi E}{\Lambda^2} $ 
& $ \ell_{4,5}~g^4\frac{f_\pi^2}{\Lambda^2} $ \\
& & & & &  \\
\hline
& & & & & \\
$ {\cal L}_{6,7,10} $
& /
& /
& /
& /
& /  \\
& & & & &  \\
\hline
& & & & &  \\
$ {\cal L}_{8,14} $
& / 
& $ \ell_{8,14}~g^3\frac{f_\pi E}{\Lambda^2} $
& $ \ell_{8,14}~g^2\frac{E^2}{\Lambda^2} $
& $ \ell_{8,14}~g^3\frac{f_\pi E}{\Lambda^2} $
& $ \ell_{8,14}~g^4\frac{f_\pi^2}{\Lambda^2} $ \\
& & & & & \\
\hline
& & & & & \\
$ {\cal L}_9 $
& $ \ell_9~g^2\frac{E^2}{\Lambda^2} $
& $ \ell_9~g\frac{E}{f_\pi}\frac{E^2}{\Lambda^2} $ 
& $ \ell_9~g^2\frac{E^2}{\Lambda^2} $
& $ \ell_9~g^3\frac{f_\pi E}{\Lambda^2} $
& $ \ell_9~g^4\frac{f_\pi^2}{\Lambda^2} $ \\
& & & & & \\
\hline
& & & & & \\
$ {\cal L}_{11,12} $
& /
& $ \ell_{11,12}~g\frac{E}{f_\pi}\frac{E^2}{\Lambda^2} $
& $ \ell_{11,12}~g^2\frac{E^2}{\Lambda^2} $
& $ \ell_{11,12}~g^3\frac{f_\pi E}{\Lambda^2} $
& $ \ell_{11,12}~g^4\frac{f_\pi^2}{\Lambda^2} $ \\
& & & & & \\
\hline\hline 
\end{tabular}
\end{center}
\end{table}

\newpage

\renewcommand{\baselinestretch}{1}

\begin{table}[t]   
\begin{center}

\tcaption{
Order estimates of $B$-terms 
for $~W^{\pm}W^{\pm}\rightarrow W^{\pm}W^{\pm}~$  scattering. }
\vspace{0.8cm}

{\small 
Table~IIa. Model-independent contributions. }
\vspace{0.5cm}

\small

\begin{tabular}{||c||c|c|c|c||} 
\hline\hline
& & & &  \\
${\cal L}_G +{\cal L}_F + {\cal L}^{(2)}$  
&  $~~~B_{\ell}^{(0)}~~~     $  
&  $~~~B_{\ell}^{(1)}~~~     $  
&  $~~~B_{\ell}^{(2)}~~~     $  
&  $~~~B_{\ell}^{(3)}~~~     $  \\ 
& & & &   \\
\hline\hline
& & & & \\
Tree-Level
&  $ g^2$
&  $ g^2\frac{M_W}{E} $
&  $ g^2\frac{M_W^2}{E^2} $
&  $ g^2\frac{M_W}{E} $\\
 ( ${\ell}=0$ ) & & & &\\
& & & & \\
\hline
& & & & \\
One-Loop 
& $g^2\frac{E^2}{\Lambda_0^2}$ 
& $g^3\frac{Ef_\pi}{\Lambda_0^2}$ 
& $ g^4\frac{f_{\pi}^2}{\Lambda_0^2}$  
& $g^4\frac{f_{\pi}^2}{\Lambda_0^2}\frac{M_W}{E} $\\
( ${\ell}=1$ ) & & & &\\
& & & &  \\
\hline\hline 
\end{tabular}

\vspace{1.5cm}


{\small 
Table~IIb. 
Relevant operators for model-dependent contributions.$^{(a)}$ }

\vspace{0.8cm}

\small

\begin{tabular}{||c|c|c|c||} 
\hline\hline
& & &   \\
    $O(g^2\frac{E^2}{\Lambda^2})$
&    $O(g^3\frac{Ef_\pi}{\Lambda^2})$
&    $O(g^2\frac{f_{\pi}^2}{\Lambda^2})$ 
&    $O(g^4\frac{f_{\pi}^2}{\Lambda^2})$ \\
    ( from $B^{(0)}_1$ )
&   ( from $B^{(1)}_1$ )
&   ( from $B^{(0)}_1$ )  
&   ( from $B^{(2)}_1 ~{\rm or} ~B^{(0)}_1 $ )\\
 & & &   \\
\hline
 & & & \\
  ${\cal L}_{3,4,5,9,11,12}$  
& ${\cal L}_{2,3,4,5,8,9,11,12,14}$ 
& ${\cal L}^{(2)\prime}$ 
& \parbox[t]{4.28cm}{
${\cal L}_{1\sim 5,8,9,11\sim 14} ~~(B^{(2)}_1)$ \\
${\cal L}_{1,2,8,13,14} ~~(B^{(0)}_1)$ \\ 
${\cal L}_{2\sim 5,8,9,11,12,14} ~~(B^{(0)}_1)~^{(b)}$ } \\
& & & \\
\hline\hline 
\end{tabular}
\end{center}
\end{table}

\begin{table}[t]
\begin{center}
\vspace{0.4cm}
\begin{tabular}{l}
{\footnotesize 
$^{(a)}$ We list the relevant operators for each order of $B$-terms.} \\
{\footnotesize 
$^{(b)}$ Here $B^{(0)}_1$ is contributed by $ T_1[2\pi^\pm ,2v^\pm ] $.}\\
\end{tabular}
\end{center}
\end{table}


\newpage
\begin{table}[t]  
\begin{center}
\tcaption{
Estimates of amplitudes for $~q\bar{q}^{(\prime )}
\rightarrow V^aV^b~$: Model-independent contributions. }
\vspace{0.8cm}

{\small 
Table~IIIa.  ~For $~q\bar{q}^{(\prime )}\rightarrow W^{+}W^{-}~,
                ~W^{\pm}Z ~$.  }
\vspace{0.5cm}

\small

\begin{tabular}{||c||c|c|c||c|c||} 
\hline\hline
& & & & & \\
$~~{\cal L}_G +{\cal L}_F + {\cal L}^{(2)} ~~$  
&  $~~T_{\ell}[q\bar{q}^{(\prime )}\rightarrow \pi\pi ]~~    $  
&  $~T_{\ell}[q\bar{q}^{(\prime )}\rightarrow \pi V_T]~ $  
&  $~T_{\ell}[q\bar{q}^{(\prime )}\rightarrow V_T V_T ]~$ 
&  $B^{(0)}_{\ell}$
&  $B^{(1)}_{\ell}$  \\ 
& & & & & \\
\hline\hline
& & & & & \\
Tree-Level ( $\ell = 0$ )
&  $ g^2 $
&  $ e^2g \frac{f_\pi}{E} $
&  $ g^2 $
&  $ g^2\frac{M_W^2}{E^2} $
&  $ g^2\frac{M_W}{E}$  \\
& & & & & \\
\hline
& & & & & \\
One-Loop ( $\ell = 1$ )
& $g^2\frac{E^2}{\Lambda_0^2} $
& $g^3\frac{f_\pi E}{\Lambda_0^2} $
& $g^4\frac{f_\pi^2}{\Lambda_0^2} $
& $g^3\frac{f_\pi M_W}{\Lambda_0^2}$ 
& $g^4\frac{f_\pi^2}{\Lambda^2_0}\frac{M_W}{E}$  \\
& & & & & \\
\hline\hline 
\end{tabular}

\vspace{1.8cm}

{\small 
Table~IIIb. ~For $~q\bar{q}\rightarrow ZZ~$. }
\vspace{0.5cm}
\small

\begin{tabular}{||c||c|c|c||c|c||} 
\hline\hline
& & & & &  \\
$~~{\cal L}_G +{\cal L}_F + {\cal L}^{(2)} ~~$  
&  $~~T_{\ell}[q\bar{q}\rightarrow \pi\pi ]~~     $  
&  $~T_{\ell}[q\bar{q}\rightarrow \pi Z_T]~ $  
&  $~T_{\ell}[q\bar{q}\rightarrow Z_T Z_T ]~$ 
&  $B^{(0)}_{\ell}$
&  $B^{(1)}_{\ell}$  \\ 
& & & & &  \\
\hline\hline
& & & & & \\
Tree-Level ( $\ell = 0$ )
&  /
&  /
&  $g^2$
&  $g^2\frac{M_W^2}{E^2}$
&  $g^2\frac{M_W}{E}$  \\
& & & & & \\
\hline
& & & & &\\
One-Loop ( $\ell = 1$ )
& $g^2\frac{E^2}{\Lambda_0^2} $
& $g^3\frac{f_\pi E}{\Lambda_0^2} $
& $g^4\frac{f_\pi^2}{\Lambda_0^2} $
& $g^3\frac{f_\pi M_W}{\Lambda_0^2}$ 
& $g^4\frac{f_\pi^2}{\Lambda^2_0}\frac{M_W}{E}$  \\
& & & & & \\
\hline\hline 
\end{tabular}
\end{center}
\end{table}

\vspace{0.8cm}



\newpage

\addtolength{\oddsidemargin}{-2.5cm}

\begin{table}[t]  
\begin{center}
\tcaption{
 Estimates of amplitudes for $~q\bar{q}^{(\prime )}\rightarrow V^aV^b~$: 
 Model-dependent contributions.$^{(a)}$  }

\vspace{0.8cm}

{\small 
Table~IVa.  ~For $~q\bar{q}\rightarrow W^{+}W^{-}~ $. }
\vspace{0.5cm}

\small

\begin{tabular}{||c||c|c|c||c|c||} 
\hline\hline
& & & & & \\
~Operators~
&  $~~T_1[q\bar{q}\rightarrow \pi\pi ]~~ $  
&  $~T_1[q\bar{q}\rightarrow \pi V_T]~ $  
&  $~T_1[q\bar{q}\rightarrow V_T V_T ]~$ 
&  $B^{(0)}_{1}$
&  $B^{(1)}_{1}$  \\  
& & & & & \\
\hline\hline
& & & & & \\
   $ {\cal L}^{(2)\prime} $ 
&  $\ell_0 ~g^2\frac{f_\pi^2}{\Lambda^2} $
&  $\ell_0 ~g^3\frac{f_\pi^3}{E\Lambda^2} $
&  / 
&  $g^2\frac{f_\pi^2}{\Lambda^2}\frac{M_W^2}{E^2}$
&  /  \\
& & & & & \\
\hline
& & & & & \\
   $ {\cal L}_{1,13} $ 
& / 
&  $\ell_{1,13}~ e^2g\frac{f_\pi E}{\Lambda^2} $ 
&  $\ell_{1,13}~ e^2g^2\frac{f_\pi^2}{\Lambda^2} $ 
&  $e^2g^2\frac{f_\pi^2}{\Lambda^2}$
&  $e^2g^2\frac{f_\pi^2}{\Lambda^2}\frac{M_W}{E}$\\
& & & & & \\
\hline
& & & & & \\
$ {\cal L}_2 $ 
& $\ell_2~ e^2\frac{E^2}{\Lambda^2} $
& $\ell_2~ e^2 g\frac{f_\pi E}{\Lambda^2} $ 
& $\ell_2~ e^2g^2\frac{f_\pi^2}{\Lambda^2} $ 
&  $e^2g^2\frac{f_\pi^2}{\Lambda^2}$
&  $e^2g^2\frac{f_\pi^2}{\Lambda^2}\frac{M_W}{E}$\\
& & & & & \\
\hline
& & & & & \\
$ {\cal L}_3 $ 
& $\ell_3~ g^2\frac{E^2}{\Lambda^2} $
& $\ell_3~ g^3\frac{f_\pi E}{\Lambda^2} $ 
& $\ell_3~ g^4\frac{f_\pi^2}{\Lambda^2} $  
&  $g^4\frac{f_\pi^2}{\Lambda^2}$
&  $g^4\frac{f_\pi^2}{\Lambda^2}\frac{M_W}{E}$\\
& & & & & \\
\hline
& & & & & \\
$ {\cal L}_{8,14} $ 
& /
& $\ell_{8,14}~ g^3\frac{f_\pi E}{\Lambda^2} $ 
& $\ell_{8,14}~ g^4\frac{f_\pi^2}{\Lambda^2} $  
&  $g^4\frac{f_\pi^2}{\Lambda^2}$
&  $g^4\frac{f_\pi^2}{\Lambda^2}\frac{M_W}{E}$ \\
& & & & & \\
\hline
& & & & & \\
$ {\cal L}_9 $ 
& $\ell_9~ g^2\frac{E^2}{\Lambda^2} $ 
& $\ell_9~ g^3\frac{f_\pi E}{\Lambda^2} $ 
& $\ell_9~ g^4\frac{f_\pi^2}{\Lambda^2} $ 
&  $g^4\frac{f_\pi^2}{\Lambda^2}$
&  $g^4\frac{f_\pi^2}{\Lambda^2}\frac{M_W}{E}$ \\
& & & & & \\
\hline
& & & & & \\
$ {\cal L}_{11,12} $ 
& /
& $\ell_{11,12} ~g^3\frac{f_\pi E}{\Lambda^2} $ 
& $\ell_{11,12} ~g^4\frac{f_\pi^2}{\Lambda^2} $  
&  $g^4\frac{f_\pi^2}{\Lambda^2}$
&  $g^4\frac{f_\pi^2}{\Lambda^2}\frac{M_W}{E}$\\
& & & & & \\
\hline\hline 
\end{tabular}
\end{center}
\end{table}

\vspace{1.0cm}
\begin{table}[t]
\begin{center}
\begin{tabular}{l}
{\footnotesize 
$^{(a)}$ Here we only consider the
 light quarks ( $q \neq t$ ) whose Yukawa coupling $~y_q \approx 0~$.
At tree level,}\\
{\footnotesize  $~q\bar{q}\rightarrow ZZ~$ contains no 
model-dependent contribution and the operators $~{\cal L}_{4,5,6,7,10}~$ 
do not contribute to }\\
{\footnotesize   $~q\bar{q}^{(\prime )}\rightarrow W^+W^-,~W^{\pm}Z~$. }  
\end{tabular}
\end{center}
\end{table}

\addtolength{\oddsidemargin}{2cm}
\newpage
\addtolength{\oddsidemargin}{1.6cm}
\addtolength{\topmargin}{1.0cm}

\begin{table}[t]  
\begin{center}

{\small 
 Table~IVb. ~For $~q\bar{q}^{\prime}\rightarrow W^{\pm}Z~ $. }
\vspace{0.5cm}

\vspace{0.8cm}
\small

\begin{tabular}{||c||c|c|c||c|c||} 
\hline\hline
& & & & &  \\
~Operators~
&  $~~T_1[q\bar{q}^{\prime}\rightarrow \pi^{\pm}\pi^0 ]~~$  
&  $~T_1[q\bar{q}^{\prime}\rightarrow \pi V_T]~ $  
&  $~T_1[q\bar{q}^{\prime}\rightarrow W_T^{\pm} Z_T ]~$ 
&  $B^{(0)}_{1}$
&  $B^{(1)}_{1}$  \\  
& & & & &  \\
\hline\hline
& & & & &  \\
   $ {\cal L}^{(2)\prime} $ 
&  $\ell_0 ~g^2\frac{f_\pi^2}{\Lambda^2} $
&  $\ell_0 ~g^3\frac{f_\pi^3}{E\Lambda^2} $
& /  
&  $g^2\frac{f_\pi^2}{\Lambda^2}\frac{M_W^2}{E^2}$
&  / \\
& & & & &  \\
\hline
& & & & &  \\
   $ {\cal L}_{1,13} $ 
&  / 
&  $\ell_{1,13}~ e^2g\frac{f_\pi E}{\Lambda^2} $ 
&  $\ell_{1,13}~ e^2g^2 \frac{f_\pi^2}{\Lambda^2} $ 
&  $e^2g^2\frac{f_\pi^2}{\Lambda^2}$
&  $e^2g^2\frac{f_\pi^2}{\Lambda^2}\frac{M_W}{E}$\\
& & & & &  \\
\hline
& & & & &  \\
$ {\cal L}_2 $ 
& /
& $\ell_2~ e^2g \frac{f_\pi E}{\Lambda^2} $ 
& $\ell_2~ e^2g^2\frac{f_\pi^2}{\Lambda^2} $ 
&  $e^2g^2\frac{f_\pi^2}{\Lambda^2}$
&  $e^2g^2\frac{f_\pi^2}{\Lambda^2}\frac{M_W}{E}$\\
& & & & &  \\
\hline
& & & & &  \\
$ {\cal L}_3 $ 
& $\ell_3~ g^2\frac{E^2}{\Lambda^2} $
& $\ell_3~ g^3\frac{f_\pi E}{\Lambda^2} $ 
& $\ell_3~ g^4\frac{f_\pi^2}{\Lambda^2} $  
&  $g^4\frac{f_\pi^2}{\Lambda^2}$
&  $g^4\frac{f_\pi^2}{\Lambda^2}\frac{M_W}{E}$\\
& & & & &  \\
\hline
& & & & &  \\
$ {\cal L}_{8,14} $ 
& /
& $\ell_{8,14}~ g^3\frac{f_\pi E}{\Lambda^2} $ 
& $\ell_{8,14}~ g^4\frac{f_\pi^2}{\Lambda^2} $  
&  $g^4\frac{f_\pi^2}{\Lambda^2}$
&  $g^4\frac{f_\pi^2}{\Lambda^2}\frac{M_W}{E}$\\
& & & & &  \\
\hline
& & & & &  \\
$ {\cal L}_9 $ 
& /
& $\ell_9~ g^3\frac{f_\pi E}{\Lambda^2} $ 
& $\ell_9~ g^4\frac{f_\pi^2}{\Lambda^2} $  
&  $g^4\frac{f_\pi^2}{\Lambda^2}$
&  $g^4\frac{f_\pi^2}{\Lambda^2}\frac{M_W}{E}$\\
& & & & &  \\
\hline
& & & & &  \\
$ {\cal L}_{11,12} $ 
& $\ell_{11,12}~ g^2\frac{E^2}{\Lambda^2} $ 
& $\ell_{11,12} ~g^3\frac{f_\pi E}{\Lambda^2} $ 
& $\ell_{11,12} ~g^4\frac{f_\pi^2}{\Lambda^2} $ 
&  $g^4\frac{f_\pi^2}{\Lambda^2}$
&  $g^4\frac{f_\pi^2}{\Lambda^2}\frac{M_W}{E}$  \\
& & & & &  \\
\hline\hline 
\end{tabular}
\end{center}
\end{table}


\addtolength{\textheight}{0.5cm}
\addtolength{\topmargin}{-1.5cm}
\addtolength{\oddsidemargin}{-0.6cm}
\addtolength{\oddsidemargin}{-0.8cm}
\evensidemargin=\oddsidemargin
\newpage


\tabcolsep 1pt
\begin{table}[t]  

\tcaption{ Global classification of 
sensitivities to probing direct and indirect EWSB\\ 
information from effective operators at the level 
of $S$-matrix elements (A).$^{(a)}$  }

\begin{center}
\vspace{0.5cm}

\small

\begin{tabular}{||c||c|c|c||} 
\hline\hline
& & &  \\
Required Precision
&  Relevant Operators
&  Relevant Amplitudes
&  MI or MD $^{(b)}$ \\
& & & ? \\
\hline\hline
   $O\left(\frac{E^2}{f_{\pi}^2}\right)$ 
&  ${\cal L}_{\rm MI}~(\equiv 
   {\cal L}_{\rm G}+{\cal L}_{\rm F}+{\cal L}^{(2)}) $
&  $ T_0[4V_L] (\neq T_0[4Z_L]) $
&  MI \\
\hline
\parbox[t]{2.2cm}{ 
~~\\
~~\\
$O\left(\frac{E^2}{f_\pi^2}
 \frac{E^2}{\Lambda^2},~g\frac{E}{f_\pi}\right)$\\
~~\\
~~ }
&  \parbox[t]{2.8cm}{ 
  ${\cal L}_{4,5}$\\
  ${\cal L}_{6,7}$\\
  ${\cal L}_{10}$\\
  ${\cal L}_{\rm MI}$\\
  ${\cal L}_{\rm MI}$ }
&  \parbox[t]{5.0cm}{ 
  $T_1[4V_L]$\\
  $T_1[2Z_L,2W_L],~T_1[4Z_L]$\\
  $T_1[4Z_L]$\\
  $T_0[3V_L,V_T] ~(\neq T_0[3Z_L,Z_T])$\\
  $T_1[4V_L]$ }
&  \parbox[t]{0.8cm}{ 
  MD\\
  MD\\
  MD\\
  MI\\
  MI }\\
\hline
\parbox[t]{2.2cm}{ 
~~\\
~~\\
$O\left(g\frac{E}{f_\pi}\frac{E^2}{\Lambda^2},~g^2\right)$\\
~~\\
~~\\
~~  }
& \parbox[t]{2.8cm}{ 
  $ {\cal L}_{3,4,5,9,11,12} $\\
  $ {\cal L}_{2,3,4,5,6,7,9,11,12} $\\
  $ {\cal L}_{3,4,5,6,7,10} $\\
  $ {\cal L}_{\rm MI} $\\ 
  $ {\cal L}_{\rm MI} $\\
  $ {\cal L}_{\rm MI} $} 
&  \parbox[t]{5.8cm}{ 
  $T_1[3W_L,W_T]$\\
  $T_1[2W_L,Z_L,Z_T],~T_1[2Z_L,W_L,W_T]$\\
  $T_1[3Z_L,Z_T]$\\
  $T_0[2V_L,2V_T],~T_0[4V_T]~~^{(c)}$\\
  $T_1[3V_L,V_T]$\\
  $B^{(0)}_0 \simeq T_0[3\pi ,v]~(\neq T_0[3\pi^0 ,v^0])$  }
&   \parbox[t]{0.8cm}{ 
  MD\\
  MD\\
  MD\\
  MI\\
  MI\\
  MI }\\
\hline
  $O\left(\frac{E^2}{\Lambda^2}\right)$
& ${\cal L}^{(2)\prime}$ 
& $T_1[4W_L],~T_1[2W_L,2Z_L]$
& MD \\
\hline
   \parbox[t]{2.5cm}{
        ~~\\
~~\\
~~\\
$O\left(g^2\frac{E^2}{\Lambda^2},~g^3\frac{f_\pi}{E}\right)$ \\ 
~~\\          
~~\\
~~ }
&  \parbox[t]{3.0cm}{ 
   ${\cal L}_{\rm MI}$\\
   ${\cal L}_{2,3,9}$\\
   ${\cal L}_{3,11,12}$\\
   ${\cal L}_{2,3,4,5,8,9,11,12,14}$\\
   ${\cal L}_{1\sim 9,11\sim 14}$ \\
   ${\cal L}_{4,5,6,7,10}$ \\
   ${\cal L}_{{\rm MI},2,3,4,5,6,7,9\sim 12}$ }
&  \parbox[t]{6.3cm}{  
   $T_0[V_L,3V_T],T_1[2V_L,2V_T], B^{(1,3)}_0~~^{(c,d)} $\\
   $T_1[4W_L]$\\
   $T_1[2Z_L,2W_L]$\\
   $T_1[2W_L,2W_T]$\\
   $T_1[2V_L,2V_T]~~~^{(e)}$\\
   $T_1[2Z_L,2Z_T]$\\
{\footnotesize  $B^{(0)}_1 \simeq T_1[3\pi ,v]~~^{(f,g)}$}  }
&  \parbox[t]{2.0cm}{ 
  MI\\
  MD\\
  MD\\
  MD\\
  MD\\
  MD\\
  MI $+$ MD  }\\
\hline
\parbox[t]{2.75cm}{
~~\\
~~\\
  $O\left(g^3\frac{Ef_\pi}{\Lambda^2},~g^4\frac{f_\pi^2}{E^2}\right)$\\
~~\\
~~}
& \parbox[t]{3.3cm}{ 
${\cal L}_{{\rm MI},1,2,3,8,9,11\sim 14}$ \\
${\cal L}_{4,5}$\\  
${\cal L}_{6,7,10}$ \\
${\cal L}_{2\sim 5,8,9,11,12,14}$\\
${\cal L}_{\rm MI}$   }
& \parbox[t]{5.5cm}{ 
$T_1[V_L,3V_T]~(\neq T_1[Z_L,3Z_T])$\\
$T_1[V_L,3V_T]$\\
$T_1[V_L,3V_T]~(\neq T_1[W_L,3W_T])~~^{(g)}$ \\
{\footnotesize $B^{(1)}_1\simeq T_1[2\pi ,V_T,v]$} \\
{\footnotesize $B^{(2)}_0\simeq T_0[2V_T,2v]$}$~~^{(c,h)}$  }
& \parbox[t]{2.0cm}{ MI$+$MD \\
                     MD\\ 
                     MD\\
                     MD\\
                     MI }\\
\hline
\parbox[t]{2.0cm}{ 
~~\\
~~\\
~~\\
~~\\
~~\\
  $O\left((g^2,g^4)\frac{f_\pi^2}{\Lambda^2}\right)$\\
~~\\
~~\\
~~\\
~~\\
~~}
& \parbox[t]{3.1cm}{ 
  ${\cal L}^{(2)\prime} $ \\
  ${\cal L}_1$\\
  ${\cal L}_{{\rm MI},1\sim 5,8,9,11\sim 14}$\\
  ${\cal L}_{{\rm MI},1\sim 9,11\sim 14}$\\
  ${\cal L}_{{\rm MI},1,4,5,6,7,10}$\\
  ${\cal L}_{1,2,8,13,14}$\\
  ${\cal L}_{{\rm MI},1\sim 9,11\sim 14}$\\
  ${\cal L}_{{\rm MI},4,5,6,7,10}$\\
  ${\cal L}_{{\rm MI},1\sim 5,8,9,11\sim 14}$\\
  ${\cal L}_{{\rm MI},1\sim 9,11\sim 14}$\\
  ${\cal L}_{{\rm MI},4,5,6,7,10}$ } 
& 
\parbox[t]{6.4cm}{
$T_1[2V_L,2V_T],B^{(0)}_1\simeq T_1[3\pi ,v]~~^{(c)}$ \\
  $ T_1[2W_L,2W_T]~~^{(i)} $\\
  $ T_1[4W_T] $\\
  $ T_1[4V_T]~(\neq T_1[4W_T],T_1[4Z_T]) $\\
  $ T_1[4Z_T] $\\
{\footnotesize  $ B_1^{(0)}\simeq T_1[3\pi ,v]~~^{(c,j)} $\\
  $ B_1^{(0)}\simeq T_1[2\pi ,2v]~~^{(c,k)} $\\
  $ B_1^{(0)}\simeq 
                      T_1[2\pi ,2v](\neq T_1[2\pi^\pm ,2v^\pm ])~^{(g)} $ \\
  $ B_1^{(2)}\simeq T_1[\pi^\pm ,2W_T,v^\pm ] $\\
  $ B_1^{(2)}\neq T_1[\pi^\pm ,2W_T,v^\pm ],
     T_1[\pi^0 ,2Z_T,v^0] $\\
  $ B_1^{(2)}\simeq T_1[\pi^0 ,2Z_T,v^0] $ } }
&  
\parbox[t]{2.0cm}{ 
  MD\\
  MD\\
  MI$+$MD\\
  MI$+$MD\\
  MI$+$MD\\
  MD\\
  MI$+$MD\\
  MI$+$MD\\
  MI$+$MD\\
  MI$+$MD\\
  MI$+$MD  }\\
& & &  \\
\hline\hline 
\end{tabular}
\end{center}
\end{table}


\addtolength{\topmargin}{1.8cm}

\newpage
\addtolength{\topmargin}{-1.8cm}


\tabcolsep 1pt
\begin{table}[t]  

\tcaption{Global classification 
of sensitivities to probing direct and indirect EWSB\\ 
information from effective operators at the level 
of $S$-matrix elements (B). }

\vspace{0.5cm}

\small

\begin{tabular}{||c||c|c|c||} 
\hline\hline
& & &  \\
~Required Precision~
&  Relevant Operators
&  Relevant Amplitudes
&  MI or MD $^{(b)}$ \\
& & & ? \\
\hline\hline
& & & \\
     $O(g^2)$  
&    $~{\cal L}_{\rm MI}~(\equiv 
       {\cal L}_{\rm G}+{\cal L}_{\rm F}+{\cal L}^{(2)}) ~$
& $T_0[q\bar{q};V_LV_L],~ T_0[q\bar{q};V_TV_T]$
& MI \\
& & & \\
\hline
& & & \\
   \parbox[t]{2.5cm}{
        ~~\\
        ~~\\
$O\left(g^2\frac{E^2}{\Lambda^2},~g^3\frac{f_\pi}{E}\right)$ \\ 
        ~~\\          
        ~~ }
&  \parbox[t]{3.0cm}{ 
   ${\cal L}_{2,3,9}$\\
   ${\cal L}_{3,11,12}$\\
   ${\cal L}_{\rm MI}$\\ 
   ${\cal L}_{\rm MI}$\\
   ${\cal L}_{\rm MI}$ }
&  \parbox[t]{4.0cm}{  
   $T_1[q\bar{q};W_LW_L] $\\
   $T_1[q\bar{q};W_LZ_L] $ \\
   $T_0[q\bar{q};V_LV_T] $\\
   $T_1[q\bar{q};V_LV_L] $\\
 {\footnotesize  $B_0^{(1)}$}$\simeq T_0[q\bar{q};V_T,v]$ }
&  \parbox[t]{0.8cm}{ 
  MD\\
  MD\\
  MI\\
  MI\\
  MI }\\
& & & \\
\hline
& & & \\
\parbox[t]{2.5cm}{
~~\\
  $O\left(g^3\frac{Ef_\pi}{\Lambda^2},
   ~g^4\frac{f_\pi^2}{E^2}\right)$\\
~~   }
& \parbox[t]{3.3cm}{ 
${\cal L}_{1,2,3,8,9,11\sim 14}$\\
${\cal L}_{\rm MI}$ \\
${\cal L}_{\rm MI}$  }
& \parbox[t]{4.0cm}{ 
$T_1[q\bar{q};V_LV_T]$\\  
$T_1[q\bar{q};V_LV_T]$\\
{\footnotesize $B_0^{(0)}$}$\simeq T_0[q\bar{q};2v]~^{(c)}$  }
& \parbox[t]{0.8cm}{ MD \\
                     MI \\
                     MI }\\
& & & \\
\hline
& & & \\
\parbox[t]{2.0cm}{ 
~~\\
  $O\left((g^2,g^4)\frac{f_\pi^2}{\Lambda^2}\right)$\\
~~}
& \parbox[t]{3.1cm}{ 
  ${\cal L}^{(2)\prime} $ \\
  ${\cal L}_{1,2,3,8,9,11\sim 14}$\\
  ${\cal L}_{\rm MI}$ } 
& 
\parbox[t]{5.4cm}{$ T_1[q\bar{q};V_LV_L]$\\  
                  $ T_1[q\bar{q};V_TV_T]$,~
 {\footnotesize $B_1^{(0)}$}$\simeq T_1[q\bar{q};\pi ,v]$ \\  
  $ T_1[q\bar{q};V_TV_T]$,~
 {\footnotesize $B_1^{(0)}$}$\simeq T_1[q\bar{q};\pi ,v]$   }
&  \parbox[t]{0.8cm}{ 
  MD\\
  MD\\
  MI  }\\
& & &  \\
\hline\hline 
\end{tabular}
\end{table}
\begin{table}[t]
\begin{center}

\vspace{0.4cm}
\begin{tabular}{l}
{\footnotesize 
$^{(a)}$ The contributions from $~{\cal L}_{1,2,13}~$ are always associated
with a factor of $~\sin^2\theta_W~$, unless specified otherwise.}\\
{\footnotesize 
~~~$~{\cal L}_{4,5,6,7,10}~$ do not contribute to the 
processes considered in this table. 
Also, for contributions to the $B$-term}\\
{\footnotesize 
~~~~in a given $V_L$-amplitude, 
we list them separately with the $B$-term specified.}\\
{\footnotesize  $^{(b)}$ MI~$=$~model-independent, MD~$=$~model-dependent.
~~}\\
{\footnotesize  $^{(c)}$ Here, $~B_0^{(0)}~$ is dominated by 
$~T_0[q\bar{q};2v]~$ since $~T_0[q\bar{q};\pi,v]~$ contains a 
suppressing factor $\sin^2\theta_W$ as can be}\\
{\footnotesize 
~~~~deduced from $~T_0[q\bar{q};\pi V_T]~$ (cf. Table~3a) 
times the factor $~v^\mu = O\left(\frac{M_W}{E}\right)~$.}\\
\end{tabular}
\end{center}
\end{table}



\tcaption{ 
~{\small Probing the EWSB Sector at High Energy Colliders:
A Global Classification for the NLO Bosonic Operators}  \\[0.3cm]
 (~Notations: ~$\surd =~$Leading contributions, 
$~\triangle =~$Sub-leading contributions,~ 
and ~$\bot =~$Low-energy contributions.~
~Notes:~ $^{\dagger}$Here, $~{\cal L}_{13}$ or $~{\cal L}_{14}~$ 
does not contribute at $~O(1/\Lambda^2)~$.   ~~ $^\ddagger$At LHC($14$),  
$W^+W^+\ra W^+W^+$ should also be included.~)  }
 
\small
\vspace{0.3cm}

\begin{sideways}
\begin{tabular}{||c||c|c|c|c|c|c|c|c|c|c||c||c||} 
\hline\hline
& & & & & & & & & & & & \\
 Operators 
& $ {\cal L}^{(2)\prime} $ 
& $ {\cal L}_{1,13} $ 
& $ {\cal L}_2 $
& $ {\cal L}_3 $
& $ {\cal L}_{4,5} $
& $ {\cal L}_{6,7} $ 
& $ {\cal L}_{8,14} $ 
& $ {\cal L}_{9} $
& $ {\cal L}_{10} $
& $ {\cal L}_{11,12} $
& $T_1~\parallel  ~B$ 
& Processes \\
& & & & & & & & & & & & \\
\hline\hline
 LEP-I (S,T,U) 
& $\bot$ 
& $\bot~^\dagger$
&  
& 
& 
& 
& $\bot~^\dagger$
& 
&
&
& $g^4\frac{f^2_\pi}{\Lambda^2}$ 
& $e^-e^+\ra Z \ra f\bar{f}$\\ 
\hline
  LEP-II
& $\bot$ 
& $\bot$  
& $\bot$  
& $\bot$  
&  
& 
& $\bot$  
& $\bot$ 
&
& $\bot$  
& $g^4\frac{f^2_\pi}{\Lambda^2}$
& $e^-e^+ \ra W^-W^+$\\
\hline
  LC($0.5$)/LHC($14$)
& 
& 
& $\surd$
& $\surd$
& 
& 
& 
& $\surd$
&
& 
& $g^2\frac{E^2}{\Lambda^2} \parallel g^2\frac{M_W^2}{E^2}$
& $f \bar f\ra W^-W^+ /(LL)$\\  
& 
& $\triangle$
& $\triangle$
& $\triangle$
& 
& 
& $\triangle$
& $\triangle$
&
& $\triangle$
& $g^3\frac{Ef_\pi}{\Lambda^2} \parallel g^2\frac{M_W}{E} $ 
& $f \bar f\ra W^-W^+/(LT) $\\  
\hline
& 
& 
& 
& $\surd$
& $\surd$
& $\surd$
& 
& $\surd$
&
& $\surd$
& $g^2\frac{1}{f_\pi}\frac{E^2}{\Lambda^2}
  \| g^3\frac{M_W}{E^2} $
& $f \bar f\ra W^-W^+Z /(LLL) $\\
& 
& $\triangle$ 
& $\triangle$
& $\triangle$
& $\triangle$
& $\triangle$
& $\triangle$
& $\triangle$
&
& $\triangle$
& $g^3\frac{E}{\Lambda^2}\parallel g^3\frac{M_W^2}{E^3}$ 
& $f \bar f\ra W^- W^+ Z /(LLT)  $\\
& 
& 
&  
& $\surd$
& $\surd$
& $\surd$
& 
& 
& $\surd$
& 
& $g^2\frac{1}{f_\pi}\frac{E^2}{\Lambda^2}\parallel 
  g^3\frac{M_W}{\Lambda^2}$
& $f \bar f \ra ZZZ /(LLL) $\\
& 
& 
& 
& 
& $\triangle$
& $\triangle$
& 
& 
& $\triangle$
& 
& $g^3\frac{E}{\Lambda^2}\parallel
   g^3\frac{f_\pi}{\Lambda^2}\frac{M_W}{E}$ 
& $f \bar f \ra ZZZ  /(LLT)  $\\
 ~LC($1.5$)/LHC($14$)~ 
& 
& 
& 
& 
& $\surd$
& 
& 
&
& 
& 
& $\frac{E^2}{f_\pi^2}\frac{E^2}{\Lambda^2}\parallel g^2$ 
& $W^-W^\pm \ra W^-W^\pm /(LLLL)~^\ddagger$\\
&
& 
& 
& $\triangle$
& $\triangle$
& 
& 
& $\triangle$
&
& $\triangle$
& $g\frac{E}{f_\pi}\frac{E^2}{\Lambda^2}\parallel g^2\frac{M_W}{E}$ 
& $W^-W^\pm\ra W^-W^\pm /(LLLT)~^\ddagger$ \\
& 
& 
& 
& 
& $\surd$
& $\surd$
&
& 
&
& 
& $\frac{E^2}{f_\pi^2}\frac{E^2}{\Lambda^2}\parallel g^2 $
& $W^-W^+ \ra ZZ ~\&~{\rm perm.}/(LLLL)$ \\
&
& 
& $\triangle$
& $\triangle$
& $\triangle$
& $\triangle$
& 
& $\triangle$
&
& $\triangle$
& $g\frac{E}{f_\pi}\frac{E^2}{\Lambda^2}\parallel g^2\frac{M_W}{E}$ 
& $W^-W^+ \ra ZZ ~\&~{\rm perm.} /(LLLT)$ \\
& 
& 
& 
& 
& $\surd$
& $\surd$
& 
& 
& $\surd$ 
&  
& $\frac{E^2}{f_\pi^2}\frac{E^2}{\Lambda^2}\parallel
   g^2\frac{E^2}{\Lambda^2} $
& $ZZ\ra ZZ /(LLLL) $\\
&
& 
& 
& $\triangle$
& $\triangle$
& $\triangle$
&  
&
& $\triangle$
&
& $g\frac{E}{f_\pi}\frac{E^2}{\Lambda^2}\parallel
  g^2\frac{M_WE}{\Lambda^2}$ 
& $ZZ\ra ZZ /(LLLT) $\\
\hline
& 
& 
& 
& $\surd$
& 
& 
& 
&
& 
& $\surd$
& $g^2\frac{E^2}{\Lambda^2} \parallel  g^2\frac{M^2_W}{E^2}$
& $q\bar{q'}\ra W^\pm Z /(LL) $\\
& 
& $\triangle$
& $\triangle$
& $\triangle$
& 
& 
& $\triangle$
& $\triangle$
&
& $\triangle$
& $g^3\frac{Ef_\pi}{\Lambda^2}\parallel g^2\frac{M_W}{E}$ 
& $q\bar{q'}\ra W^\pm Z /(LT) $\\
 LHC($14$)
& 
& 
& 
& $\surd$
& $\surd$
& 
& 
& $\surd$
&
& $\surd$
& $g^2\frac{1}{f_\pi}\frac{E^2}{\Lambda^2}\parallel g^3\frac{M_W}{E^2}$
& $q \bar{q'}\ra W^-W^+W^\pm /(LLL) $\\
& 
& 
& $\triangle$
& $\triangle$
& $\triangle$
&
& $\triangle$
& $\triangle$
&
& $\triangle$
& $g^3\frac{E}{\Lambda^2}\parallel g^3\frac{M_W^2}{E^3}$ 
& $q \bar{q'} \ra W^- W^+W^\pm  /(LLT)  $\\
& 
& 
& 
& $\surd$
& $\surd$
& $\surd$
& 
& 
&
& $\surd$
& $g^2\frac{1}{f_\pi}\frac{E^2}{\Lambda^2}\parallel g^3\frac{M_W}{E^2}$
& $q \bar{q'}\ra W^\pm ZZ /(LLL) $\\
& 
& $\triangle$
& $\triangle$
& $\triangle$
& $\triangle$
& $\triangle$
& $\triangle$
& $\triangle$
&
& $\triangle$
& $g^3\frac{E}{\Lambda^2}\parallel  g^3\frac{M_W^2}{E^3}$ 
& $q \bar{q'} \ra W^\pm ZZ  /(LLT)  $\\
\hline
 LC($e^-\gamma$)
& 
& $\surd$
& $\surd$
& $\surd$
& 
& 
& $\surd$
& $\surd$
&
& $\surd$
& $eg^2\frac{E}{\Lambda^2}\parallel eg^2\frac{M_W^2}{E^3}$ 
& $e^-\gamma \ra \nu_e W^-Z,e^-WW /(LL)$\\
\hline
& 
& $\surd$
& $\surd$
& $\surd$
& 
& 
& $\surd$
& $\surd$
&
& 
& $e^2\frac{E^2}{\Lambda^2}\parallel e^2\frac{M_W^2}{E^2}$ 
& $\gamma \gamma \ra W^- W^+ /(LL)$\\
LC($\gamma \gamma $)
& 
& $\triangle$
& $\triangle$
& $\triangle$
& 
& 
& $\triangle$
& $\triangle$
&
& 
& $e^2g\frac{Ef_\pi}{\Lambda^2}\parallel e^2\frac{M_W}{E}$  
& $\gamma\gamma \ra W^-W^+ /(LT)$\\
& & & & & & & & & & & & \\
\hline\hline 
\end{tabular}
\end{sideways}
\clearpage


\newpage
\begin{figure}
\epsfig{file=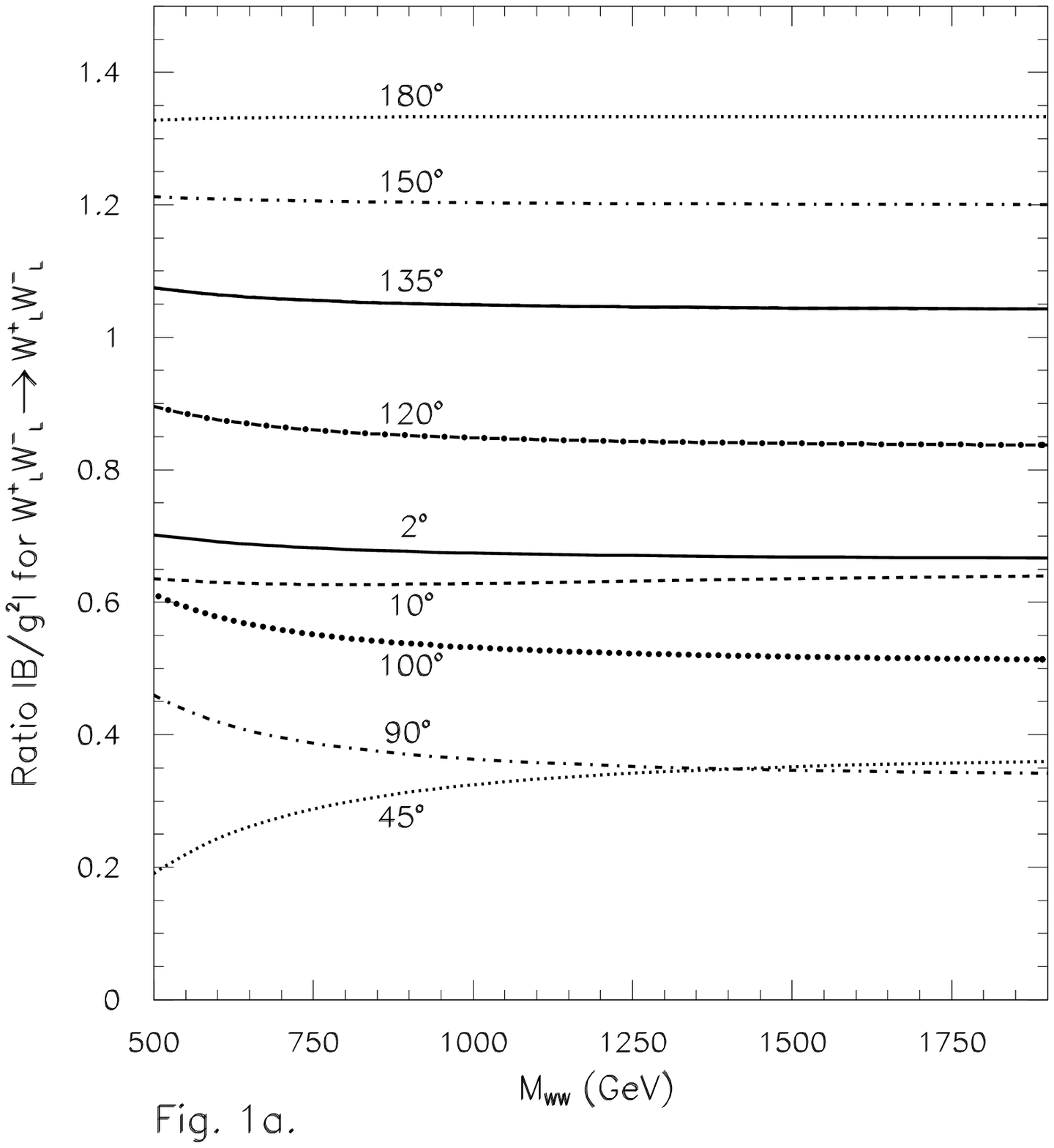,height=22cm,width=15.3cm}
\end{figure}

\newpage
\begin{figure}
\epsfig{file=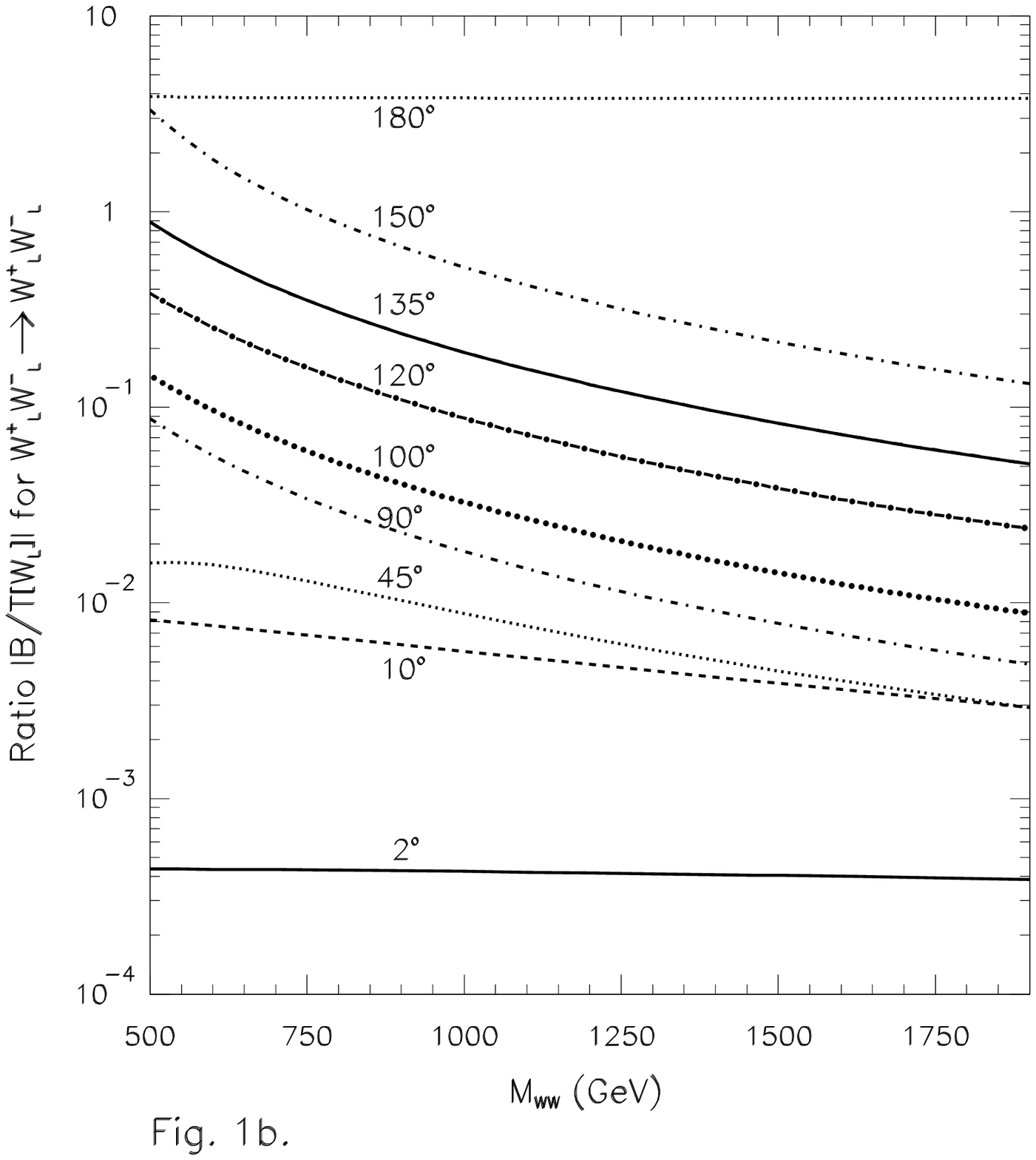,height=22cm,width=15.3cm}
\end{figure}

\newpage
\begin{figure}
\epsfig{file=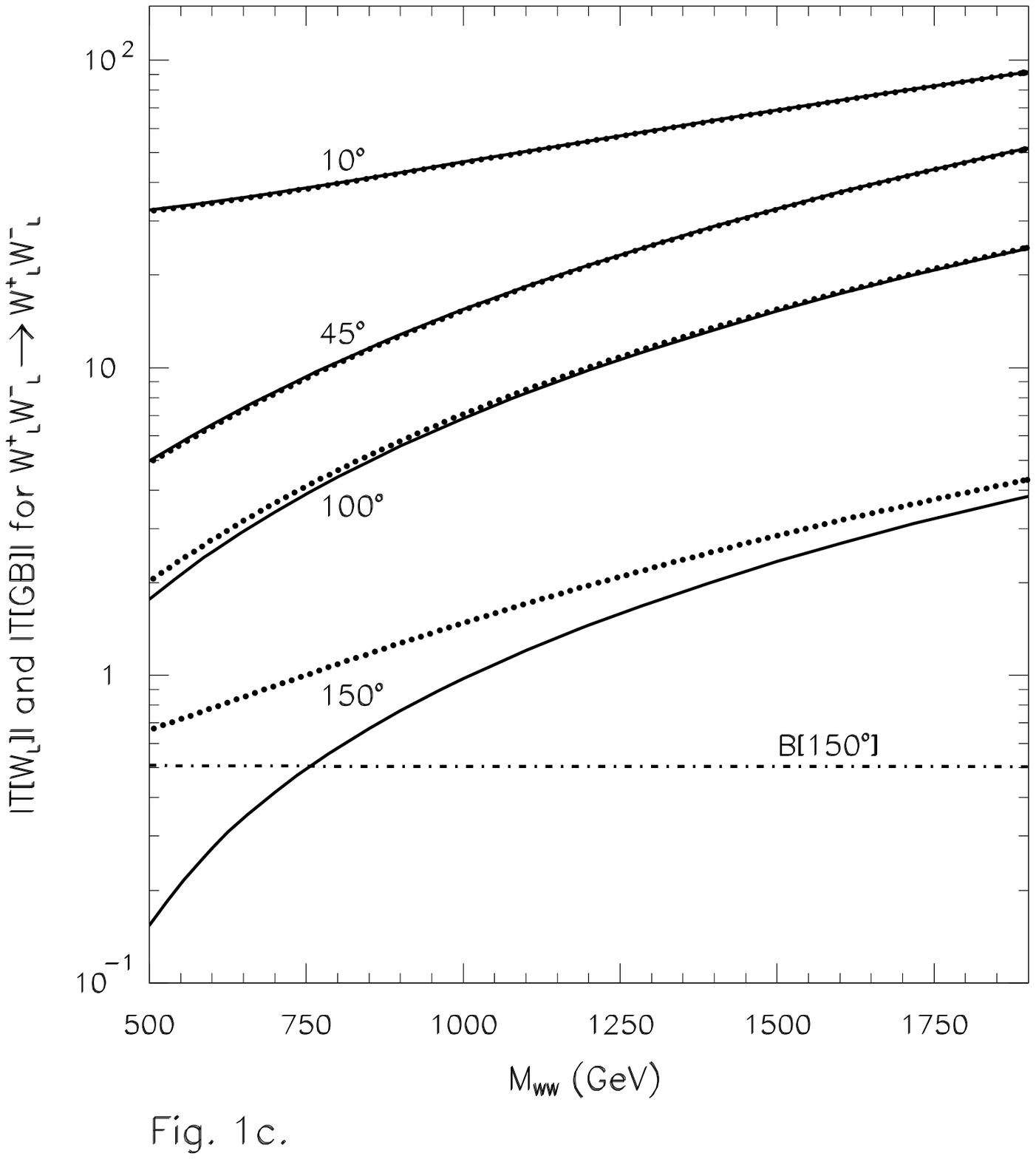,height=22cm,width=15.3cm}
\end{figure}

\newpage
\begin{figure}
\epsfig{file=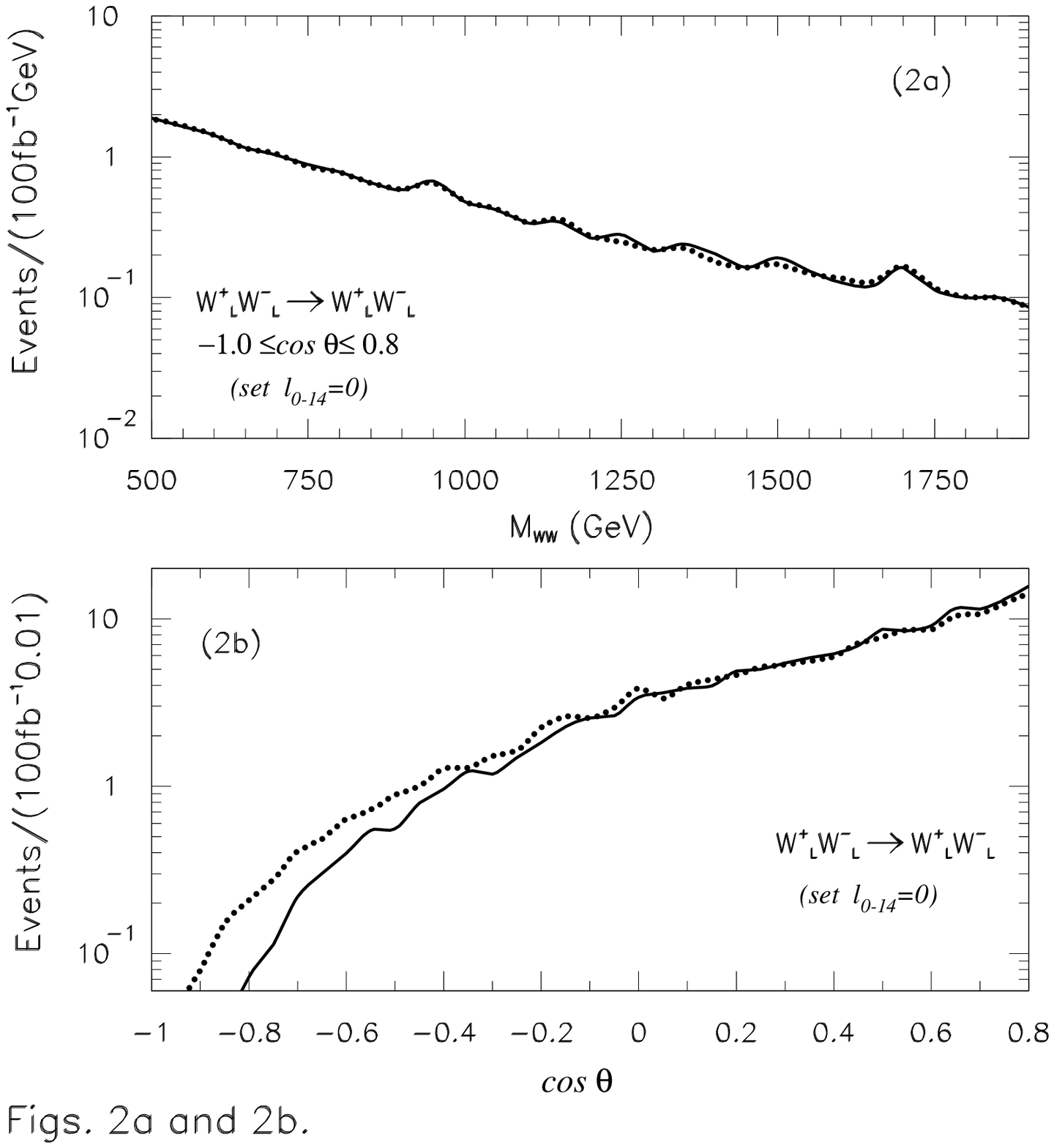,height=22cm,width=15.3cm}
\end{figure}

\newpage
\begin{figure}
\epsfig{file=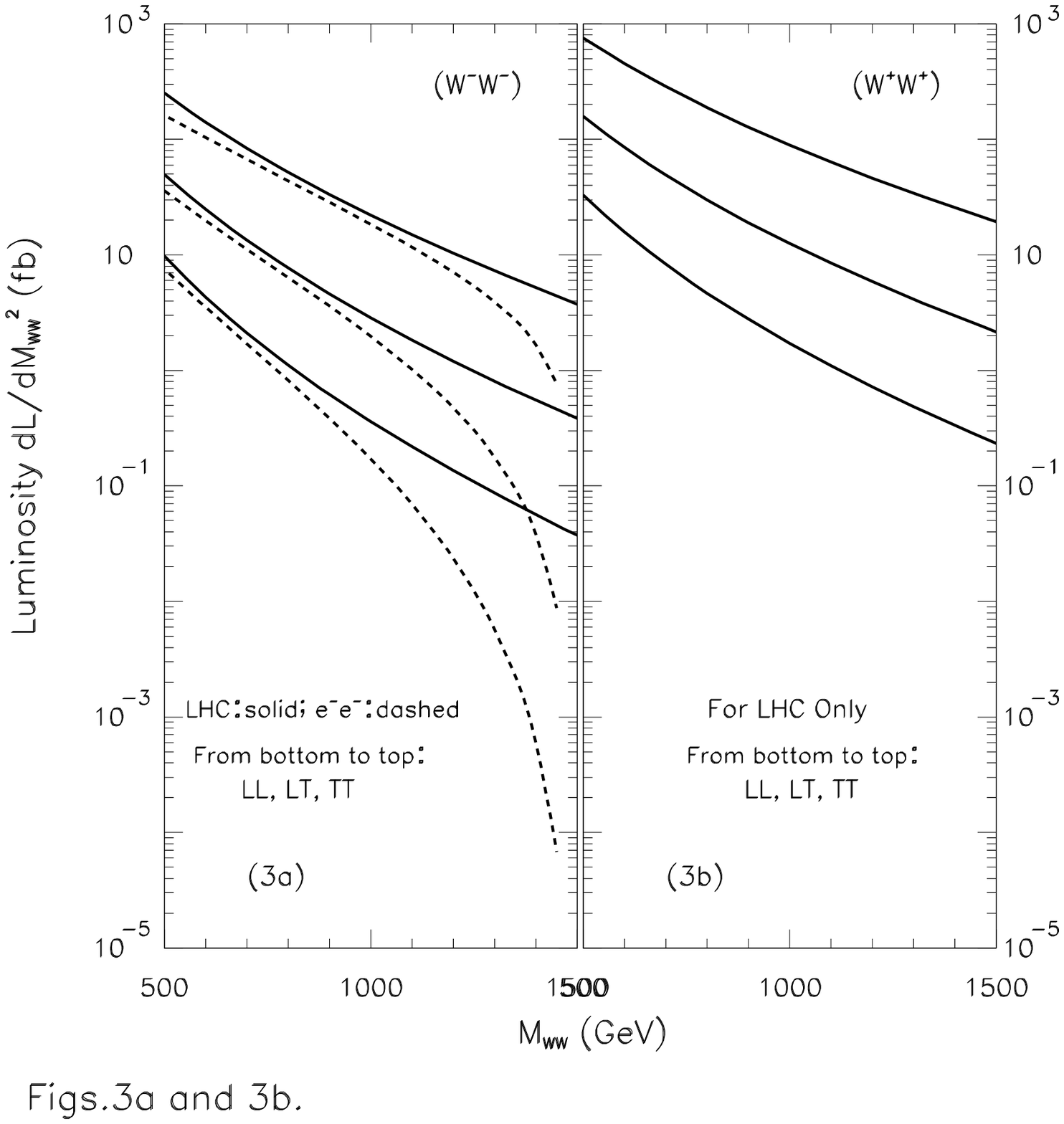,height=22cm,width=15.3cm}
\end{figure}

\newpage
\begin{figure}
\epsfig{file=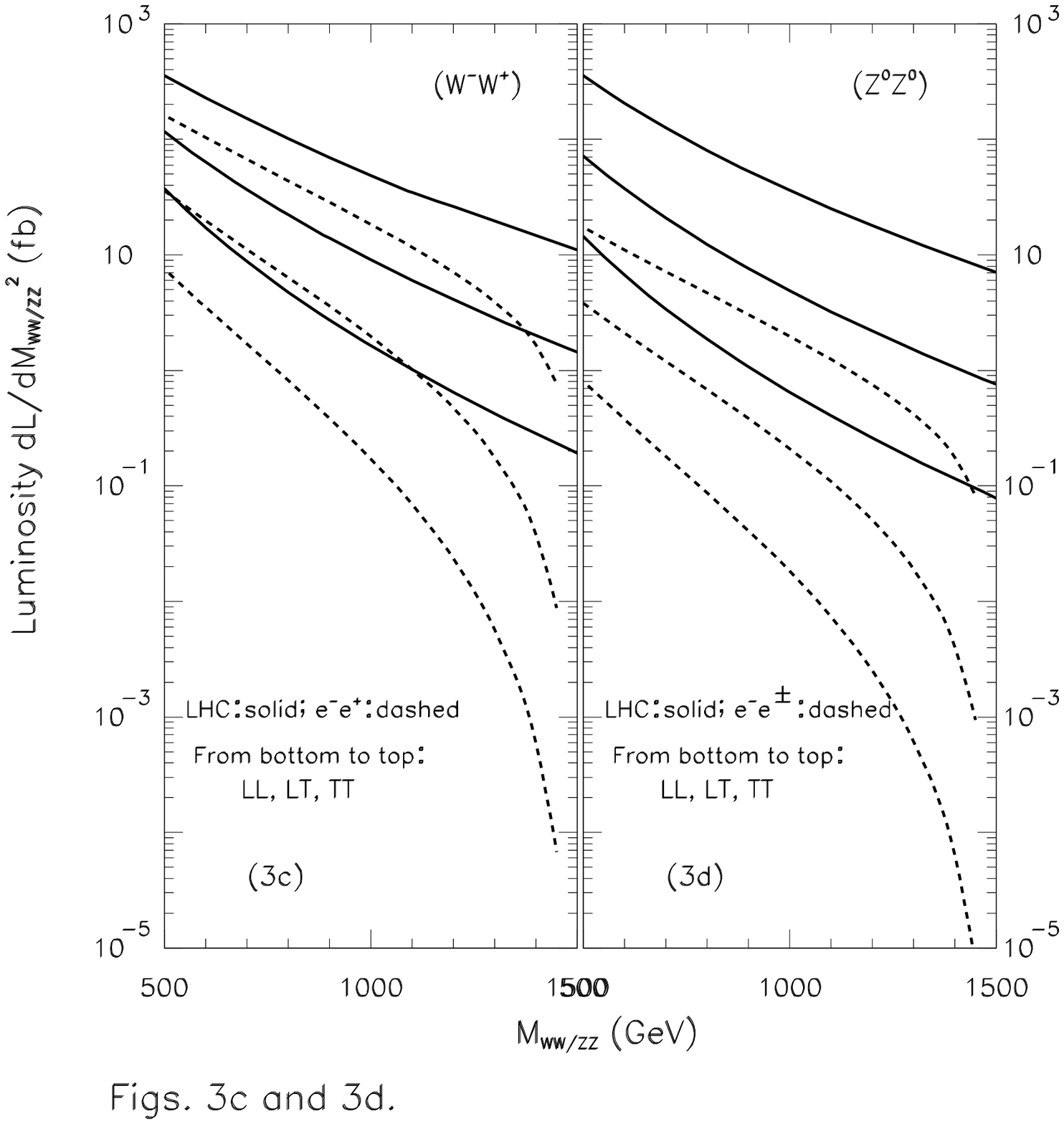,height=22cm,width=15.3cm}
\end{figure}

\newpage
\begin{figure}
\epsfig{file=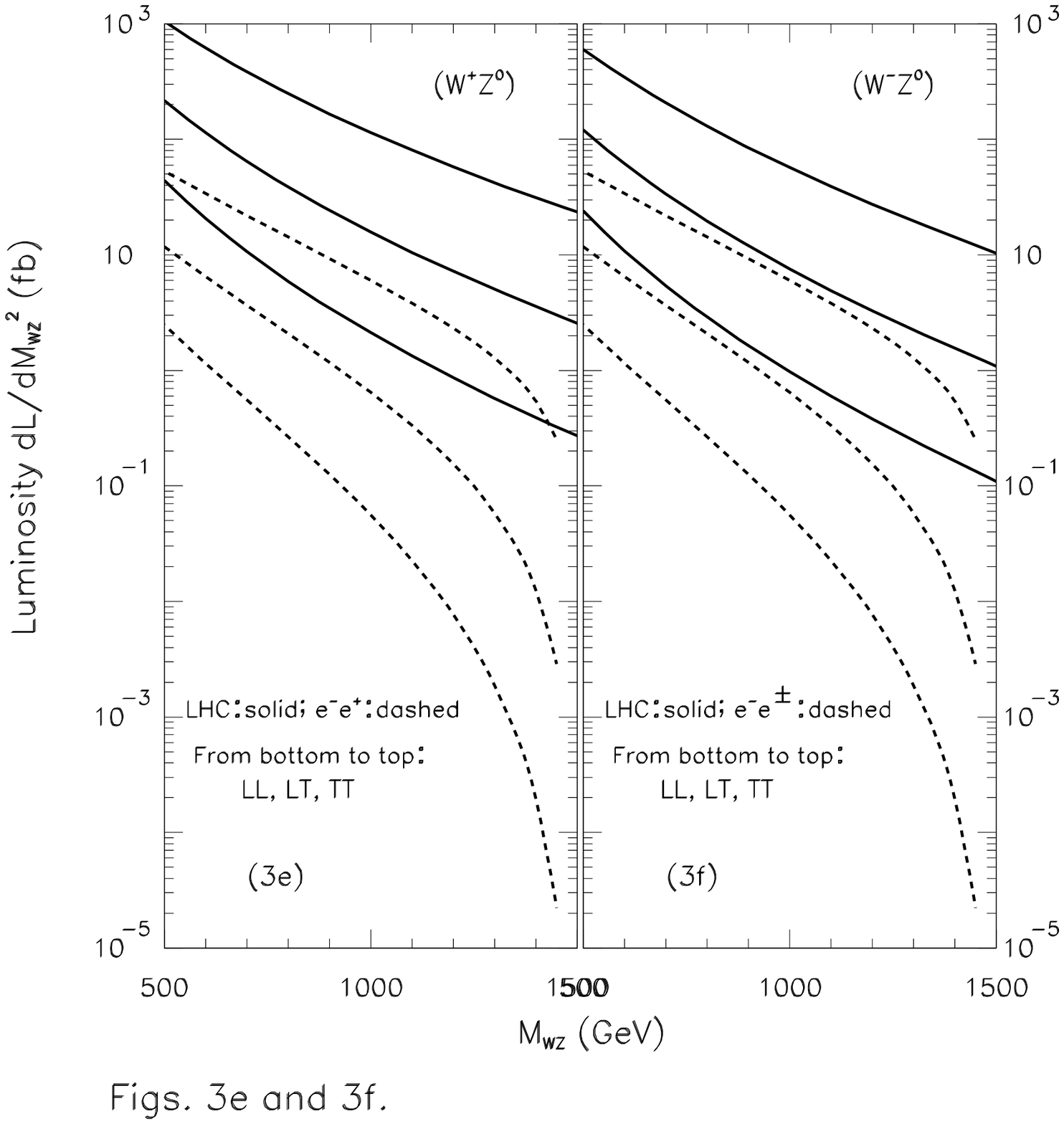,height=22cm,width=15.3cm}
\end{figure}

\newpage
\begin{figure}
\epsfig{file=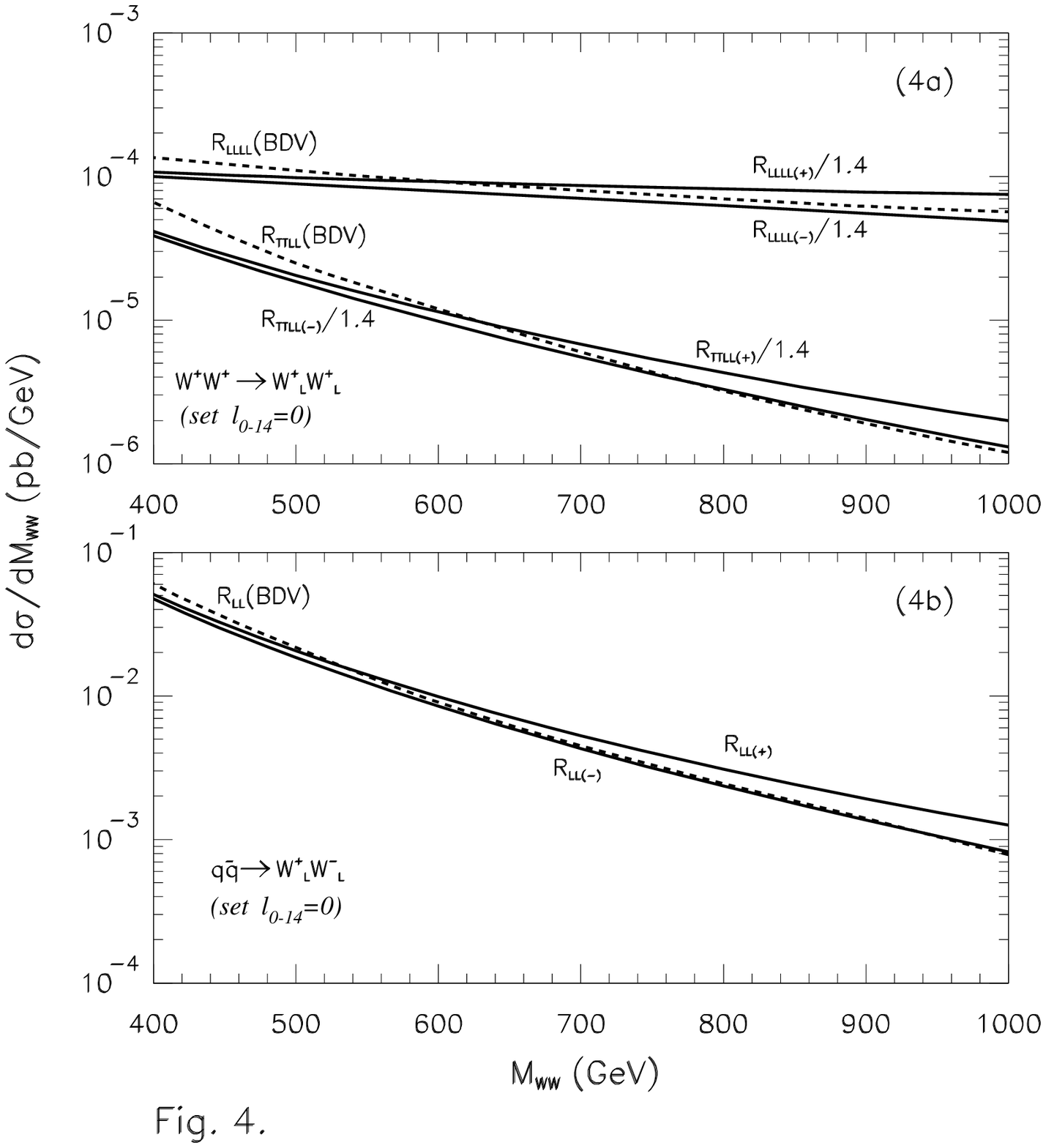,height=22cm,width=15.3cm}
\end{figure}

\newpage
\begin{figure}
\epsfig{file=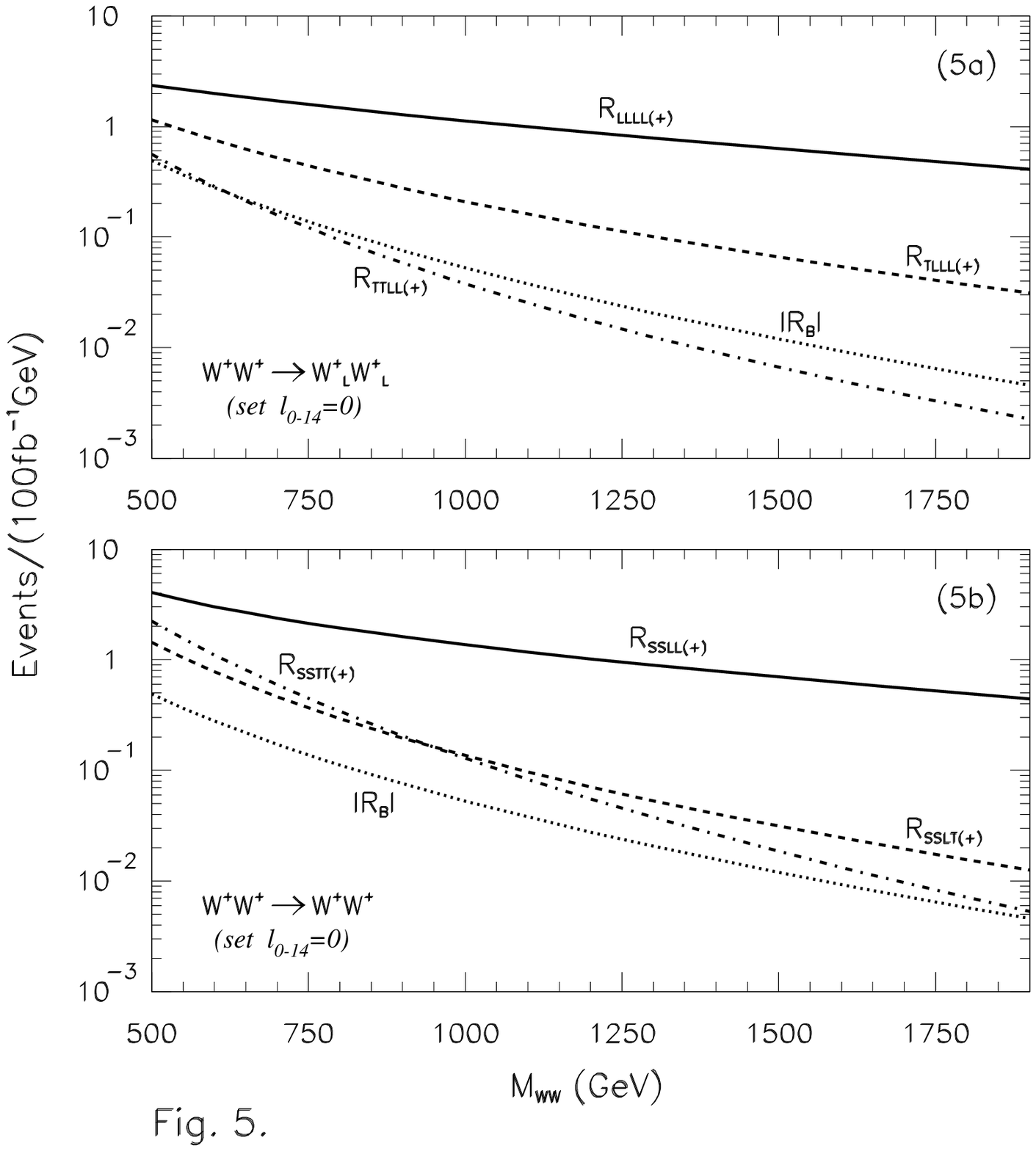,height=22cm,width=15.3cm}
\end{figure}

\newpage
\begin{figure}
\epsfig{file=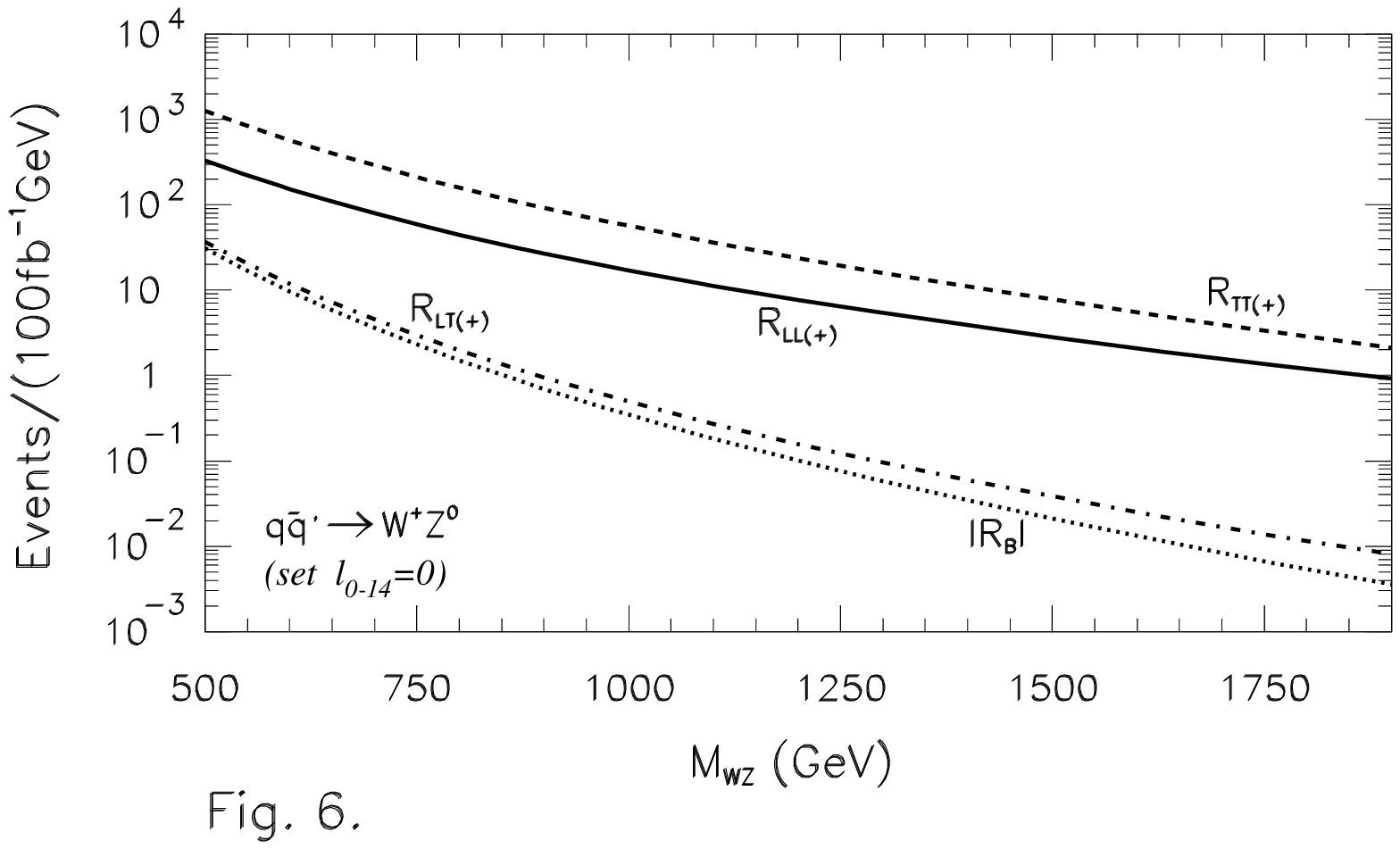,height=22cm,width=15.3cm}
\end{figure}

\newpage
\begin{figure}
\epsfig{file=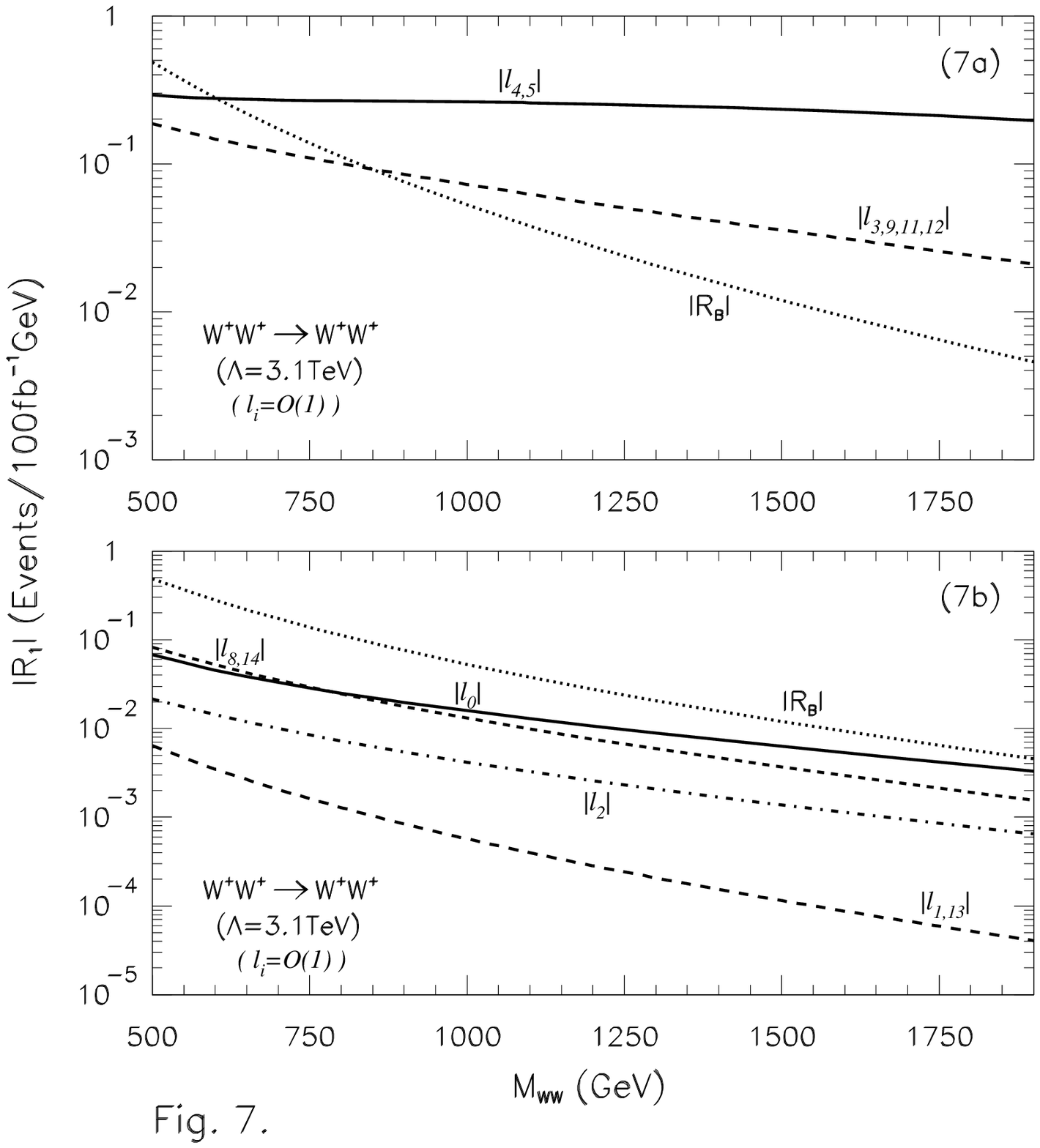,height=22cm,width=15.3cm}
\end{figure}

\newpage
\begin{figure}
\epsfig{file=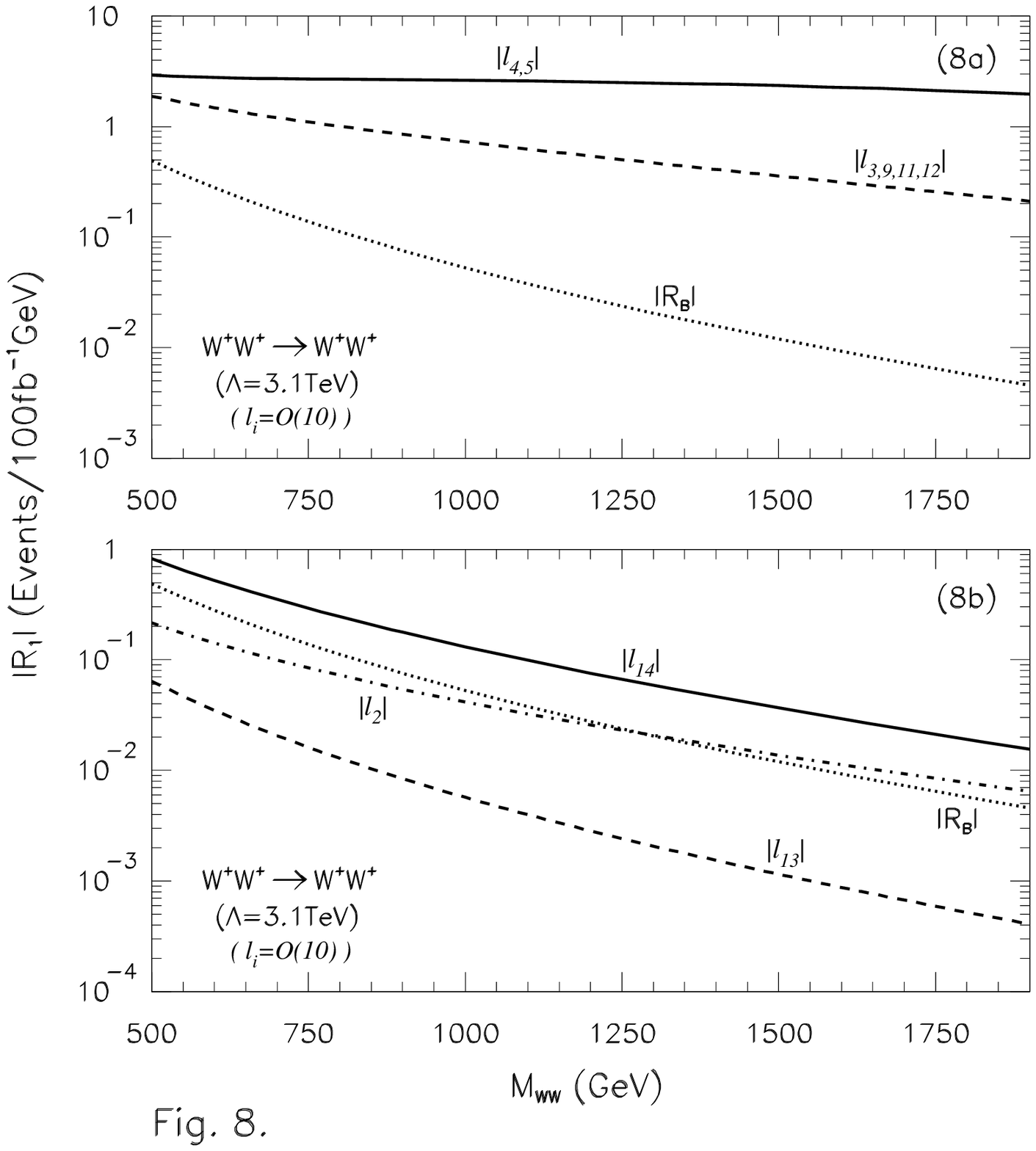,height=22cm,width=15.3cm}
\end{figure}

\newpage
\begin{figure}
\epsfig{file=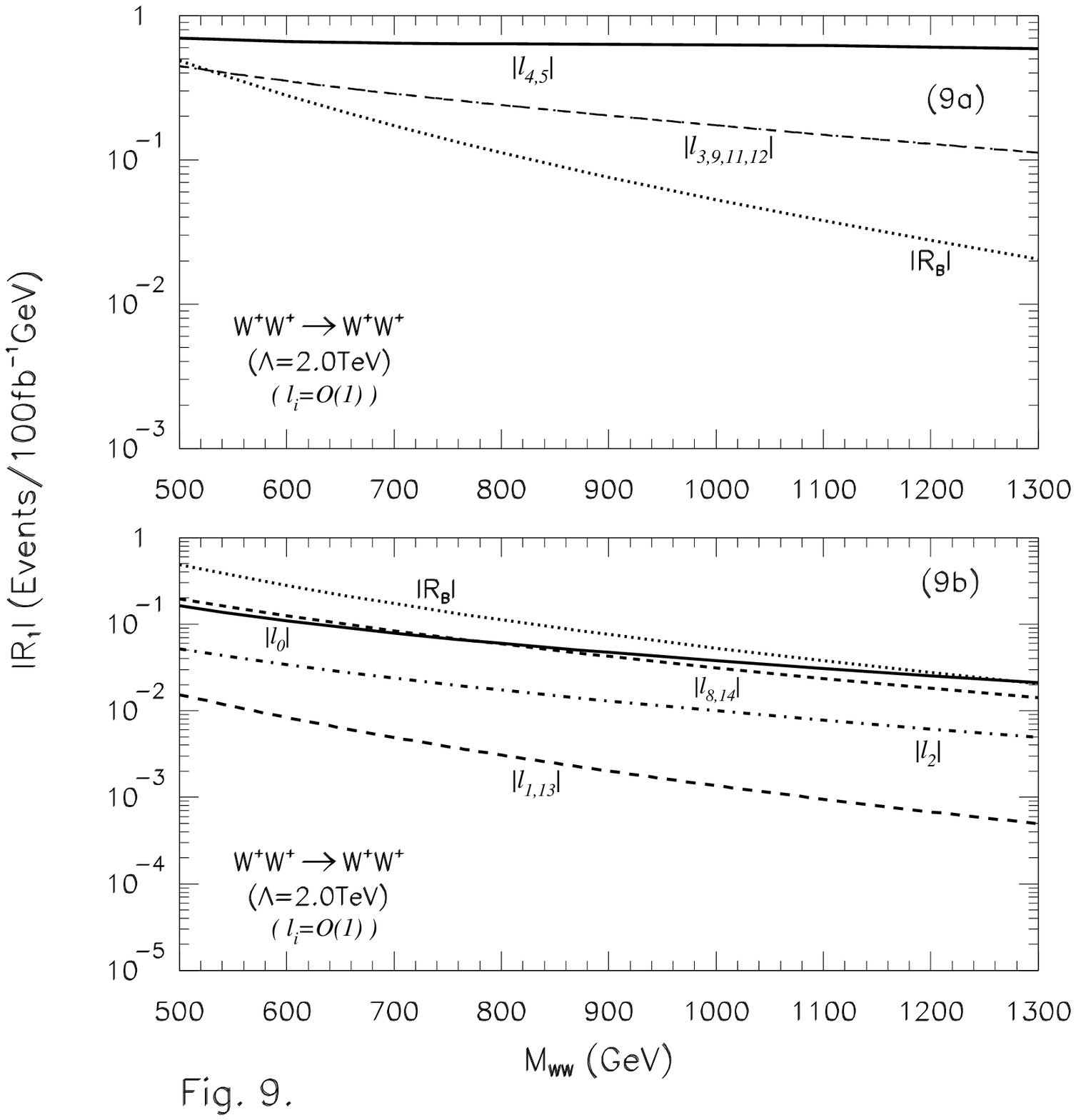,height=22cm,width=15.3cm}
\end{figure}

\newpage
\begin{figure}
\epsfig{file=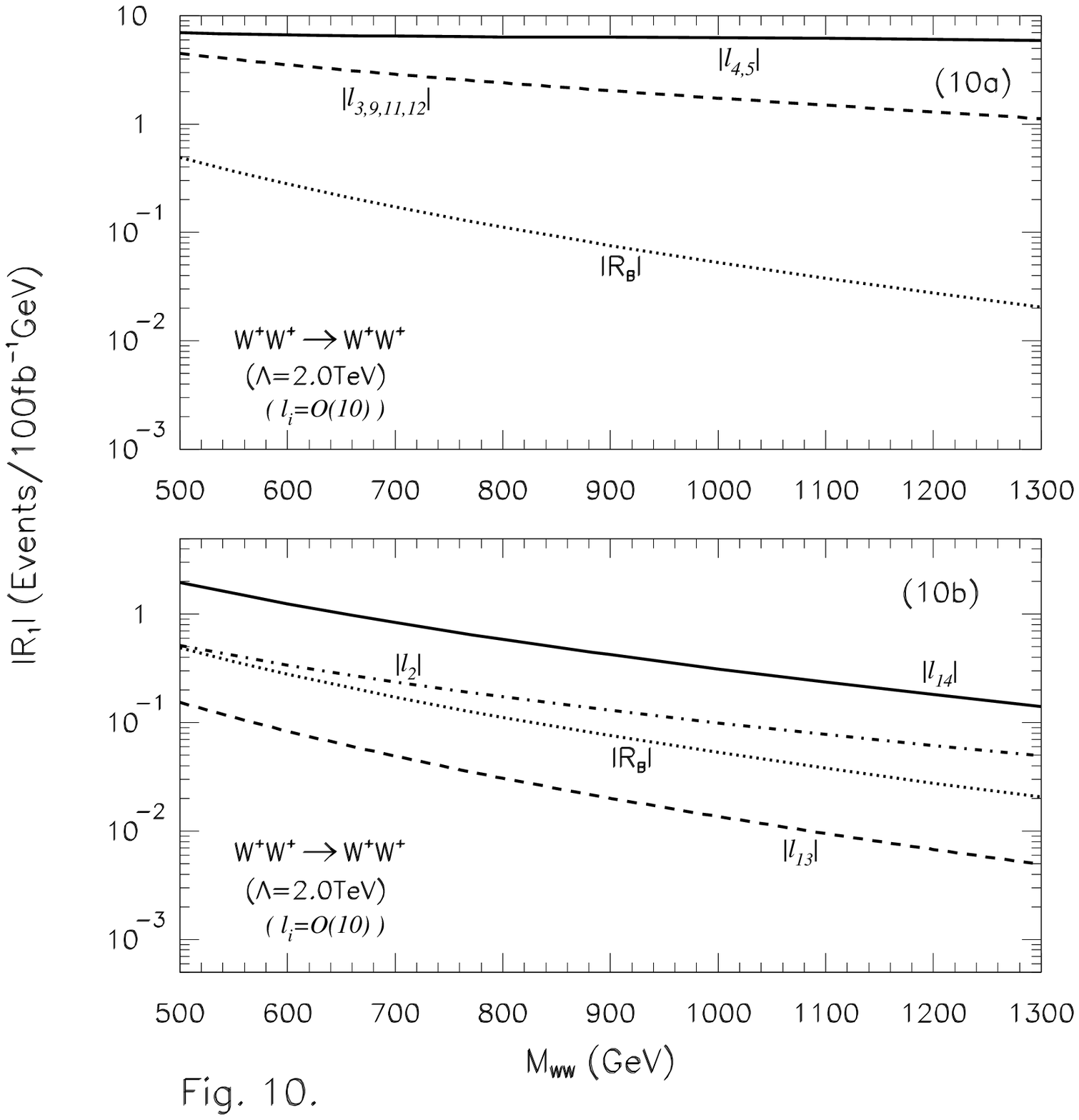,height=22cm,width=15.3cm}
\end{figure}

\newpage
\begin{figure}
\epsfig{file=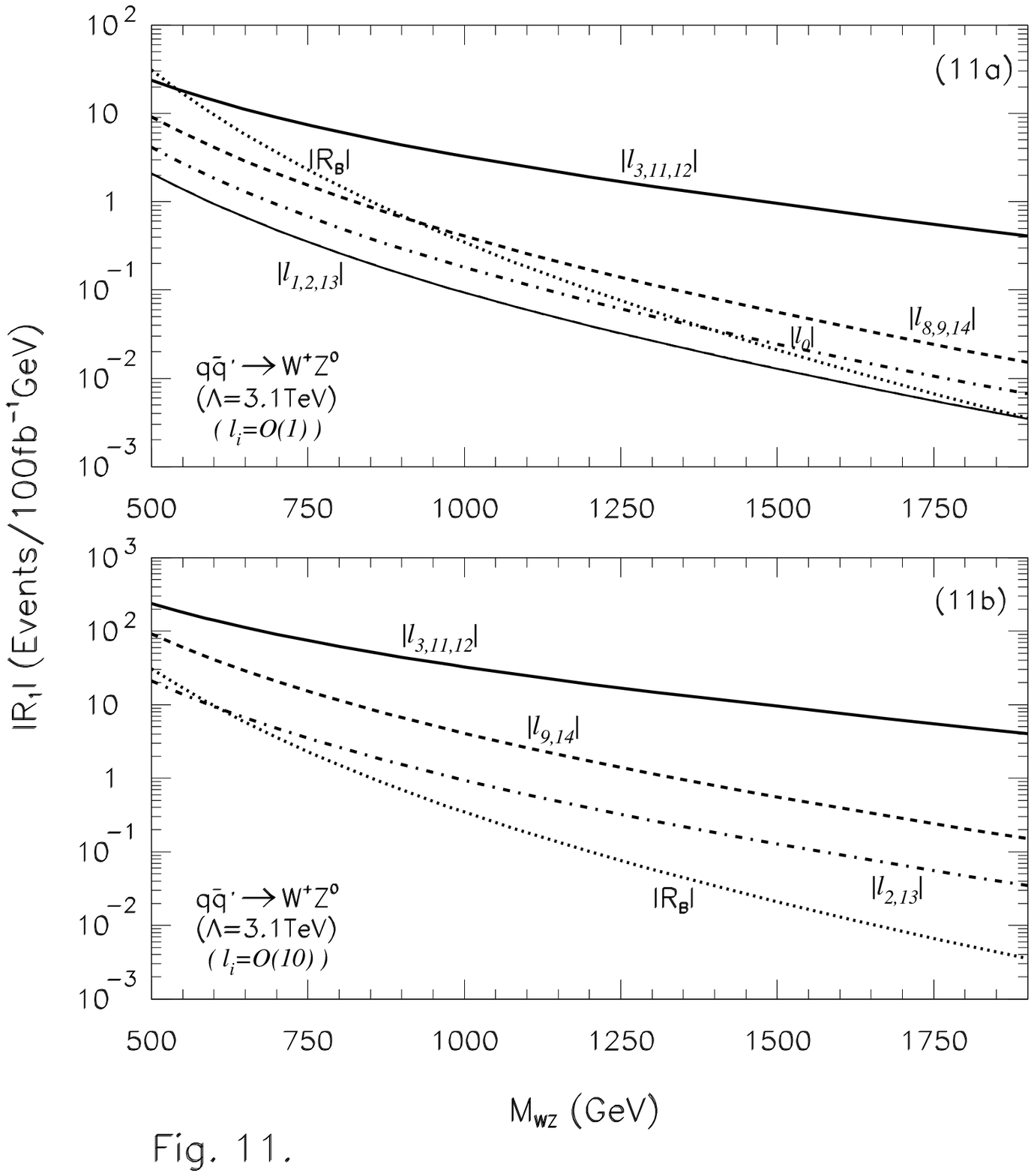,height=22cm,width=15.3cm}
\end{figure}

\newpage
\begin{figure}
\epsfig{file=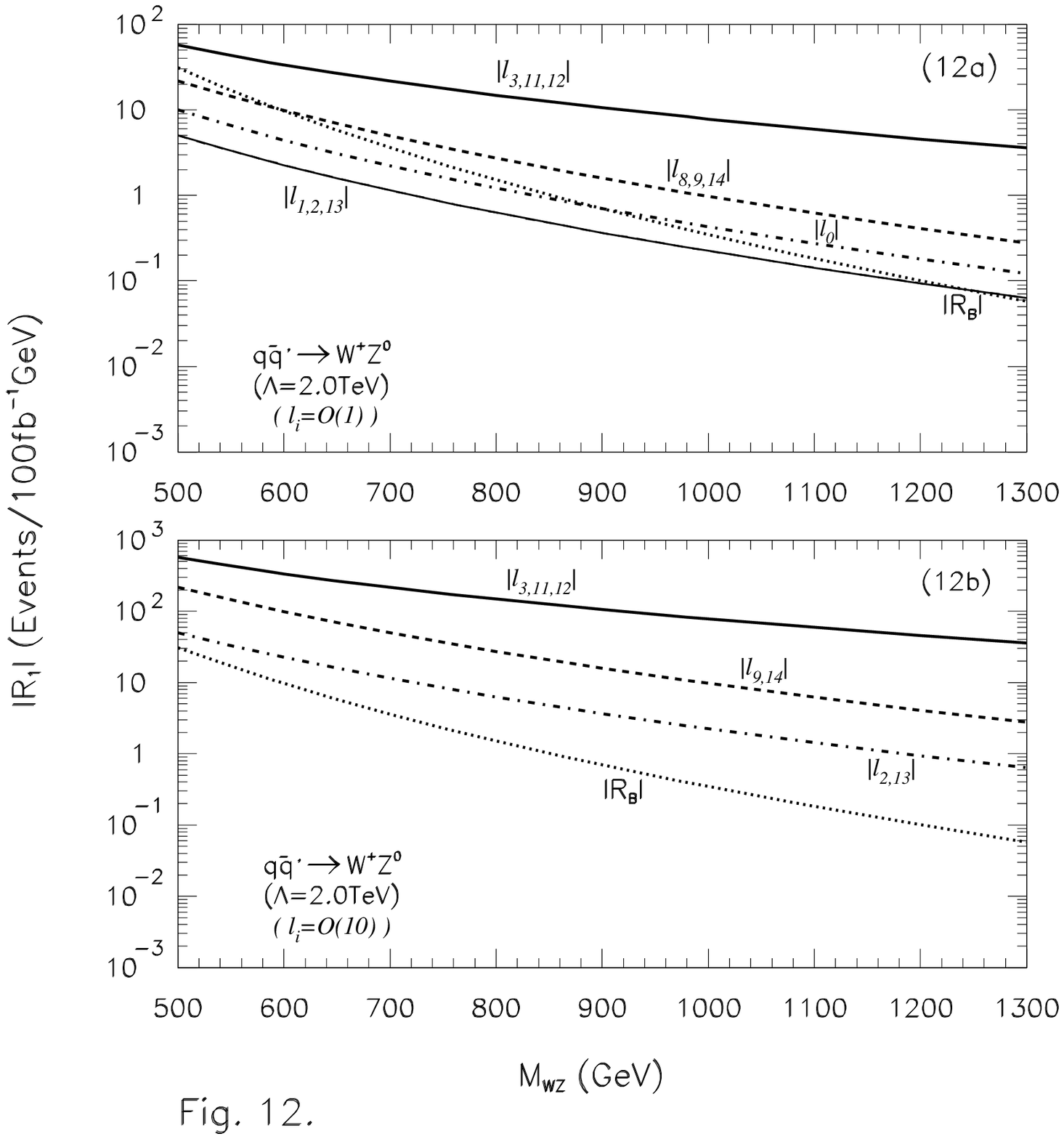,height=22cm,width=15.3cm}
\end{figure}


\noindent
{\bf References}
\begin{thebibliography}{200}

\bibitem{screening}
M. Veltman, {\it Acta. Phys. Pol.} {\bf B8}, 475 (1977).

\bibitem{LEP2}
 M. Carena and P.M. Zerwas (Conv.), {\it Higgs Physics At LEP2~,}
CERN-96-01, Report on Physics at LEP2, Vol.~1,
Ed. G. Altarelli et al, CERN.

\bibitem{discovery}
J. Gunion, et al, hep-ph/9703330, and Haber, et al, hep-ph/9703391, 
Summary Reports, June~25--July~12, 1996, Snowmass, CO.

\bibitem{wwlhc}
J. Bagger, V. Barger, K. Cheung, J. Gunion, T. Han,
G.A. Ladinsky, R. Rosenfeld, C.--P. Yuan,
{\it Phys. Rev.} {\bf D49}, 1246 (1994); ~{\bf D52}, 3878 (1995). 

\bibitem{wwnlc}
 V. Barger, J.F. Beacom, K. Cheung, T. Han, 
{\it Phys. Rev.} {\bf D50}, 6704 (1993).

\bibitem{lcws93}  
E.g., 
{\it Physics and Experiments with Linear $e^+e^-$ Colliders}, 
  Waikoloa, USA, 1993, Ed. F.A. Harris et al; and M.S. Chanowitz,
{\it Physics at High Energy $\gamma$-$\gamma$ Colliders,} 
{\it Nucl. Instr. $\&$ Meth.} {\bf A355}, 42 (1995);
 H. Murayama and M.E. Peskin,
{\it Ann. Rev. Nucl. Part. Sci.}, {\bf 46}, 533 (1996) and 
hep-ex/9606003;  S. Kuhlman et al 
(NLC ZDR Design Group and NLC Physics Working Group), 
SLAC-R-0485 and hep-ex/9605011. 

\bibitem{peskin-snow}
E.g., M. Peskin, Talk at the Snowmass Conference (June, 1996), and the
working group summary report, hep-ph/9704217.

\bibitem{wei}
S. Weinberg, {\it Physica} {\bf 96A}, 327 (1979).

\bibitem{global1}
H.-J. He, Y.-P. Kuang, and C.-P. Yuan,
{\it Phys. Rev.}       {\bf D55}, 3038 (1997) and hep-ph/9611316;
{\it Mod. Phys. Lett.} {\bf A11}, 3061 (1996) and hep-ph/9509278.

\bibitem{global2}
H.-J. He, Y.-P. Kuang and C.-P. Yuan, 
{\it Phys. Lett.} {\bf B382}, 149 (1996) and hep-ph/9604309;~
and MSUHEP-51201,  Published in Proc. of International Workshop on
{\it Physics and Experiments with Linear Colliders,} pp.~510-519, 
September 8-12, 1995, Iwate, Japan, and
in Proc. of Fermilab Linear Collider Workshop: 
 {\it Physics with High Energy $e^+e^-$ Colliders,}  
November 16-18, 1995, Batavia, USA.

\bibitem{et3}
H.-J. He, Y.-P. Kuang, and C.-P. Yuan,  
{\it Phys. Rev.} {\bf D51}, 6463 (1995); and\\
 hep-ph/9503359, Published in Proc. International Symposium on
{\it Beyond The Standard Model IV,} pp.~610, Eds. J.F. Gunion, T. Han,
J. Ohnemus, December 13-18, 1994, Tahoe, California, USA.

\bibitem{app}
T. Appelquist and C. Bernard, {\it Phys. Rev.} {\bf D22}, 200 (1980);
A.C. Longhitano, {\it Nucl. Phys.} {\bf B188}, 118 (1981);
T. Appelquist and G.-H. Wu, {\it Phys. Rev.} {\bf D48}, 3235 (1993);
{\bf D51}, 240 (1995);  and references therein.

\bibitem{peccei}
E. Malkawi and C.-P. Yuan, {\it Phys. Rev.} {\bf D50}, 4462 (1994);
F. Larios, E. Malkawi, and C.-P. Yuan, hep-ph/9609482;
R.D. Peccei, S. Peris and X. Zhang, 
{\it Nucl. Phys.} {\bf B349}, 305 (1991); and references therein.

\bibitem{georgi}
H. Georgi, {\it Weak Interaction and Modern Particle Theory}, 
Benjamin Publishing Company, 1984;  
A. Manohar and H. Georgi, {\it Nucl. Phys.} {\bf B234}, 189 (1984).

\bibitem{eom}
See: C. Arzt, {\it Phys. Lett.} {\bf B342}, 189 (1995); 
and references therein.

\bibitem{review}
For recent reviews, 
H. Georgi, {\it ~Ann. Rev. Nucl. $\&$ Part. Sci.,} {\bf 43}, 209 (1994);
F. Feruglio, {\it Inter. J. Mod. Phys.} {\bf A8}, 4937 (1993).

\bibitem{STU}
M. Peskin and T. Takeuchi, {\it Phys. Rev. Lett.} {\bf 65}, 964 (1990).

\bibitem{PDG}
P. Langacker and J. Erler,  hep-ph/9703428 (updated); and
{\it Phys. Rev.} {\bf D54}, 103 (1996).

\bibitem{warsaw}
A. Blondel, Plenary talk at 28th International Conference on High Energy 
Physics, Warsaw, July, 1996;
W. de Boer, A. Dabelstein, W. Hollik, W. M\"{o}sle, U. Schwickerath,
IEKP-KA/96-08 and hep-ph/9609209 (revised version).

\bibitem{long}
A.C. Longhitano, in Ref.~\cite{app}.

\bibitem{TGC}
H. Aihara, et al, hep-ph/9503425, published in {\it 
Electroweak Symmetry Breaking and Beyond the Standard Model}, Eds.
T.L. Barklow, et al; 
U. Baur, T. Han and J. Ohnemus, {\it Phys. Rev.} {\bf D53}, 1098 (1996);
and references therein.

\bibitem{TGC1}
T.L. Barklow, et al, DPF-Report, hep-ph/9505296; and 
G. Gounaris, J.L. Kneur, and D. Zeppenfeld, hep-ph/9601233; 
and references therein.

\bibitem{app2}
T. Appelquist and G.-H. Wu, in Ref.~\cite{app}. 

\bibitem{dv}
S. Dawson and G. Valencia, {\it Nucl. Phys.} {\bf B439}, 3 (1995).

\bibitem{ebo}
A. Brunstein, O.J.P. Eboli, M.C. Gonzalez-Garcia, 
{\it Phys. Lett.} {\bf B375}, 233 (1996); 
O.J.P. Eboli, et al, {\it Phys. Lett.} {\bf B339}, 119 (1995).

\bibitem{hveltman}
See, {\it e.g.}, H. Veltman, in Ref.~\cite{et-extra} 
and C. Grosse-Knette, in Ref.~\cite{gk}.

\bibitem{georgi1}
H. Georgi, {\it ~Ann. Rev. Nucl. $\&$ Part. Sci.,} {\bf 43}, 209 (1994);
C.P. Burgess and D. London, {\it Phys. Rev.}  {\bf D48}, 4337 (1993);
{\it Phys. Rev. Lett.} {\bf 69}, 3428 (1992).

\bibitem{let}
M.S. Chanowitz, H. Georgi and M. Golden, {\it Phys. Rev. Lett.} {\bf 57},
616 (1987); {\it Phys. Rev.} {\bf D36}, 1490 (1987).


\bibitem{linear}
W. Buchm\"{u}ller and D. Wyler, {\it Nucl. Phys.} {\bf B268}, 621 (1986).

\bibitem{HK}
P.W. Higgs, {\it Phy. Rev. Lett.} {\bf 13},  508(1964); 
            {\it Phys. Rev.}      {\bf 145}, 1156(1966);
G.S. Guralnik, C.R. Hagen, and T.W.B. Kibble, 
{\it Phys. Rev. Lett.} {\bf 13}, 585 (1984);
T.W.B. Kibble, {\it Phys. Rev.} {\bf 155}, 1554 (1967).

\bibitem{et-tree}                                                        
J.M. Cornwall, D.N. Levin, and G. Tiktopoulos, 
{\it Phys. Rev.}  {\bf D10}, 1145 (1974);\\
C.E. Vayonakis, {\it Lett. Nuovo. Cimento} {\bf 17}, 383 (1976);\\  
B.W. Lee, C. Quigg, and H. Thacker, 
{\it Phys. Rev.} {\bf D16}, 1519 (1977).

\bibitem{et1}
M.S. Chanowitz and M.K. Gaillard, {\it Nucl. Phys.} {\bf B261}, 379 (1985);
G.J. Gounaris, R. K\"{o}gerler, and H. Neufeld, 
{\it Phys. Rev.} {\bf D34}, 3257 (1986).

\bibitem{YY-BS}
Y.-P. Yao and C.-P. Yuan, {\it Phys. Rev.}  {\bf D38}, 2237 (1988);\\
J. Bagger and C. Schmidt, {\it Phys. Rev.}  {\bf D41}, 264 (1990).

\bibitem{et2}
H.-J. He, Y.-P. Kuang, and X. Li, 
{\it Phys. Rev. Lett.} {\bf 69}, 2619 (1992);\\
{\it Phys. Rev.} {\bf D49}, 4842 (1994); 
{\it Phys. Lett.} {\bf B329}, 278 (1994).

\bibitem{et-new}
H.-J. He and W.B. Kilgore, 
 {\it Phys. Rev.} {\bf D55}, 1515 (1997) and hep-ph/9609326.

\bibitem{et-extra}
E.g., 
S.D. Willenbrock, {\it Ann. Phys.} {\bf 186}, 15 (1988);
H. Veltman, {\it Phys. Rev.} {\bf D41}, 2294 (1990);
W.B. Kilgore, {\it Phys. Lett.} {\bf B294}, 257 (1992);
A. Dobado, J.R. Pelaez, {\it ibid}, 
{\bf B329}, 469 (1994) (Erratum, B335, 554);
C. Grosse-Knetter, I. Kuss, {\it Z. Phys.} {\bf C95}, 66 (1995);
J.F. Donoghue, J. Tandean, {\it Phys. Lett.} {\bf B361}, 69 (1995);
J. Horejsi, PRA-HEP 95/9;
A. Denner, S. Dittmaier, BI-TP-96/02 and hep-ph/9603341.

\bibitem{etIR}
T. Torma, {\it Phys. Rev.} {\bf D54}, 2168 (1996) and hep-ph/9510217.

\bibitem{BRST}
C. Becchi, A. Rouet, B. Stora, 
{\it Comm. Math. Phys.} {\bf 42}, 127 (1975); 
{\it Ann. Phys.} {\bf 98}, 287 (1976).

\bibitem{STID}
G.J. Gounaris, R. K\"{o}gerler, and H. Neufeld, in Ref.~\cite{et1}.

\bibitem{Higgs-ET}
H.-J. He, Y.-P. Kuang, and X. Li, {\it Phys. Rev.} {\bf D49}, 4842 (1994).

\bibitem{1/N} 
E.g., M.B. Einhorn, {\it Nucl. Phys.} {\bf B246}, 75 (1984);
and references therein.

\bibitem{gk}
See, {\it e.g.}, C. Grosse-Knetter, BI-TP/25; 
      D. Espriu et al, UE-CBM-PF-94/61.

\bibitem{effective-W}
R.N. Cahn and S. Dawson, {\it Phys. Lett.}
{\bf B136}, 196 (1984), {\bf B138}, 464(E) (1984); \\
M.S. Chanowitz and M.K. Gaillard, {\it Phys. Lett.} {\bf B142},
85 (1984);\\
G.L. Kane, W.W. Repko and W.R. Rolnick, {\it Phys. Lett.}
 {\bf B148}, 367 (1984); \\
S. Dawson, {\it Nucl. Phys.} {\bf B249}, 427 (1985);\\
J. Lindfors, {\it Z. Phys.} {\bf C28}, 427 (1985);\\
W.B. Rolnick, {\it Nucl. Phys.} {\bf B274}, 171 (1986);\\
P.W. Johnson, F.I. Olness and W.-K. Tung, {\it Phys. Rev.}
 {\bf D36}, 291 (1987);\\
Z. Kunszt and D.E. Soper, {\it Nucl. Phys.} {\bf B296}, 253 (1988);\\
A. Abbasabadi, W.W. Repko, D.A. Dicus and R. Vega,
{\it Phys. Rev.} {\bf D38}, 2770 (1988);
S. Dawson, {\it Phys. Lett.} {\bf B217}, 347 (1989); \\
 S. Cortese and R. Petronzio,  {\it Phys. Lett.} {\bf B276}, 203 (1992);\\
I. Kuss and H. Spiesberger, {\it Phys. Rev.} {\bf D53}, 6078 (1996).

\bibitem{mike-EWA}
M.S. Chanowitz, private communications, and 
{\it Phys. Lett.} {\bf B373}, 141 (1996) and hep-ph/9512358.

\bibitem{mike-ww}
M.S. Chanowitz and W.B. Kilgore, {\it Phys. Lett.} {\bf B322}, 147 (1994).

\bibitem{wwww}
C.-P. Yuan, published in {\it Perspectives on Higgs Physics,}
edited by Gordon L. Kane, World Scientific, 1992;~
G.L.  Kane and C.-P. Yuan, {\it Phys. Rev.} {\bf D40}, 2231 (1989).

\bibitem{eAback}
K. Cheung, S. Dawson, T. Han, 
G. Valencia, {\it Phys. Rev.} {\bf D51}, 5 (1995).

\bibitem{DPF}
E.g., M. Golden, T. Han, G. Valencia, hep-ph/9511206, October, 1995.

\bibitem{desy}
H.-J. He, DESY-97-037 and invited talk in the proceedings of 
`` {\it The Higgs Puzzle---What can 
we learn from LEPII, LHC, NLC, and FMC?} '', 
Ringberg Castle, Munich, Dec.~8-13, 1996, Ed. Bernd Kniehl; 
E. Boos, H.-J. He, W. Kilian, A. Pukhov, C.-P. Yuan, and P.M. Zerwas, 
DESY-96-256.

\bibitem{nda}
H. Georgi, {\it Weak Interaction and Modern Particle Theory}, 
(Benjamin Publishing Company, 1984); 
A. Manohar and H. Georgi, {\it Nucl. Phys.} {\bf B234}, 189 (1984);\\
H. Georgi, {\it Ann. Rev. Nucl. $\&$ Part. Sci.} {\bf 43}, 209 (1994).

\bibitem{gunion}
J.F. Gunion, J. Kalinowksi, A. Tofighi-Niaki, {\it Phys. Rev. Lett.}
 {\bf 57}, 2351 (1986).

\bibitem{EPA}
E.g., S.J. Brodsky and P.M. Zerwas,
`` {\it High Energy Photon-Photon Collisions} '',
{\it Nucl. Instr. $\&$ Meth.} {\bf A355}, 19 (1995). 
V.M. Budnev, et al, {\it Phys. Rep.} {\bf C15}, 182 (1975).

\bibitem{BDV}
J. Bagger, S. Dawson and G. Valencia,  
{\it Nucl. Phys.} {\bf B399}, 364 (1993).

\bibitem{lcllll}
For studies on the purely hadronic mode of $WW$ at lepton colliders, see \\
V. Barger, K. Cheung, T. Han, R.J.N. Phillips,
{\it Phys. Rev.} {\bf D52}, 3815 (1995);\\
V. Barger, M.S. Berger, J.F. Gunion, T. Han,
{\it Phys. Rev.} {\bf D55}, 142 (1997).
 
\bibitem{light}
For a review, see:
J.F. Gunion, H.E. Haber, G.L. Kane and S. Dawson, 
{\it The Higgs Hunter's Guide}, (Addison-Wesley Pub. Company, 1990);
and references therein.

\bibitem{ZZllnn}
R.N. Cahn and M.S. Chanowitz, {\it Phys. Rev. Lett.} 
{\bf 56}, 1327 (1986);\\
U. Baur and E.W.N. Glover, {\it Phys. Rev.} {\bf D44}, 99 (1991), and
references therein.

\bibitem{llll}
For studies on the purely leptonic mode of $WW$ at hadron colliders, see \\
J. Bagger, V. Barger, K. Cheung, J. Gunion, T. Han, 
G. Ladinsky, R. Rosenfeld and C.--P. Yuan, 
 {\it Phys. Rev.} {\bf D49}, 1246 (1994); 
 {\it Phys. Rev.} {\bf D52}, 3878 (1995). 

\bibitem{wpwpll}
M.S. Chanowitz and M. Golden, {\it Phys. Rev. Lett.} {\bf 61} (1988)
1053; {\bf 63}, 466 (1989) (E);
V. Barger, K. Cheung, T. Han and R.J.N. Phillips, {\it Phys. Rev.} 
{\bf D42}, 3052 (1990), and references therein.

\bibitem{wpwpew}
R. Vega and D.A. Dicus, {\it Nucl. Phys.} {\bf B329}, 533 (1990).

\bibitem{gluonex}
D. Dicus and R. Vega, {\it Phys. Lett.} {\bf 217}, 194 (1989).

\bibitem{qqww}
K.O. Mikaelian, M.A. Samuel, D. Brown, {\it Nuovo. Cim. Lett.} {\bf 27},
211 (1980).

\bibitem{ggww}
C. Kao and D.A. Dicus, {\it Phys. Rev.} {\bf D43}, 1555 (1991).

\bibitem{wpwmew}
D.A. Dicus and R. Vega, {\it Phys. Rev. Lett.} {\bf 57}, 1110 (1986);
J.F. Gunion, J. Kalinowski and A. Tofighi-Niaki, 
{\it Phys. Rev. Lett.} {\bf 57}, 2351 (1986).

\bibitem{wwjet}
U. Baur, E.W.N. Glover and J.J. van der Bij, {\it Nucl. Phys.} 
{\bf B318}, 106 (1989);\\
V. Barger, T. Han, J. Ohnemus and D. Zeppenfeld, 
{\it Phys. Rev.} {\bf D41}, 2782 (1990), and references therein.

\bibitem{wtwoj}
S.D. Ellis, R. Kleiss and W.J. Stirling, {\it Phys. Lett.} {\bf 154B},
435 (1985); \\
J.F. Gunion, Z. Kunszt and M. Soldate, {\it Phys. Lett.} {\bf 163B}, 389
(1985); {\bf 168B}, 427 (1986) (E).


\bibitem{wthreej}
F.A. Berends, W.T. Giele, H. Kuijf, R. Kleiss and 
W.J. Stirling, {\it Phys. Lett.} {\bf B224}, 237 (1989);
V. Barger, T. Han, J. Ohnemus and D. Zeppenfeld, 
{\it Phys. Rev. Lett.} {\bf 62}, 1971 (1989); 
{\it Phys. Rev.} {\bf D40}, 2888 (1989); {\bf D41}, 1715 (1990) (E),
and references therein.

\bibitem{ttbaro}
B.C. Combridge, {\it Phys. Scr.} {\bf 20}, 5 (1979).

\bibitem{ttbarqcd}
P. Nason, S. Dawson and R.K. Ellis,
{\it Nucl. Phys.} {\bf B303}, 607 (1988); {\bf B327}, 49 (1989);
W. Beenakker, H. Kuijf, W.L. van Neerven and J. Smith,
{\it Phys. Rev.} {\bf D40}, 54 (1989);
R. Meng, G.A. Schuler, J. Smith and W.L. van Neerven,
{\it Nucl. Phys.} {\bf B339}, 325 (1990).

\bibitem{ttbar}
R. Kauffman and C.--P. Yuan, {\it Phys. Rev.} {\bf D42}, 956 (1990).

\bibitem{wt}
G.A. Ladinsky and C.--P. Yuan, {\it Phys. Rev.} {\bf D43}, 789 (1991).

\bibitem{ttbarjet}
V. Barger, K. Cheung, T. Han and R. Phillips, {\it Phys. Rev.} {\bf
D42}, 3052 (1990); \\
D. Dicus, J.F. Gunion, L.H. Orr and R. Vega,
 {\it Nucl. Phys.} {\bf B377}, 31 (1992), and references therein.

\bibitem{qqwz}
R.W. Brown, D. Sahdev, K.O. Mikaelian, 
{\it Phys. Rev.} {\bf D20}, 1164 (1979).

\bibitem{zttbar}
 V. Barger, K. Cheung, T. Han, A. Stange and D. Zeppenfeld,
{\it Phys. Rev.} {\bf D46}, 2028 (1991), and references therein.

\bibitem{aleph}
W. Ash et al, {\it Phys. Rev. Lett.} {\bf 58}, 1080 (1987);\\
OPAL Collaboration (R. Akers et al.), {\it Phys. Lett.}
{\bf B327}, 411 (1994);\\
ALEPH Collaboration (D. Buskulic et al.), 
{\it Z. Phys.} {\bf C71}, 357 (1996).


\bibitem{wwpt}
R.N.~Cahn, S.D. Ellis, R. Kleiss, W.J. Stirling,
 {\it Phys. Rev.} {\bf D35}, 1626 (1987).

\bibitem{tag}
V.~Barger, T.~Han, and R.J.~Phillips, 
{\it Phys. Rev.} {\bf D37}, 2005 (1988);\\
R.~Kleiss and W.J.~Stirling, {\it Phys. Lett.} {\bf B200}, 193 (1988);\\
U.~Baur and E.W.~Glover, {\it Nucl. Phys.} {\bf B347}, 12 (1990);\\
D.~Dicus, J.~Gunion, L.~Orr, and R.~Vega,  
{\it Nucl. Phys.} {\bf B377}, 31 (1992).

\bibitem{veto}
V.~Barger, K.~Cheung, T.~Han, and R.J.~Phillips,
{\it Phys.~Rev.} {\bf D42}, 3052 (1990);\\
D.~Dicus, J.~Gunion, and R.~Vega, {\it Phys.~Lett.} {\bf B258},  475 (1991). 

\bibitem{multi}
J.~F.~Gunion, G.~L.~Kane, H.F.-W.~Sadrozinski, A.~Seiden, A.J.~Weinstein, 
and C.--P.~Yuan, {\it Phys. Rev.} {\bf D40}, 2223 (1989).

\bibitem{kane}
G.L.~Kane and C.--P.~Yuan, {\it Phys.~Rev.} {\bf D40}, 2231 (1989).

\bibitem{gap}
V. Barger, R.J.N. Phillips, D. Zeppenfeld,
{\it Phys. Lett.} {\bf 346B}, 106 (1995), and references therein.

\bibitem{toppol}
G.L. Kane, G.A. Ladinsky and C.--P. Yuan, 
{\it Phys. Rev.} {\bf D45}, 124 (1992). 

\bibitem{trivial}
R. Dashen and H. Neuberger, 
{\it Phys. Rev. Lett.} {\bf 50}, 1897 (1983);\\
M. Lindner, {\it Z. Phys.} {\bf C31}, 295 (1986), and references therein.

\bibitem{chiralone}
S. Weinberg, {\it Phys. Rev.} {\bf 166}, 1568 (1968);\\
S. Coleman, J. Wess and B. Zumino, 
{\it Phys. Rev.} {\bf 177}, 2239 (1969);\\
C. Callan, S. Coleman, J. Wess and B. Zumino, 
{\it Phys. Rev.} {\bf 177}, 2247 (1969).

\bibitem{chiral}
J. Gasser and H. Leutwyler, {\it Ann. Phys.} {\bf 158}, 142 (1984); 
{\it Nucl. Phys.} {\bf B250}, 465 (1985).

\bibitem{techni}
E. Farhi and L. Susskind, {\it Phys. Rep.} {\bf 74}, 277 (1981);\\
For a review of the current status of technicolor models, see \\ 
Kenneth Lane, {\it Technicolor}, talk given at International Conference on 
the History of Original Ideas and Basic
Discoveries in Particle Physics, Erice, Italy, 1994, 
e-Print Archive: hep-ph/9501249. 

\bibitem{velt}
H. Veltman and M. Veltman, {\it Acta. Phys. Polon.} {\bf B22}, 669 (1991).

\bibitem{neural}
B. Denby, in La Londe-les-Maures 1992, Proceedings, 
New computing techniques in physics research II, pp. 287-325.
and FERMILAB-Conf-92-121, 1992.

\bibitem{sdcgem}
Some studies beyond the parton level were listed in
Ref.~\cite{kane}. For more references, see, for example,\\ 
Solenoidal Detector Collaboration, {\it 
Technical Design Report}, SDC-92-201, 1992; \\
GEM Collaboration, {\it Letter of Intent}, SSCL-SR-1184, 1991;\\
{\it The CMS detector and physics at the LHC},
by CMS Collaboration (D. Denegri for the collaboration),
CERN-PPE-95-183, Nov 1995,
published in Valencia Elementary Particle Physics
1995, pp. 160-182;\\
{\it The ATLAS detector for the LHC},
by M.A. Parker, CAMBRIDGE-HEP-92-10, 1992,
published in Dallas HEP 1992, pp. 1837-1841.

\end{thebibliography}
\end{document}
\end